\documentclass[10pt]{article}

\usepackage{amssymb}
\usepackage{amsmath}

\newtheorem{theorem}{Theorem}

\newtheorem{corollary}[theorem]{Corollary}

\newtheorem{proposition}{Proposition}

\begin{document}


\title
{Uniqueness for inverse boundary value problems by Dirichlet-to
-Neumann map on subboundaries}

\author{
O.~Yu.~Imanuvilov, \thanks{ Department of Mathematics, Colorado
State University, 101 Weber Building, Fort Collins, CO 80523-1874,
U.S.A. E-mail: {\tt oleg@math.colostate.edu},}\,
M.~Yamamoto\thanks{ Department of Mathematical Sciences, University
of Tokyo, Komaba, Meguro, Tokyo 153, Japan e-mail:
myama@ms.u-tokyo.ac.jp}\,}

\date{}

\maketitle
\begin{abstract}
We consider inverse boundary value problems for elliptic equations
of second order of determining coefficients by Dirichlet-to-Neumann map
on subboundaries, that is, the mapping from Dirichlet data supported
on $\partial\Omega\setminus \Gamma_-$ to Neumann data on
$\partial\Omega\setminus \Gamma_+$.
First we prove uniqueness results in three dimensions under some conditions
such as $\overline{\Gamma_+ \cup \Gamma_-} = \partial\Omega$.
Next we survey uniqueness results in two dimensions for various elliptic
systems for arbitrarily given $\Gamma_- = \Gamma_+$.
Our proof is based on complex geometric optics
solutions which are constructed by a Carleman estimate.
\end{abstract}

\section{Introduction}
Let $\Omega \subset \Bbb R^n$, $n=2,3,...,$  be a bounded domain with
$\partial\Omega\in C^2$.  Let $\nu = \nu(x)$ be the outward unit
normal vector to $\partial\Omega$ at $x$ and $\frac{\partial
v}{\partial\nu} = \nabla v \cdot \nu$.  We consider the conductivity
equation:
\begin{equation}\label{(0.1)}
\nabla \cdot (\gamma\nabla v) =0 \quad \mbox{in}\quad \Omega
\end{equation}
with the Dirichlet boundary condition
\begin{equation}\label{(0.2)}
v = f \quad \mbox{on}\quad \partial\Omega.
\end{equation}
We assume suitable regularity for the conductivity $\gamma$ and
the boundary condition $ f$.  For example, let $\gamma$ belong
to the space $C^2(\overline{\Omega})$ and be a strictly positive function
on $\overline{\Omega}$ and $f \in
H^{\frac{1}{2}}(\partial\Omega)$.  Then there exists a unique
solution $v \in H^1(\Omega)$ to (\ref{(0.1)}) and (\ref{(0.2)}) and
we can define $\gamma\frac{\partial v}{\partial\nu} \in
H^{-\frac{1}{2}}(\partial\Omega)$ (e.g., Lions and Magenes \cite{LM}).
We call the map $f \mapsto \gamma\frac{\partial u}{\partial\nu}$ the
Dirichlet-to-Neumann map.  Since the Dirichlet-to-Neumann map
depends on the conductivity $\gamma$, we denote it by $\Lambda_{\gamma}$:
$$
\Lambda_{\gamma}f =\gamma\frac{\partial v}{\partial\nu}, \quad
\mathcal{D}(\Lambda_{\gamma}) =H^{\frac{1}{2}}(\partial\Omega).
$$
This article is concerned with the following boundary value problem:
\\
{\bf Inverse boundary value problem}. {\it Determine the conductivity
$\gamma(x)$  from the Dirichlet-to-Neumann map $\Lambda_{\gamma}$.}
\\

This inverse problem is a theoretical basis for example, for the
electrical impedance tomography and often called
Calder\'on's  problem.  As for applications, we refer for example to
Cheney, Isaacson, and Newell \cite{CIN},  Holder \cite{H},
Issacson, M\"uller, and Siltanen \cite{IMS}, Jordana, Gasulla, and
Pall\`as-Areny \cite{JGP}.

 In this formulation, the information
for the inverse problem is the operator
$\Lambda_{\gamma}$ itself, which is all the pairs of Dirichlet
input data $f$ and the corresponding derivatives
$\gamma\frac{\partial v}{\partial\nu}$.  In other
words, the formulation of the inverse problem requires infinitely
many repeats of input-output manipulations.

Since the operator $\Lambda_{\gamma}:
H^{\frac{1}{2}}(\partial\Omega) \longrightarrow
H^{-\frac{1}{2}}(\partial\Omega)$ is bounded for fixed $\gamma$ and
$C^{\infty}(\partial\Omega)$ is dense in
$H^{\frac{1}{2}}(\partial\Omega)$, we can restrict $f \in
C^{\infty}(\partial\Omega)$ for $\Lambda_{\gamma}$. More precisely,
for $\gamma_1, \gamma_2 \ge 0, \in C^2(\overline\Omega)$, we have
$$
\{ (f, \Lambda_{\gamma_1}f); \thinspace f \in
C^{\infty}(\partial\Omega) \} =\{ (f, \Lambda_{\gamma_2}f);
\thinspace f \in C^{\infty}(\partial\Omega) \}
$$
if and only if
$$
 \{ (f, \Lambda_{\gamma_1}f); \thinspace
f \in H^{\frac{1}{2}}(\partial\Omega)\} = \{ (f,
\Lambda_{\gamma_2}f); \thinspace f \in
H^{\frac{1}{2}}(\partial\Omega)\}.
$$
Throughout this article, we always consider that Dirichlet input
data $f$ from $H^{\frac{1}{2}}(\partial\Omega)$.

{\bf Inverse problem for stationary equations}\\
In general, an inverse problem of determining a spatially varying
coefficient in partial differential equation in a spatially
$n$-dimensional bounded domain, is called a coefficient inverse
problem. For coefficient inverse problems for time-dependent
differential equations such as parabolic equations and hyperbolic
equations, one can take extra time-dependent boundary data at most
finite times after changing suitable initial values to establish the
stability and the uniqueness. In this case, one unknown coefficient
depends on $n$ spatial coordinates, while the extra data depend on
the time variable and $n-1$ spatial variables (because data are
limited to the boundary of the spatial domain), and so the data
depend on $n$ variables.  Thus the data are not under-determinative
and we can expect the uniqueness.  In fact, Bukhgeim and Klibanov
\cite{BK} proposed a methodology for proving the uniqueness.   Their
method is based on a Carleman estimate which is a weighted
$L^2$-estimate of solutions to partial differential equations.
Later Imanuvilov and Yamamoto established global stability results
(\cite{IY1998}, \cite{IY2001}, \cite{IY2001-1}: here we refer only
to these three papers and \cite{Ya}, and there are many remarkable
works but we do not describe because our main topics in this article
are different).  This formulation requires finite times
(sometimes single) of repeats of
observations. On the other hand, for a coefficient inverse problem
for  time-independent partial differential equations, there are no
compensating variable such as time, and the current formulation by
the inverse boundary value problem is the main theoretical setting
of a coefficient inverse problem, even though it requires infinitely
many times of measurements. There are various realizations for
numerical computation but this is not our scope here.

{\bf Mathematical subjects.}\\
For the inverse boundary value problem, we state mathematical
subjects as follows:
\begin{itemize}
\item
{\bf Uniqueness.}\\
Does $\Lambda_{\gamma_1} = \Lambda_{\gamma_2}$ imply $\gamma_1 =
\gamma_2$?
\item
{\bf Stability.}\\
If $\Lambda_{\gamma_1} - \Lambda_{\gamma_2}$ is small in view of a
suitable norm, then can we conclude that $\gamma_1 - \gamma_2$ is
small in some norm?
\item
{\bf Reconstruction.}\\
Given a data set from $\{ (f, \Lambda_{\gamma}f); \thinspace f \in
H^{\frac{1}{2}}(\partial\Omega)\}$, establish an algorithm for finding
$\gamma$. Here we assume that the data set really comes from some
conductivity $\gamma$, and do not discuss the existence of $\gamma$
yielding the data set.
\end{itemize}

For the uniqueness, there is a large difference between the two
dimensional case (i.e., $n=2$) and the higher dimensional case
(i.e., $n\ge 3$), in view of the degree of freedom of data.  More
precisely, we can explain as follows.  In $n$-dimensional case, an
unknown function depends on $n$ number variables. On the other
hand, our boundary data are the set $\{ (f, \Lambda_{\gamma}f);
\thinspace f \in H^{\frac{1}{2}}(\partial\Omega)\}$ and so are pairs
of Dirichlet data and Neumann data which both depend on $(n-1)$
number of variables and it can considered as data depending on
$2(n-1)$ variables.  Therefore
\\
{\bf Case $n=2$}: we have $2(n-1)=n$. The inverse problem is
formally determining.
\\
{\bf Case $n\ge 2$}: we have $2(n-1)>n$.  The inverse problem is
overdetermining.
\\
\vspace{0.2cm}

Naturally the methods for cases $n=2$ and $n\ge 3$ are different.
Moreover, as we will explain in later sections, in the two
dimensional case, one can construct a larger set of complex
geometrical optics solutions, which are the key ingredient for the
uniqueness of the inverse boundary value problem.
\\

{\bf Inverse boundary value problem for Schr\"odinger equations}\\
We consider the second order elliptic operator of the form
$$
L_q(x,D)u=\Delta u + q(x)u = 0, \quad  \mbox{in}\quad \Omega,
$$
which is called a Schr\"odinger equation with potential $q$.
Then we note that the
uniqueness for determination of conductivity can be derived from
the uniqueness for determination of  some potential $q$  for the Schr\"odinger
equation from the following Dirichlet-to-Neumann map
\begin{equation}\label{*}
\Lambda (q) f =  \frac{\partial u} {\partial
\nu}\vert_{\partial\Omega},
\end{equation}
where
\begin{equation}\label{**} L_q(x,D)u=
0\hbox{ in }\Omega,\,\,u\vert_{\partial\Omega}=f,\quad \ u\in
H^1(\Omega).
\end{equation}

We note that if $0$ is not an eigenvalue of the Schr\"odinger equation,
then the operator $\Lambda (q):
H^{\frac{1}{2}}(\partial\Omega)\rightarrow
H^{-\frac{1}{2}}(\partial\Omega)$ is correctly defined. Then the uniqueness in
determining $\gamma$ in the conductivity equation by
$\Lambda_{\gamma}$ follows from the uniqueness  of determination for the
potential $q$ by the Dirichlet-to-Neumann map $\Lambda (q)$:
Indeed, if $\gamma_1,\gamma_2\in C^2(\overline{\Omega})$ be strictly positive functions in $\overline \Omega$ and $\Lambda_{\gamma_1}=\Lambda_{\gamma_2}$,
then it is known that (see e.g., Alessandrini \cite{Ale})
\begin{equation}\label{(0.4)}
\gamma_1 = \gamma_2, \quad \frac{\partial\gamma_1}{\partial\nu} =
\frac{\partial\gamma_2}{\partial\nu} \quad \mbox{on $\partial\Omega$},
\end{equation}

Let $v$ solve the conductivity equation
$$
\nabla \cdot (\gamma\nabla v) =0 \quad \mbox{in $\Omega$}.
$$
Then, setting
$q=-\frac{\Delta\sqrt{\gamma}}{\sqrt{\gamma}}$ and $\thinspace
u=\sqrt{\gamma}v$, we have
$$
L_{q}(x,D)u=\Delta u + q(x)u =0 \quad \mbox{in $\Omega$}.
$$
This observation combined with (\ref{(0.4)}) implies
$$
\Lambda \left(-\frac{\Delta\sqrt{\gamma_1}}{\sqrt{\gamma_1}}\right)
=\Lambda\left(-\frac{\Delta\sqrt{\gamma_2}}{\sqrt{\gamma_2}}\right).
$$

Therefore, by the uniqueness result of $q$ by the Dirichlet-to-Neumann map
(\ref{*}) and (\ref{**}), we can derive the equality
$\frac{\Delta\sqrt{\gamma_1}}{\sqrt{\gamma_1}}
= \frac{\Delta\sqrt{\gamma_2}}{\sqrt{\gamma_2}}$, that is,
\begin{equation}\label{(0.3)}
\Delta w - \frac{\Delta\sqrt{\gamma_2}}{\sqrt{\gamma_2}}w = 0
\quad\mbox{in $\Omega$}
\end{equation}
with $w =\sqrt{\gamma_1} - \sqrt{\gamma_2}$. By (\ref{(0.4)}) we
have
\begin{equation}\label{***}
w\vert_{\partial\Omega}=\frac{\partial w}{\partial \nu}\vert_{\partial\Omega}=0.
\end{equation}
Applying to (\ref{(0.3)}) and (\ref{***}) the classical unique
continuation for the second order
elliptic operator (see e.g., Chapter XXVIII, \S 28.3 of \cite{Ho1},
Corollary 2.9, Chapter XIV of \cite{Taylor}), we obtain
$\gamma_1\equiv\gamma_2.$
Here for the
inverse conductivity problem, compared to the regularity of $q$ in
the Schr\"odinger equation, we need to increase the regularity
assumptions by order $2$ for unknown coefficients. Therefore in the succeeding
parts, we often consider the inverse boundary value problem for
Schr\"odinger type of equations.
\\

{\bf Existing results on the uniqueness.}\\
As for researches on the uniqueness, we can point out two main
streams.
\begin{itemize}
\item
Dirichlet-to-Neumann map on subboundaries: reduction of subboundaries
where we consider Dirichlet inputs and observations of Neumann data.
\item
relaxation of regularity of unknown coefficients.
\end{itemize}

Now, according to these streams, we list some important works on the
uniqueness mainly before 2010.  Here we do not intend any complete lists.
\\
First of all, for the case of the Dirichlet-to-Neumann map on the
whole boundary, we refer to
\\
{\bf Case $n\ge 3$}:\\
Sylvester and Uhlmann  \cite{SU} proved the uniqueness for $\gamma
\in C^2(\overline{\Omega})$, and  P\"aiv\"arinta,
Panchenko and Uhlmann \cite{PPU} for
$\gamma \in W^{\frac{3}{2},\infty}(\Omega)$,
Brown and Torres \cite{BT} for $\gamma \in W^{\frac{3}{2},p}(\Omega)$
with $p>2n$.
Recently Haberman and Tataru \cite{HT} proved the local
uniqueness within Lipschitz conductivities $\gamma$ under the condition
that $\left\Vert \frac{\nabla\gamma}
{\gamma}\right\Vert_{L^{\infty}(\Omega)}$ is sufficiently small.
\\
{\bf Case $n=2$:}\\
Nachman \cite{N} proved the uniqueness for $\gamma\in
C^2(\overline{\Omega})$.  Finally Astala and P\"aiv\"arinta
\cite{AP} established the uniqueness within $\gamma \in
L^{\infty}(\Omega)$, which is a very sharp uniqueness result, but no
corresponding results are known for the case $n\ge 3$. On the other
hand, Bukhgeim \cite{Bu} proves the uniqueness for the Schr\"odinger
equations within $q \in L^{\infty}(\Omega)$.
\\
\vspace{0.2cm}

In contrast to the above works, we state the formulation with
Dirichlet-to-Neumann map on subboundaries.
Let  $\Gamma_+, \Gamma_- \subset
\partial\Omega$ be subboundaries. We set

\begin{equation}\label{(0.5)}
\Lambda (q,\Gamma_-,\Gamma_+)f= \frac{\partial
u}{\partial\nu}\vert_{\partial\Omega\setminus\Gamma_+},
\end{equation}
where
\begin{equation}
\Delta u + qu =0 \quad \mbox{in}\quad \Omega, \quad u\vert_{
\Gamma_-} = 0,\quad u\vert_{\partial\Omega\setminus \Gamma_-} = f.
\end{equation}
We regard $\partial\Omega\setminus \Gamma_-$ and
$\partial\Omega\setminus \Gamma_+$ as input subboundary and output
subboundary.

As for the conductivity equation, we can define the
Dirichlet-to-Neumann map on subboundaries by
$$
\Gamma_{\gamma}f = \frac{\partial u}{\partial\nu}\vert_{\partial\Omega
\setminus \Gamma_+},
$$
where $\nabla\cdot(\gamma\nabla u) = 0$ in $\Omega$ and
$u\vert_{\partial\Omega\setminus \Gamma_-}=f$,
$u\vert_{\Gamma_-}=0$.

We note that the Dirichlet-to-Neumann map on the whole boundary
corresponds to the
case $\Gamma_+ = \Gamma_- = \emptyset$. It is natural to discuss
the uniqueness by $\Lambda(q,\Gamma_-,\Gamma_+)$ with larger $\Gamma_+,
\Gamma_-$.
We list main works published before 2010.\\
{\bf Case $n \ge 3$}:\\
Bukhgeim and Uhlmann \cite{BuU} proved the uniqueness within the
class of $\gamma \in C^2(\overline{\Omega})$ if $\Gamma_- =
\emptyset$ and $\Gamma_+$ is some  specific part of
$\partial\Omega$. Isakov \cite{I} proved the uniqueness if
$\Gamma_+$ and $\Gamma_-$ are included in planes or spheres. Knudsen
\cite{K} improved the uniqueness by Bukhgeim and Uhlmann \cite{BuU}
to the class of $\gamma \in
C^{\alpha+\frac{3}{2}}(\overline{\Omega})$ with some $\alpha>0$.
Finally we refer to Kenig,  Sj\"ostrand  and Uhlmann \cite{KSU}
which established the uniqueness for some specially defined  sets
$\Gamma_\pm.$
\\
{\bf Case $n = 2$}:\\
Imanuvilov, Uhlmann and Yamamoto \cite{IUY}
first proved the uniqueness by Dirichlet-to-Neumann map on arbitrarily
given subboundary provided that $\Gamma_+ = \Gamma_-$.
\\
\vspace{0.2cm}

The proposes of this article are\\
(i) to provide uniqueness results in the three dimensional case by a
simpler argument and the uniqueness is sharper than e.g., Kenig {\it
et al.} \cite{KSU}.
\\
(ii) to simplify the existing proofs: we do not need advanced tools
for example from the microlocal analysis (see \cite{KSU}).
\\
(iii) to describe uniqueness results by Dirichlet-to-Neumann map on
subboundaries which have been recently obtained by the authors and
colleagues.

In the succeeding sections, we will give more detailed references
related to the topics of this article.
\\
\vspace{0.3cm}

{\bf Key to the proof of the uniqueness.}\\
For convenience, here we explain the key to the proof which has been
an essential idea since Sylvester and Uhlmann  \cite{SU}.
\begin{itemize}
\item
We consider a family $u_1=u_1(\tau)(x)$, $\tau > 0$ of solutions
which are parameterized by $\tau > 0$:
\begin{equation}\label{(0.6)}
L_{q_1}(x,D)u_1=\Delta u_1 + q_1 u_1 =0 \quad \mbox{in $\Omega$}, \quad
u_1\vert_{\Gamma_-} =0.
\end{equation}
\item
For $u_1(\tau)$, we construct the solution $u_2 = u_2(\tau)$ to
$$
L_{q_2}(x,D)u_2=\Delta u_2 + q_2 u_2 = 0 \quad \mbox{in $\Omega$},\quad
u_2\vert_{\partial\Omega} = u_1\vert_{\partial\Omega}.
$$
By $\Lambda (q_1,\Gamma_-,\Gamma_+) = \Lambda(q_2,\Gamma_-,\Gamma_+)$, we have
$\frac{\partial u_1}{\partial\nu} =\frac{\partial u_2}{\partial\nu}$
on $\partial\Omega\setminus\Gamma_+$. Setting $u = u_1-u_2$, we obtain
\begin{equation}\label{(0.7)}
L_{q_2}(x,D)u_2=\Delta u + q_2u = (q_2-q_1)u_1 \quad \mbox{in $\Omega,$}
\end{equation}
\begin{equation}\label{(0.8)}
u\vert_{\partial\Omega} =0, \quad \frac{\partial
u}{\partial\nu}\vert _{\partial\Omega\setminus\Gamma_+} =0.
\end{equation}
\item
We consider another family $v=v(\tau)(x)$, $\tau > 0$ of solutions
which are parameterized by $\tau > 0$:
\begin{equation}\label{(0.9)}
L_{q_2}(x,D)v=\Delta v + q_2v =0 \quad \mbox{in $\Omega$}, \quad
v\vert_{\Gamma_+} =0.
\end{equation}
\item
Multiplying (\ref{(0.7)}) with $v(\tau)$ and using (\ref{(0.6)}),
(\ref{(0.8)}) and (\ref{(0.9)}), we obtain
$$
0=\int_{\Omega} v L_{q_2}(x,D)u dx = \int_{\Omega} (q_2-q_1)vu_1 dx,
$$
that is,
$$
\int_{\Omega}(q_1-q_2)(x)v(\tau)(x)u_1(\tau)(x)dx=0
$$
for all $\tau > 0$.
\end{itemize}
Thus the uniqueness is reduced to the completeness of products of
solutions, that is, the density in $L^2(\Omega)$ of
$\{ v(\tau)u_1(\tau)\}_{\tau>0}$.
The works on the uniqueness in $\Omega$ since \cite{SU}
have relied on how to construct the
families $u_1(\tau), v(\tau)$, $\tau > 0$ satisfying such
completeness. For it, the idea for $u_1(\tau)$ and $v(\tau)$ is
solutions to the elliptic equations under consideration in the form:
\begin{equation}\label{(0.11)}
e^{\tau\Phi(x)}(a + o(1)) \quad \mbox{as $\tau\to\infty$
}
                                            \end{equation}
with suitably chosen phase function $\Phi$ and solutions to the
transport equation $a.$  Solution in the form of
(\ref{(0.11)}) are called (complex) geometric optics solutions.
The paper by Kenig, Sj\"ostrand and Uhlmann \cite{KSU} uses a
Carleman estimate for constructing complex geometric optics
solutions in higher dimensions and see also Bukhgeim \cite{Bu}.
We can understand a Carleman estimate as weighted $L^2$-estimate.

Careful choices of weight functions in Carleman estimates yield the
uniqueness results for various elliptic systems by
Dirichlet-to-Neumann map limited on some subboundaries.  We have
developed such relevant Carleman estimates mainly in two dimensions,
and in the succeeding sections, we clarify such Carleman estimates.

The article is composed of the following sections.
\begin{itemize}
\item
\S1. Introduction.
\item
\S2. A key Carleman estimate for the proof of the uniqueness in
 \S3.
\item
\S3. Uniqueness in the three dimensional case: We demonstrate how
our method can produce the uniqueness with Dirichlet-to-Neumann map
on subboundaries, and the proof for the uniqueness is concise and based on
Carleman estimate and the Radon transform.  Moreover
our uniqueness generalizes the results by Bukhgeim and Uhlmann
\cite{BuU}, Kenig, Sj\"ostrand and Uhlmann \cite{KSU}.
\item
\S4. Survey on two dimensional inverse boundary value problems by
Dirichlet-to-Neumann map on an arbitrary subboundary.
\item
\S5 Calder\'on's problem for semilinear elliptic equations.
\item \S6 {Uniqueness by Dirichlet-to-Neumann maps for
Lam\'e equations and the Navier-Stokes equations.}
\item
\S7. Appendix.
\end{itemize}

This section is closed with explanations of notations which are
used throughout this paper.

{\bf Notations.}
Let $x = (x_1,..., x_n) \in \Bbb R^n$ and $x' = (x_1, x_2, ...,
x_{n-1})$, and $\Bbb S^{n-1} = \{ x\in \Bbb R^n; \thinspace
\vert x\vert = 1 \}$.
Henceforth let ${\Bbb N}_+ = \Bbb N \cup \{0\}$,
$\partial_x^\beta = \partial_{x_1}^{\beta_1}
\partial_{x_2}^{\beta_2}\cdots \partial_{x_n}^{\beta_n}$, $\beta=(\beta_1,\dots,\beta_n) \in (\Bbb N_+)^n$
and $\vert \beta\vert=\beta_1+\dots+\beta_n$. We set  $i=\sqrt{-1}$,
$x_1, x_2 \in {\Bbb R}^1$, $z=x_1+ix_2$, $\overline{z}$ denotes the
complex conjugate of $z \in \Bbb C$. We identify $x = (x_1,x_2) \in
{\Bbb R}^2$ with $z = x_1 +ix_2 \in {\Bbb C}$ and
$\xi=(\xi_1,\xi_2)$ with $\zeta=\xi_1+i\xi_2$. $\partial_z = \frac
12(\partial_{x_1}-i\partial_{x_2})$, $\partial_{\overline z}=
\frac12(\partial_{x_1}+i\partial_{x_2})$, $D = (D_1,\dots,
D_n)=\left( \frac{1}{i}\partial_{x_1}, \frac{1}{i}\partial_{x_2},
\dots , \frac{1}{i}\partial_{x_n}\right),
D^\beta=D_1^{\beta_1}\cdots D_n^{\beta_n}$ $\partial_\zeta=\frac
12(\partial_{\xi_1}-i\partial_{\xi_2}),
\partial_{\overline \zeta}=
\frac12(\partial_{\xi_1}+i\partial_{\xi_2}).$ Denote by
$B(x,\delta)$ a ball centered at $x$ of radius $\delta.$ For a
normed space $X$, by $o_X(\frac{1}{\tau^\kappa})$ we denote a
function $f(\tau,\cdot)$ such that $ \Vert
f(\tau,\cdot)\Vert_X=o(\frac{1}{\tau^\kappa})\quad \mbox{as}
\,\,\vert \tau\vert\rightarrow +\infty$. By $H_\phi$ we denote the
Hessian matrix  with the entries $\frac{\partial^2\phi}{\partial
x_i\partial x_j}.$ The tangential derivative on the boundary is
given by $\partial_{\vec\tau}=\nu_2\frac{\partial}{\partial x_1}
-\nu_1\frac{\partial}{\partial x_2}$, where $\nu=(\nu_1, \nu_2)$ is
the unit outer normal to $\partial\Omega$. For the functions
$p(x,\xi)$ and $q(x,\xi)$ we define the Poisson bracket:
$\{p,q\}(x,\xi)=\sum_{i=1}^n\frac{\partial p}{\partial
x_i}\frac{\partial q}{\partial \xi_i}-\frac{\partial p}{\partial
\xi_i}\frac{\partial q}{\partial x_i}.$

Let $P(x,D)=\sum_{\vert \beta\vert\le k}a_\beta (x) D^\beta$ be the
differential operator of order $k.$ Denote by
$p(x,\xi)=\sum_{\vert\beta\vert=k}a_k(x)\xi^\beta$ the principal symbol
of this operator.

We call $b(x_1,x_2)$ antiholomorphic if $(\frac{\partial}{\partial
x_1}-i\frac{\partial }{\partial x_2})b=0.$  In the Sobolev space
$H^k(\Omega)$ we introduce the following norm
$$
\Vert u\Vert_{H^{k,\tau}(\Omega)} = (\Vert u\Vert_{H^k(\Omega)}^2 +
\vert \tau\vert^{2k}\Vert u\Vert^2_{L^2(\Omega)})^{\frac{1}{2}}.
$$
For any strictly positive function $\rho$ in $\Omega$, we introduce
the normed space $L^2_{\rho}(\Omega)=\{ w(x)\vert \int_\Omega \rho
w^2dx <\infty\}$ with the norm $\Vert
w\Vert_{L^2_\rho(\Omega)}=(\int_\Omega \rho w^2dx)^\frac 12.$

For the operators $L_+$ and $L_-$, we denote by $[L_+,L_-]$ the
commutator of these operators: $[L_+,L_-]=L_+L_--L_-L_+.$

\section{Carleman estimates for the Schr\"odinger equation}

As we mentioned in the introduction, in order to construct the
remaining term in complex geometric optics solution we will use the
technique  based on Carleman estimates.
The Carleman estimate itself was introduced by Carleman \cite{Carl}
for the purpose of proving the uniqueness  for the Cauchy problem for
the system of elliptic equations.  The Carleman estimate is some a priori
estimate depending on
parameter  and some weight function. In this section we concentrate
 on the case of the second-order elliptic operator whose principal
 part is the Laplace operator. For the general theory of Carleman
 estimates see  e.g. \cite{Ho1}.

 Consider the second order elliptic equation in domain
 \begin{equation}\label{E1}
P(x,D)u= \Delta u+\sum_{j=1}^nb_j\frac{\partial u}{\partial
 x_j}+cu=f\quad\mbox{in}\,\,\Omega, \quad u\vert_{\partial
 \Omega}=0.
 \end{equation}

The principal symbol of this operator is $ p(x,\xi)=-\vert
\xi\vert^2$ where $\xi=(\xi_1,\dots, \xi_n).$
\\
{\bf Definition 1.} {\it We say that the function $\varphi$ is
 pseudoconvex with respect to the principal symbol $p$ if $\nabla \varphi\ne 0$ on
 $\overline \Omega$ and
 \begin{eqnarray}\label{weak}
 \{ \overline p(x,\xi-i\tau \nabla \varphi(x)), p(x,\xi+i\tau\nabla \varphi(x))\}/i\tau=2\sum_{i,j=1}^n
 \frac{\partial^2\varphi}{\partial x_i\partial
 x_j}p^{(j)}(x,\zeta)\overline{p^{(k)}(x,\zeta)}> 0\nonumber\\
 \quad \mbox{on}\,\, \{(x,\xi,\tau)\in \Omega\times
(\Bbb R^n\setminus\{0\}) \times \Bbb R_+^1\vert
p(x,\xi+i\tau\nabla \varphi)=0\},\quad
 \zeta=\xi+i\tau\nabla\varphi.
 \end{eqnarray}
}

The construction of the pseudoconvex function for the second-order
elliptic operator is a very easy task: if $\phi\in C^2(\overline \Omega)$
and $\nabla \phi(x)\ne 0$ for any $x$ from $\overline\Omega$ then for all
sufficiently large positive $\lambda$ the function $\varphi=
e^{\lambda\phi}$ is pseudoconvex with respect to a principal symbol of
this operator.

We introduce the following subsets of the boundary of domain
$\Omega$:
\begin{eqnarray*}
&& \partial \Omega_-=\{x\in
\partial\Omega\vert\frac{\partial \varphi}{\partial\nu}(x)< 0\},
\quad \partial \Omega_0=Int \{x\in
\partial\Omega\vert\frac{\partial \varphi}{\partial\nu}(x)=0\},\\
&& \partial \Omega_+=\{x\in
\partial\Omega\vert\frac{\partial \varphi}{\partial\nu}(x)> 0\}.
\end{eqnarray*}

A typical Carleman estimate  for the Schr\"odinger equation
(see e.g. \cite{I2}) is given by the following proposition:

\begin{proposition} Let $b_j, c\in L^\infty(\Omega)$ and $\varphi$ be
pseudoconvex function with respect to the principal symbol of the
operator $P(x,D).$  Then there exist constants $\tau_0$ and $C$
independent of $\tau$ such that for all $\tau\ge \tau_0$ the
following estimate holds true:
\begin{eqnarray}\label{Lolo1}
\tau\Vert u
e^{\tau\varphi}\Vert^2_{H^{1,\tau}(\Omega)}
+ \int_{\partial\Omega}\vert
\frac{\partial u}{\partial \nu}\vert^2e^{2\tau\varphi}d\sigma
+\tau\int_{\partial\Omega_-\cup \partial
\Omega_0}\vert\frac{\partial \varphi}{\partial\nu}\vert\vert
\frac{\partial u}{\partial \nu}\vert^2 e^{2\tau\varphi}d\sigma
\nonumber\\
\le C\left(\Vert
(P(x,D)u)e^{\tau\varphi}\Vert^2_{L^2(\Omega)}+
\tau\int_{\partial\Omega_+}\vert \frac{\partial u}{\partial
\nu}\vert^2 e^{2\tau\varphi}d\sigma\right)
\end{eqnarray}
for $u$ satisfying (\ref{E1}).
\end{proposition}

For the construction of the complex geometric optics solutions for the
operator $P(x,D)$  we will use as a weight function the real part of
the function $\Phi$ which solves the Eikonal equation.

\begin{proposition}\label{gopnikk}
Let a function $\Phi(x)=\varphi+i\psi\in
C^2(\overline\Omega)$ be a solution to the Eikonal equation
\begin{equation}\label{iii-}
(\nabla \Phi,\nabla \Phi)=0\quad \mbox{on}\,\,\Omega.
\end{equation}
Then $(x,\tau\nabla\psi(x),\tau)$ belongs to the set
$\{(x,\xi,\tau)\in \Omega\times (\Bbb R^n\setminus\{0\})
\times \Bbb R_+^1\vert p(x,\xi+i\tau\nabla \varphi)=0\}$ and
  \begin{equation}\label{iiii}\{ \overline
p(x,\tau\nabla\psi-i\tau \nabla \varphi(x)), p(x,\tau\nabla
\psi+i\tau\nabla \varphi(x))\}/i\tau=0.
 \end{equation}
\end{proposition}

{\bf Proof.} The Eikonal equation (\ref{iii-}) is equivalent to the
following two equalities
\begin{equation}\label{Eq3}
\sum_{j=1}^n\frac{\partial \varphi}{\partial x_j}\frac{\partial
\varphi}{\partial x_j}=\sum_{j=1}^n\frac{\partial \psi}{\partial
x_j}\frac{\partial \psi}{\partial x_j}\quad\mbox{and}\quad
\sum_{j=1}^n\frac{\partial \varphi}{\partial x_j}\frac{\partial
\psi}{\partial x_j}=0.
\end{equation}
Equations  (\ref{Eq3}) immediately imply
\begin{equation}\label{Eq2}
(x,\tau\nabla \psi(x),\tau)\in \{(x,\xi,\tau)\in \Omega\times
(\Bbb  R^n\setminus\{0\}) \times \Bbb R_+^1\vert
p(x,\xi+i\tau\nabla \varphi)=0\}.
\end{equation}
Differentiating equations (\ref{Eq3}) with respect to $x_k$ and then taking
the sum over $k$, we have
 \begin{equation}\label{EE}
 H_\varphi \nabla \varphi =H_\psi \nabla \psi, \quad H_\varphi \nabla
 \psi=-H_\psi \nabla
 \varphi.
 \end{equation}
 The right-hand side of (\ref{weak}) can be written as
\begin{equation}\label{EEE}
\{ \overline p(x,\xi-i\tau \nabla \varphi(x)), p(x,\xi+i\tau\nabla
\varphi(x))\}/i\tau=
 4((H_\varphi\nabla\xi,\nabla\xi)+\tau^2 (H_\varphi\nabla
 \varphi,\nabla\varphi)).
 \end{equation}
Hence, using the equalities (\ref{EE}) we rewrite (\ref{EEE}) for
$\xi=\tau\nabla\psi(x)$ as
\begin{eqnarray}\label{EEEE}
\tau^2(H_\varphi \nabla\psi,\nabla\psi)+\tau^2 (H_\varphi\nabla
\varphi,\nabla\varphi)=-\tau^2(H_\psi\nabla\varphi,\nabla\psi)
+\tau^2
(H_\varphi\nabla \varphi,\nabla\varphi)=\nonumber\\
-\tau^2(\nabla\varphi,H_\psi\nabla\psi)+\tau^2 (H_\varphi\nabla
\varphi,\nabla\varphi)=-\tau^2(\nabla\varphi,H_\varphi\nabla\varphi)+\tau^2
(H_\varphi\nabla \varphi,\nabla\varphi)=0.
\end{eqnarray}
Equalities (\ref{EEE}) and (\ref{EEEE}) imply (\ref{iiii})
immediately. $\blacksquare$

Proposition \ref{gopnikk} implies that the real part of a solution
of the Eikonal equation does not satisfy the pseudoconvexity
condition (\ref{weak}).  Thus we relax this pseudoconvexity condition
as follows:
\\
{\bf Definition 2.} {\it We say that the function $\varphi$ is weakly
 pseudoconvex with respect to the symbol $p$ if $\nabla \varphi\ne 0$ on
 $\overline \Omega$ and
 \begin{eqnarray}\label{ogogo}
 \{ \overline p(x,\xi-i\tau \nabla \varphi(x)),
p(x,\xi+i\tau\nabla \varphi(x))\}/i\tau=2\sum_{i,j=1}^n
 \frac{\partial^2\varphi}{\partial x_i\partial
 x_j}p^{(j)}(x,\zeta)\overline{p^{(k)}(x,\zeta)}\ge 0\nonumber\\
 \quad \mbox{on}\,\, \{(x,\xi,\tau)\in \Omega\times
(\Bbb R^n\setminus\{0\})\times \Bbb R_+^1\vert
p(x,\xi+i\tau\nabla \varphi)=0\},\quad
\zeta=\xi+i\tau\nabla\varphi.
\end{eqnarray}
}
If the real part  of a solution of the Eikonal equation is weakly
pseudoconvex with respect to the principal symbol of the elliptic
operator, then one can construct the complex geometric optics solution for
large parameter $\tau>0$.   In some cases we
need to construct the complex geometric optics solutions for the
large negative values of parameter $\tau$ for the same weight
function as well. In \cite {KSU}, in order to deal with this
situation, the notion of the {\it limiting Carleman
weight} was introduced.
\\
{\bf Definition 3.} {\it We say that the function $\varphi$ is a
limiting Carleman weight for the operator $P(x,D)$
 if $\nabla \varphi\ne 0$ on
 $\overline \Omega$ and
 \begin{eqnarray}
 \{ \overline p(x,\xi-i\tau \nabla \varphi(x)), p(x,\xi+i\tau\nabla \varphi(x))\}/i\tau=2\sum_{i,j=1}^n
 \frac{\partial^2\varphi}{\partial x_i\partial
 x_j}p^{(j)}(x,\zeta)\overline{p^{(k)}(x,\zeta)}=0\nonumber\\
 \quad \mbox{on}\,\, \{(x,\xi,\tau)\in \Omega\times
(\Bbb R^n\setminus\{0\}) \times \Bbb R_+^1\vert
p(x,\xi+i\tau\nabla \varphi)=0\},\quad
 \zeta=\xi+i\tau\nabla\varphi.
 \end{eqnarray}
}

Obviously any limiting Carleman estimate for the operator $P(x,D)$
is weakly pseudoconvex  with respect to the principal symbol of this
operator.

Another important property of the limiting Carleman weight is the
following.

\begin{proposition}(\cite{KSU}) Let $\Omega\subset \Bbb R^3$ be a symply connected domain with the smooth boundary and $\varphi$ be a limiting Carleman weight in
$\Omega$. Then there exist a family of functions $\psi$ such that the
function $\Phi=\varphi+i\psi$ solves the Eikonal equation in
$\Omega.$
\end{proposition}

Later we will use the following limiting Carleman weights;

{\bf Example 1 of the limiting Carleman weights.} Let $ \vec v_1,
\vec v_2\in \Bbb R^3$ be to vectors such that $\vert \vec
v_1\vert=\vert \vec v_2\vert\ne 0$ and $(\vec v_1,\vec v_2)=0$. The
function $\Phi=(\vec v_1,x)+i(\vec v_2,x)$ solves the Eikonal
equation and the function $\varphi=(\vec v_1,x)$ is the limiting
Carleman weight.

{\bf Example 2 of the limiting Carleman weights.} Let $\Phi=\ln
r+i\theta$ where $r,\varphi,\theta$ are the spherical coordinates.
The function $\Phi$ is a solution to the Eikonal equation and the
function $\ln(r)$ is a limiting Carleman weight, provided that the
origin and the domain $\Omega$ can be separated by some plane.

In two-dimensional case we can give the complete description of the
 solutions of the Eikonal equations.
We have
\begin{proposition}\label{gopnik11}
Let $\Omega\subset \Bbb R^2$ and a function $\Phi(x)=\varphi+i\psi\in
C^2(\overline\Omega)$ be a solution to the Eikonal equation
\begin{equation}\label{iii}
(\nabla \Phi,\nabla \Phi)=0\quad \mbox{on}\,\,\Omega.
\end{equation}
Then  the function $\Phi$ is either holomorphic or antiholomorphic
in $\Omega.$
\end{proposition}

{\bf Proof.} The short computations give the formula
$$
e^{-\tau\Phi}\Delta
e^{\tau\Phi}=(\nabla\Phi,\nabla\Phi)\tau^2+\Delta\Phi \tau =
e^{-\tau\Phi}4\partial_z\partial_{\overline z}e^{\tau\Phi}=
4\partial_z\Phi \partial_{\overline z}\Phi \tau^2+\Delta\Phi \tau.
$$
Therefore, if $\Phi$ is a solution to the Eikonal equation,
then we obtain form the above equality that
$$
\partial_z\Phi \partial_{\overline z}\Phi=0\quad\mbox{in}\,\,\Omega.
$$
From this equality, the statement of the proposition follows
immediately.
$\blacksquare$

{\bf Example 3 of the limiting Carleman weights.} Let $n=2$ and
$\Phi$ be a holomorphic or an antiholomorphic function such that
$\nabla\Phi\ne 0$. Then the real part of  the function $\Phi$  is
the limiting Carleman weight.
\\

{\bf Example 4 of the limiting Carleman weights.} Let $\varphi$ be
the limiting Carleman weight and the function $\Phi=\varphi+i\psi$
solves the Eikonal equation. Then the function $\varphi(x/\vert
x\vert^2)$ is the limiting Carleman weight and the function
$\Phi\circ (\frac{x}{\vert x\vert^2})$ solves the Eikonal equation.
\\

We have
\\
\begin{theorem}\label{guilty}
Let $b_j, c\in L^\infty(\Omega)$ and $\varphi$ be a
weakly pseudoconvex function with respect to the principal symbol of the
operator $P(x,D).$ Then there exist constants $\tau_0$ and $C$
independent of $\tau$ such that for all $\tau\ge \tau_0$ the
following estimate holds true:
\begin{eqnarray}\label{Lolo}
\Vert u
e^{\tau\varphi}\Vert^2_{H^{1,\tau}(\Omega)}+\int_{\partial\Omega}\vert
\frac{\partial u}{\partial \nu}\vert^2e^{2\tau\varphi}d\sigma
+\tau\int_{\partial\Omega_-}\vert\frac{\partial
\varphi}{\partial\nu}\vert\vert \frac{\partial u}{\partial
\nu}\vert^2 e^{2\tau\varphi}d\sigma \nonumber\\
\le C\left(\Vert
(P(x,D)u)e^{\tau\varphi}\Vert^2_{L^2(\Omega)}+
\tau\int_{\partial\Omega_+}\vert \frac{\partial u}{\partial
\nu}\vert^2 e^{2\tau\varphi}d\sigma\right).
\end{eqnarray}

\end{theorem}

{\bf Proof.} First we recall the equality
$$
 \{ \overline p(x,\xi-i\tau \nabla \varphi(x)), p(x,\xi+i\tau\nabla
 \varphi(x))\}/i\tau=4\tau (H_\varphi(\xi,\xi)+\tau^2
 H_\varphi(\nabla\varphi,\nabla\varphi)).
$$
Hence the pseudoconvexity condition (\ref{ogogo}) is equivalent
to the following one:
 \begin{equation}\label{dracon}
 (H_\varphi(\xi,\xi)+\tau^2
 H_\varphi(\nabla\varphi,\nabla\varphi))\ge 0\quad \mbox{on}\,\, \{(x,\xi,\tau)\in \Omega\times (\Bbb R^n\setminus\{0\})\times
\Bbb R_+^1\vert \,p(x,\xi+i\tau\nabla
 \varphi)=0\}.
\end{equation}
We show that for each $x$ from $\Omega$ the polynomial $q(x,\xi,\tau)=(H_\varphi\xi,\xi)+\tau^2
(H_\varphi\nabla
 \varphi,\nabla\varphi)$ can be represented as the sum of two
 homogeneous polynomials of degree two  in variables $\xi,\tau$ such that
 \begin{equation}\label{001}
q(x,\xi,\tau)=q_0(x,\xi,\tau)+q_+(x,\xi),
 \end{equation}
where
\begin{equation}\label{002}
q_+(x,\xi)\ge 0, \quad \forall (x,\xi,\tau)\in \overline \Omega\times \Bbb
R^2\times \Bbb R^1,
\end{equation}
and
\begin{equation}\label{003}
q_0(x,\xi,\tau)=0, \quad \forall (x,\xi,\tau)\in \{(x,\xi,\tau)\in
\Omega\times (\Bbb R^n\setminus\{0\})\times
\Bbb R_+^1\vert\, p(x,\xi+i\tau\nabla \varphi)=0\}.
\end{equation}

The functions $q_+, q_0$ can be constructed in the following way.
Consider the partition of unity of the domain $\Omega$:
\begin{equation}\label{004}
\sum_{j=1}^Ke_j=1\quad\mbox{on}\,\,\Omega, \quad e_j\in
C^\infty_0(B(x_j,\delta)), \quad e_j(x)\ge 0, \quad \forall
x\in\overline\Omega.
 \end{equation}
 Consider
the symbol $r_j(x,\xi,\tau^2)=e_j((H_\varphi\xi,\xi)+\tau^2
(H_\varphi\nabla
 \varphi,\nabla\varphi)).$ Since the function $\varphi$ is assumed to
 be weakly pseudoconvex, taking into account (\ref{dracon}) and  (\ref{004})
we obtain
 \begin{equation}\label{005}
r_j(x,\xi,\tau^2)\ge 0\quad\mbox{on}\quad \{(x,\xi,\tau)\in
\Omega\times (\Bbb R^n\setminus\{0\})\times \Bbb R_+^1\vert
p(x,\xi+i\tau\nabla \varphi)=0\}.
\end{equation}
 Suppose that $\delta>0$ is so small that  $\frac{\partial
 \varphi}{\partial x_J}$ is not equal to zero on
 $\overline{B(x_J,\delta)}$ for some $J \in \{1,\dots ,n\}.$
  Consider the function
 $$\widetilde r_j(x,\xi)=\widetilde r_j(x,\xi_1,\dots, \xi_{J-1},
\xi_{J+1},\dots, \xi_n)=r_j(x,\xi_1,\dots, \xi_{J-1},
\frac{1}{\frac{\partial\varphi}{\partial x_J}}
\sum_{k=1, k\ne J}^n\xi_k\frac{\partial
 \varphi}{\partial x_k},\xi_{J+1},\dots, \xi_n,m_j(\xi)/\vert\nabla
 \varphi\vert^2),
$$
$$
m_j(\xi)=\sum_{k=1, k\ne J}^n \xi_k^2+\left (\frac{1}{\frac{\partial\varphi}{\partial x_J}}\sum_{k=1, k\ne J}^n\xi_k\frac{\partial
 \varphi}{\partial x_k}\right )^2.
$$
Observe that if $\widetilde x\in B(x_J,\delta)$ and $ (\widetilde x,\widetilde \xi,
\widetilde\tau)\in \{(x,\xi,\tau)\in
\Omega\times (\Bbb R^n\setminus\{0\})\times \Bbb R_+^1\vert
p(x,\xi+i\tau\nabla \varphi)=0\}$, then
\begin{equation}\label{obana}
 \widetilde r_j(\widetilde x,\widetilde \xi)=e_j((H_\varphi\widetilde \xi,\widetilde \xi)
+\widetilde\tau^2
(H_\varphi\nabla
 \varphi(\widetilde x),\nabla\varphi(\widetilde x))).
 \end{equation}
By (\ref{obana}) and (\ref{004}), we have
\begin{equation}\label{007}
\widetilde r_j(x,\xi)\ge 0\quad \forall (x,\xi)\in \overline\Omega\times
(\Bbb R^n\setminus\{0\}).
\end{equation}
 Next we set $q_+(x,\xi)=\sum_{j=1}^K\widetilde r_j(x,\xi).$
By (\ref{004}) and (\ref{007}), we see
 \begin{equation}\label{008}
q_+(x,\xi)\ge 0\quad \forall (x,\xi)\in \overline\Omega\times
(\Bbb R^n\setminus\{0\})
\end{equation}
and by (\ref{obana}) and (\ref{004}) we obtain
$$
q_+(x,\xi)=(H_\varphi\xi,\xi)+\tau^2
(H_\varphi\nabla
 \varphi,\nabla\varphi)\quad \mbox{ on}\quad\{(x,\xi,\tau)\in
\Omega\times (\Bbb R^n\setminus\{0\})\times
\Bbb R^1_+\vert p(x,\xi+i\tau\nabla \varphi)=0\}.$$
 Therefore
 we can take $q_0(x,\xi,\tau)=(H_\varphi \xi,\xi)+\tau^2
(H_\varphi\nabla
 \varphi,\nabla\varphi)-q_+(x,\xi,\tau).$
Consequently (\ref{001}) and (\ref{003}) hold true.

Next we claim that there exist a smooth function $m(x)$ and a smooth
function $\ell(x,\xi)$, which is a homogeneous polynomial of degree one
in $\xi$ for all $x \in \overline\Omega$ such that
\begin{equation}\label{nonsence}
q_0(x,\xi,\tau)=m(x) (\vert
\xi\vert^2-\tau^2\vert\nabla\varphi\vert^2)+\ell(x,\xi)(\xi,\nabla\varphi),
\quad \forall (x,\xi,\tau)\in
\Omega\times (\Bbb R^n\setminus\{0\})\times\Bbb R^1_+.
\end{equation}
Indeed, let us fix some point $\widehat x$ from $\overline\Omega$. Without
loss of generality, after a possible rotation, we may assume
that $\nabla\varphi$ is parallel to the vector $\vec e_1 = (1,0,0, ..., 0)$.
Consider the polynomial $q_0(\widehat x, \xi, \tau)$ on the hypersurface
$\{\xi_1=0\}.$ The set of zeros of the polynomial
$\sum_{k=2}^n\xi_k^2-\tau^2\vert \nabla\varphi(\widehat x)\vert^2$
 is the subset of zeros of
the quadratic polynomial $q_0(\widehat x, 0,\xi_2,\dots, \xi_n, \tau)$
since $$
\{(\xi,\tau)\in (\Bbb R^n\setminus\{0\})\times\Bbb R^1_+
\vert p(\widehat x,\xi+i\tau\nabla \varphi)=0\}
$$
$$
= \{(\xi,\tau)\in (\Bbb R^n\setminus\{0\})\times\Bbb R^1_+\vert\xi_1=0\}
\cap \{(\xi,\tau)\in (\Bbb R^n\setminus\{0\})\times\Bbb R^1_+\vert\sum_{k=1}^n
\xi_k^2-\tau^2\vert \nabla\varphi(\widehat x) \vert^2=0\}.
$$
The set of zeros of the polynomial
$\sum_{k=2}^n\xi_k^2-\tau^2\vert \nabla\varphi(\widehat x)\vert^2$ forms
a cone surface in $\Bbb R^n.$ The polynomial $q(\widehat x,
0,\xi_2,\dots, \xi_n, \tau)$ is a homogeneous polynomial of degree
$2.$ There are two possibilities. First this polynomial is
identically equal to zero. Then we set $m(\widehat x)=0.$
Second, the set of zeros of polynomials $q_0(\widehat x, 0,\xi_2,\dots, \xi_n,
\tau)$ and $\sum_{k=2}^n\xi_k^2-\tau^2\vert \nabla\varphi(\widehat
x)\vert^2$ are the same. Therefore there exists $m(\widehat x)$ such
that $q(\widehat x, 0,\xi_2,\dots, \xi_n, \tau)=m(\widehat
x)(\sum_{k=2}^n\xi_k^2-\tau^2\vert \nabla\varphi(\widehat x)\vert^2).$
Hence we have
\begin{equation}\label{008}
q_0(x,\xi,\tau)=m(x) (\vert
\xi\vert^2-\tau^2\vert\nabla\varphi(x)\vert^2)\quad\mbox{on}\,\,
\{(x,\xi,\tau)\in \overline\Omega\times \Bbb R^n\times \Bbb R^1_+\vert
(\xi,\nabla\varphi)=0\}.
\end{equation}
Since for each $x$ from $\Omega$ there exists $(\widehat \xi,\widehat \tau)$
such that $(x,\widehat \xi,\widehat \tau)\in \{(x,\xi,\tau)\in
\overline\Omega\times \Bbb R^n\times \Bbb R^1_+\vert
(\xi,\nabla\varphi)=0\}$ and $\vert\widehat \xi \vert^2-\widehat
\tau^2\vert\nabla\varphi( x)\vert^2\ne 0$, by (\ref{008}) the
function $m(x)$ is smooth.

Consider the polynomial $d(x,\xi,\tau)=q_0(x,\xi,\tau)-m(x) (\vert
\xi\vert^2-\tau^2\vert\nabla\varphi\vert^2)$.
Let $A(x)$ be a smooth matrix such that the first row of $A$ is equal
to $\nabla\varphi$ and $\mbox{det}\, A(x)\ne 0$ on
$\overline\Omega.$
Then we introduce the new coordinates
$\widetilde \xi=A(x)\xi$ and set $\widetilde
d(x,\xi,\tau)=d(x,A^{-1}(x)\widetilde\xi,\tau).$
In the new coordinates, the set $\{(x,\xi,\tau)\in \overline\Omega\times
\Bbb R^n\times \Bbb
R^1_+\vert (\xi,\nabla\varphi)=0\}$ is written as
$$
\{(x,\xi,\tau)\in \overline\Omega\times \Bbb R^n\times \Bbb R^1_+\vert
\widetilde\xi_1=0\}.
$$
The polynomial $\widetilde d$  is a homogeneous
polynomial of degree $2$ in the variable $(\widetilde\xi,\tau)$ for
each $x \in \Omega$ and $\widetilde d(x,\widetilde\xi,\tau)=0$
if $\widetilde \xi_1=0.$
Therefore  we can represent this polynomial in the form
$$
\widetilde d(x,\widetilde\xi,\tau)=\widetilde \xi_1
\left(\sum_{j=1}^n  b_j(x)\widetilde
\xi_j+ b_{n+1}(x)\tau\right)
$$
with smooth functions $\widetilde b_j(x).$
Then after returning to the coordinates $\xi$ we obtain
$$
d(x,\xi,\tau)=(\xi,\nabla \varphi) ((\vec
b(x),A(x)\xi)+b_{n+1}(x)\tau),\quad\vec b=(b_1,\dots, b_n).
$$

Next we need to show that the function $b_{n+1}$ is identically equal
to zero in $\Omega.$ Indeed the symbol $q_+$ is independent of $\tau$
and the symbol $q$ depends on $\tau^2.$  Hence the symbol $q_0=q-q_+$
depends smoothly on $\tau^2$. Since we have already proved that
$q_0(x,\xi,\tau)-m(x) (\vert
\xi\vert^2-\tau^2\vert\nabla\varphi\vert^2)-(\xi,\nabla \varphi)
(\vec b(x),A(x)\xi)=b_{n+1}(x)\tau(\xi,\nabla \varphi)$, we observe
that on the right-hand side of this equality, the $\tau$-dependent terms which
are of the form $c(x)\tau^2$.  Therefore $b_{n+1}\equiv 0$ and
$$
d(x,\xi,\tau)=(\xi,\nabla \varphi) (\vec b(x),A(x)\xi).
$$
The justification of the formula (\ref{nonsence}) is complete.

 Consider a function $f:\Bbb R^1\rightarrow \Bbb R^1$  such that $f'(y)\ne
0$ for all $y\in\{y\vert y=\varphi(x)\,\,x\in\Omega\}.$ We set
$\xi=f'(\varphi)\eta .$ Then
\begin{eqnarray}\label{00001}
(H_{f(\varphi)}\xi,\xi)+\tau^2 (H_{f(\varphi)}\nabla
 \varphi,\nabla\varphi)=f'(\varphi)^3((H_{\varphi}\eta,\eta)
+ \tau^2 (H_{\varphi}\nabla
 \varphi,\nabla\varphi)+\tau^2\frac{f''(\varphi)}{f'(\varphi)}\vert
 \nabla\varphi\vert^4)+f''(\varphi)(\nabla
 \varphi,\xi)^2\nonumber\\
= f'(\varphi)^3(m(x) (\vert
\eta\vert^2-\tau^2\vert\nabla\varphi\vert^2)+\ell(x,\eta)(\eta,\nabla\varphi)
+\tau^2\frac{f''(\varphi)}{f'(\varphi)}\vert
 \nabla\varphi\vert^4+q_+(x,\eta))+f''(\varphi)(\nabla \varphi,\xi)^2
 \nonumber\\
= f'(\varphi)(m(x) (\vert
\xi\vert^2-\tau^2f'(\varphi)^2\vert\nabla\varphi\vert^2)+\ell(x,\xi)(\xi,\nabla\varphi))+\tau^2
f''(\varphi){f'(\varphi)^2}\vert
 \nabla\varphi\vert^4+q_+(x,\xi))+f''(\varphi)(\nabla \varphi,\xi)^2.
\end{eqnarray}
Next we take
$$
f_{N,\tau}(s)=s+\frac{Ns^2}{\tau},
$$
where $N$ is a large positive parameter.

For the moment, assume that
\begin{equation}\label{zopadlo}
b_i=c=0,\quad\forall i\in \{1,\dots,n\}.
\end{equation}
We set
$$
P(x,D,\tau)=e^{\tau f_{N,\tau}(\varphi)}P(x,D)e^{-\tau
f_{N,\tau}(\varphi)}=\Delta-2\tau(\nabla
f_{N,\tau}(\varphi),\nabla)+\tau^2\vert\nabla
f_{N,\tau}(\varphi)\vert^2-\tau\Delta f_{N,\tau}(\varphi)
$$
and
$$
P(x,D,\tau)^*=e^{-\tau f_{N,\tau}(\varphi)}P(x,D)e^{\tau
f_{N,\tau}(\varphi)}=\Delta+2\tau(\nabla
f_{N,\tau}(\varphi),\nabla)+\tau^2\vert\nabla
f_{N,\tau}(\varphi)\vert^2+\tau\Delta f_{N,\tau}(\varphi).
$$

Using the operators $P(x,D,\tau)$ and $P(x,D,\tau)^*$, we construct two
more operators
$$\quad P_+(x,D,\tau)=\frac 12(P(x,D,\tau)+P(x,D,\tau)^*)=\Delta
+\tau^2\vert\nabla f_{N,\tau}(\varphi)\vert^2
$$
and
$$
P_-(x,D,\tau)=\frac 12(P(x,D,\tau)-P(x,D,\tau)^*)=-2\tau(\nabla
f_{N,\tau}(\varphi),\nabla)-\tau\Delta f_{N,\tau}(\varphi).$$ Let
$w=e^{\tau f_{N,\tau}(\varphi)}u.$ Then
\begin{equation}\label{0000}
P_+(x,D,\tau)w+P_-(x,D,\tau)w=P(x,D,\tau)w\quad\mbox{in}\,\,\Omega,
\quad w\vert_{\partial\Omega}=0.
\end{equation}
Taking the $L^2$-norm of the equation (\ref{0000})
we obtain
\begin{equation}\label{zoopark}
\Vert P(x,D,\tau)w\Vert^2_{L^2(\Omega)}=\Vert
P_+(x,D,\tau)w\Vert^2_{L^2(\Omega)}+2(P_+(x,D,\tau)
w,P_-(x,D,\tau)w)_{L^2(\Omega)}+\Vert
P_-(x,D,\tau)w\Vert^2_{L^2(\Omega)}.
\end{equation}

Integrating by parts the second term on the right-hand side of
(\ref{zoopark}), we have
\begin{equation}\label{NNN2}
2(P_+(x,D,\tau) w,P_-(x,D,\tau)w)_{L^2(\Omega)}=([P_+,P_-](x,D,\tau)
w,w)_{L^2(\Omega)}-4\int_{\partial\Omega}\tau\frac{\partial
f_{N,\tau}(\varphi)}{\partial\nu}\vert \frac{\partial
w}{\partial\nu}\vert^2d\sigma .
\end{equation}
The differential operator $[P_+,P_-]$ has the form $$
-4\tau\sum_{i,j=1}^n\frac{\partial^2f_{N,\tau}(\varphi)}{\partial
x_i\partial x_j}\frac{\partial^2}{\partial x_i\partial x_j}+4\tau^3
\sum_{i,j=1}^n\frac{\partial^2f_{N,\tau}(\varphi)}{\partial
x_i\partial x_j}\frac{\partial f_{N,\tau}(\varphi)}{\partial
x_i}\frac{\partial f_{N,\tau}(\varphi)}{\partial x_j}+\tau R(x,D),
$$
$$
\quad R(x,D)=-2(\nabla \Delta f_{N,\tau}(\varphi),
\nabla)-\Delta^2f_{N,\tau}(\varphi) .
$$
The principal symbol of the differential operator
$[P_+,P_-](x,D,\tau)$ is equal to
$4\tau((H_{f_{N,\tau}(\varphi)}\xi,\xi)+\tau^2
(H_{f_{N,\tau}(\varphi)}\nabla
 \varphi,\nabla\varphi)).$
Hence the representation (\ref{00001}) holds true.

Therefore we can write down the second term on the right-hand side of
(\ref{NNN2}) as
\begin{eqnarray}\label{NNN}
([P_+,P_-](x,D,\tau)
w,w)_{L^2(\Omega)}=\tau\int_\Omega(f_{N,\tau}'(\varphi)(m(x)
(-\Delta
-\tau^2f_{N,\tau}'(\varphi)^2\vert\nabla\varphi\vert^2)-\ell(x,\nabla)(\nabla\varphi,\nabla))\nonumber\\+\tau^2
f_{N,\tau}''(\varphi){(f_{N,\tau}'(\varphi))^2}\vert
 \nabla\varphi\vert^4
 -q_+(x,\nabla))-f_{N,\tau}''(\varphi)(\nabla
 \varphi,\nabla)^2)w,w)dx\ge -\frac 14\Vert  P_+(x,D,\tau)
 w\Vert^2_{L^2(\Omega)}\nonumber\\-4\tau^2\int_\Omega
 (f_{N,\tau}'(\varphi))^2m^2w^2dx-\tau\int_\Omega
 (\nabla\varphi,\nabla w)\ell(x,\nabla)^*(wf_{N,\tau}'(\varphi))dx
-\tau\int_\Omega
 f_{N,\tau}'(\varphi)
 q_+(x,\nabla)wwdx\nonumber\\
 +\int_\Omega \tau^3
f_{N,\tau}''(\varphi){(f_{N,\tau}'(\varphi))^2}\vert
 \nabla\varphi\vert^4 w^2dx+\tau\int_\Omega R(x,D)wwdx\nonumber\\
-\tau\int_\Omega (\nabla\varphi,\nabla
 w)(\nabla\varphi,\nabla
 )^*(f_{N,\tau}''(\varphi))w)dx .
\end{eqnarray}
The symbol $q_+(x,\xi)$ is a quadratic polynomial written as
$q_+(x,\xi)=\sum_{j,k=1}^n q_{jk}(x)\xi_j\xi_k.$
Hence
\begin{eqnarray}\label{N1p}
 -\tau \int_\Omega f_{N,\tau}'(\varphi)
 q_+(x,\nabla)wwdx=-\tau \int_\Omega f_{N,\tau}'(\varphi)
 \sum_{j,k=1}^n q_{jk}(x)\frac{\partial^2w}{\partial x_k\partial x_j}wdx
\nonumber\\
= \tau \int_\Omega f_{N,\tau}'(\varphi)
 \sum_{j,k=1}^n q_{jk}(x)\frac{\partial w}{\partial x_k}
\frac{\partial w}{\partial x_j}dx-\frac{\tau}{2} \int_\Omega
 \sum_{j,k=1}^n w^2\frac{\partial^2 (f_{N,\tau}'(\varphi)
q_{jk}(x))}{\partial x_k\partial x_j}dx.
 \end{eqnarray}
Let
\begin{equation}\label{dragon}
\frac{N}{\tau}\Vert\varphi\Vert_{C^0(\overline\Omega)}\le \frac{1}{10}.
 \end{equation}
 Then $ f_{N,\tau}'(\varphi)$ is a nonnegative function. By
 (\ref{002}) the function $\sum_{j,k}^nq_{jk}(x)\xi_j\xi_k$ is also
nonnegative on the set $\overline\Omega\times (\Bbb R^n\setminus\{0\})$.
Hence the integral $\int_\Omega
f_{N,\tau}'(\varphi)
 \sum_{j,k=1}^n q_{jk}(x)\frac{\partial w}{\partial x_k}
\frac{\partial w}{\partial x_j}dx$ is nonnegative.
Therefore we obtain from (\ref{N1p})
\begin{equation}\label{N1}
 -\tau \int_\Omega f_{N,\tau}'(\varphi)
 q_+(x,\nabla)wwdx \ge
 - C(\tau+N)\Vert w\Vert^2_{L^2(\Omega)}.
 \end{equation}
 By the definition of the function $f_{N,\tau}$ we can choose $N_0$
 such that for all $N\ge N_0$ we have
 \begin{equation}\label{N2}
 \int_\Omega \tau^3
f_{N,\tau}''(\varphi){(f_{N,\tau}'(\varphi))^2}\vert
 \nabla\varphi\vert^4 w^2dx-4\tau^2\int_\Omega
 (f_{N,\tau}'(\varphi))^2m^2w^2dx\ge (N\tau^2+N^3)\Vert w\Vert^2_{L^2(\Omega)}
 \end{equation}
and
\begin{equation}\label{N3}
-\tau\int_\Omega (\nabla\varphi,\nabla
 w)(\nabla\varphi,\nabla
 )^*(f_{N,\tau}''(\varphi))w)dx=\int_\Omega 2N(\nabla \varphi,\nabla w)^2dx\ge
 0.
 \end{equation}
 Integrating by parts we obtain
 \begin{eqnarray}\label{N104}
 \tau\int_\Omega R(x,D)wwdx=\tau\int_\Omega-(\nabla\Delta f_{N,\tau}(\varphi),
\nabla w^2)-\Delta^2f_{N,\tau}(\varphi)w^2dx=\tau
\int_\Omega((\Delta^2-\Delta^2)f_{N,\tau}(\varphi)) w^2dx\nonumber\\
=0.
 \end{eqnarray}
 Using the Cauchy inequality, we have
\begin{eqnarray}\label{N4}
-\tau\int_\Omega
 (\nabla w,\nabla\varphi)\ell(x,\nabla)^*(wf_{N,\tau}'(\varphi))dx=-\tau\int_\Omega
 f_{N,\tau}'(\varphi)
 (\nabla w,\nabla\varphi)\ell(x,\nabla)^*wdx\nonumber\\-\tau\int_\Omega
 (\nabla
 w,\nabla\varphi)(\ell(x,\nabla)^*(f_{N,\tau}'(\varphi)))wdx\ge
-\frac 14\Vert \tau f_{N,\tau}'(\varphi)(\nabla
w,\nabla\varphi)\Vert^2_{L^2(\Omega)}-C\Vert \nabla
w\Vert^2_{L^2(\Omega)}-C(\tau+N)\Vert w\Vert^2_{L^2(\Omega)}.
 \end{eqnarray}
 Using (\ref{N1})-(\ref{N4}) we obtain from (\ref{NNN})
 \begin{eqnarray}\label{NNN1}
([P_+,P_-](x,D,\tau) w,w)_{L^2(\Omega)}\ge -\frac 14\Vert
P_+(x,D,\tau)
 w\Vert^2_{L^2(\Omega)}-\frac 14\Vert \tau f_{N,\tau}'(\varphi)(\nabla
w,\nabla\varphi)\Vert^2_{L^2(\Omega)}-C\Vert \nabla
w\Vert^2_{L^2(\Omega)}\nonumber\\-C(\tau+N)\Vert
w\Vert^2_{L^2(\Omega)} +\tau^2 N\Vert w\Vert^2_{L^2(\Omega)}.
\end{eqnarray}
By (\ref{NNN1}) and (\ref{NNN2}), we obtain from (\ref{zoopark})
\begin{eqnarray}\label{zoopark1}
\Vert P(x,D,\tau)w\Vert^2_{L^2(\Omega)}\ge \frac 14\Vert
P_+(x,D,\tau)w\Vert^2_{L^2(\Omega)}+\Vert
P_-(x,D,\tau)w\Vert^2_{L^2(\Omega)}\nonumber\\
 -\frac 14\Vert \tau f_{N,\tau}'(\varphi)(\nabla
w,\nabla\varphi)\Vert^2_{L^2(\Omega)}-C\Vert \nabla
w\Vert^2_{L^2(\Omega)}-C(\tau+N)\Vert
w\Vert^2_{L^2(\Omega)}\nonumber\\
+\frac{3\tau^2 N}{4}\Vert
w\Vert^2_{L^2(\Omega)}-4\int_{\partial\Omega}\tau\frac{\partial
f_{N,\tau}(\varphi)}{\partial\nu}\vert \frac{\partial
w}{\partial\nu}\vert^2d\sigma .
\end{eqnarray}
Let $\tau_0$ and $N_0$ be sufficiently large numbers. Then for all
$(\tau,N)$ satisfying (\ref{dragon}) and $N\ge N_0$ and
$\tau\ge\tau_0$ we obtain from (\ref{zoopark1}) that
\begin{eqnarray}\label{zoopark2}
\Vert P(x,D,\tau)w\Vert^2_{L^2(\Omega)}\ge\frac 34\Vert
P_+(x,D,\tau)w\Vert^2_{L^2(\Omega)}+\Vert
P_-(x,D,\tau)w\Vert^2_{L^2(\Omega)}\nonumber\\
-\frac 14\Vert \tau f_{N,\tau}'(\varphi)(\nabla
w,\nabla\varphi)\Vert^2_{L^2(\Omega)}-C\Vert \nabla
w\Vert^2_{L^2(\Omega)} +\frac{\tau^2 N}{2}\Vert
w\Vert^2_{L^2(\Omega)}-4\int_{\partial\Omega}\tau\frac{\partial
f_{N,\tau}(\varphi)}{\partial\nu}\vert \frac{\partial
w}{\partial\nu}\vert^2d\sigma .
\end{eqnarray}
Taking the scalar product in $L^2(\Omega)$ of the functions
$P_+(x,D,\tau)w$ and $Nw$ and integrating by parts, we obtain the
inequality
\begin{equation}\label{zoopark3}
N\Vert \nabla w\Vert^2_{L^2(\Omega)}\le C(\Vert
P_+(x,D,\tau)w\Vert^2_{L^2(\Omega)}+\tau^2 N\Vert
w\Vert^2_{L^2(\Omega)}).
\end{equation}
Then increasing $\tau_0$ and $N_0$ again and using (\ref{zoopark3}),
from (\ref{zoopark2}) we have:
\begin{eqnarray}\label{zoopark4}
\Vert P(x,D,\tau)w\Vert^2_{L^2(\Omega)}\ge \frac 12\Vert
P_+(x,D,\tau)w\Vert^2_{L^2(\Omega)}+\Vert
P_-(x,D,\tau)w\Vert^2_{L^2(\Omega)}\nonumber\\
 -\frac 14\Vert \tau f_{N,\tau}'(\varphi)(\nabla
w,\nabla\varphi)\Vert^2_{L^2(\Omega)}+NC\Vert \nabla
w\Vert^2_{L^2(\Omega)} +\frac{\tau^2 N}{4}\Vert
w\Vert^2_{L^2(\Omega)}-4\int_{\partial\Omega}\tau\frac{\partial
f_{N,\tau}(\varphi)}{\partial\nu}\vert \frac{\partial
w}{\partial\nu}\vert^2d\sigma,
\end{eqnarray} where $(\tau,N)$ satisfies (\ref{dragon}) and
$\tau\ge\tau_0, N\ge N_0$.
Observe that
$$
\Vert \tau f_{N,\tau}'(\varphi)(\nabla
w,\nabla\varphi)\Vert^2_{L^2(\Omega)}
=\frac 14\Vert
P_-(x,D,\tau)w+\tau\Delta
f_{N,\tau}(\varphi)w\Vert^2_{L^2(\Omega)}\le \frac 12\Vert
P_-(x,D,\tau)w\Vert^2_{L^2(\Omega)}+\frac 12\Vert\tau\Delta
f_{N,\tau}(\varphi)w\Vert^2_{L^2(\Omega)}.
$$
Using this estimate in (\ref{zoopark4}), we obtain
\begin{eqnarray}\label{zoopark5}
\Vert P(x,D,\tau)w\Vert^2_{L^2(\Omega)}\ge\frac 12\Vert
P_+(x,D,\tau)w\Vert^2_{L^2(\Omega)}+\frac 12\Vert
P_-(x,D,\tau)w\Vert^2_{L^2(\Omega)}\nonumber\\
 +NC\Vert \nabla w\Vert^2_{L^2(\Omega)} +\frac{\tau^2 N}{8}\Vert
w\Vert^2_{L^2(\Omega)}-4\int_{\partial\Omega}\tau\frac{\partial
f_{N,\tau}(\varphi)}{\partial\nu}\vert \frac{\partial
w}{\partial\nu}\vert^2d\sigma,
\end{eqnarray}
where $(\tau,N)$ satisfies (\ref{dragon}) and $\tau\ge\tau_0,  N\ge
N_0. $

Now we remove the assumption (\ref{zopadlo}). Suppose that
some coefficients of the first or zeroth order terms are not identically
equal to zero. Then we have to replace
the term on the right-hand side of (\ref{zoopark4}) by
$\Vert P(x,D,\tau)w-\sum_{j=1}^n b_j
\frac{\partial w}{\partial x_j}+(\sum_{j=1}^n\tau b_j\frac{\partial
f_{N,\tau}}{\partial x_j}  -c) w\Vert^2_{L^2(\Omega)}. $
By
$$
\Vert P(x,D,\tau)w-\sum_{j=1}^n b_j \frac{\partial w}{\partial
x_j}+(\sum_{j=1}^n \tau b_j\frac{\partial f_{N,\tau}}{\partial x_j}
-c) w\Vert^2_{L^2(\Omega)}\le C(\Vert
P(x,D,\tau)w\Vert^2_{L^2(\Omega)}+\Vert
w\Vert^2_{H^{1,\tau}(\Omega)}),
$$
from (\ref{zoopark5}) we have
\begin{eqnarray}\label{zoopark6}
C(\Vert P(x,D,\tau)w\Vert^2_{L^2(\Omega)}+\Vert
w\Vert^2_{H^{1,\tau}(\Omega)})\ge\frac 12\Vert
P_+(x,D,\tau)w\Vert^2_{L^2(\Omega)}+\frac 12\Vert
P_-(x,D,\tau)w\Vert^2_{L^2(\Omega)}\nonumber\\
 +N C\Vert \nabla w\Vert^2_{L^2(\Omega)} +\frac{\tau^2 N}{2}\Vert
w\Vert^2_{L^2(\Omega)}-\int_{\partial\Omega}\tau\frac{\partial
f_{N,\tau}(\varphi)}{\partial\nu}\vert \frac{\partial
w}{\partial\nu}\vert^2d\sigma,
\end{eqnarray}
The term $\Vert w\Vert^2_{H^{1,\tau}(\Omega)}$ on the left-hand side
can be absorbed into the term $N C\Vert \nabla
w\Vert^2_{L^2(\Omega)}$ on the left-hand side. Therefore even
without assumption (\ref{zopadlo}) we still have
(\ref{zoopark5}). Now we fix a parameter $N=2N_0$ in (\ref{zoopark5}).

Finally we estimate the normal derivative of the function $w$ on the
boundary. Let $\rho\in C^2(\overline\Omega)$ satisfy
\begin{equation}\label{gopnik}
(\rho,\vec\nu)>0\quad\mbox{on}\,\,
\partial\Omega.
\end{equation}

Taking the scalar product of the function $P_+(x,D,\tau)w$ and
$(\nabla\rho,\nabla w)$ in $L^2(\Omega)$, we obtain
\begin{eqnarray}\label{verka}
\int_\Omega P_+(x,D,\tau)w(\nabla\rho,\nabla w)dx=-\int_\Omega\left(
\frac 12(\nabla\rho,\nabla\vert\nabla
w\vert^2)+\sum_{k,j=1}^n\frac{\partial w}{\partial
x_j}\frac{\partial^2 \rho}{\partial x_j\partial x_k}\frac{\partial
w}{\partial x_k}+\vert \nabla f_{N,\tau}(\varphi)\vert^2(\nabla
\rho,\nabla\frac{\vert w\vert^2}{2})\right )dx\nonumber\\
+\int_{\partial\Omega}\frac{\partial
w}{\partial\nu}(\nabla\rho,\nabla w)d\sigma=\int_\Omega\left(\frac
12 \Delta\rho\vert\nabla w\vert^2-\sum_{k,j=1}^n\frac{\partial
w}{\partial x_j}\frac{\partial^2 \rho}{\partial x_j\partial
x_k}\frac{\partial w}{\partial x_k}+\mbox{div}\,(\vert \nabla
f_{N,\tau}(\varphi)\vert^2\nabla
\rho)\frac{\vert w\vert^2}{2})\right )dx\nonumber\\
+\int_{\partial\Omega}\frac{\partial
w}{\partial\nu}(\nabla\rho,\nabla w)d\sigma-\frac 12
\int_{\partial\Omega}(\nabla\rho,\vec \nu)\vert\nabla
w\vert^2d\sigma\nonumber\\
= \int_\Omega\left(\frac 12 \Delta\rho\vert\nabla
w\vert^2-\sum_{k,j=1}^n\frac{\partial w}{\partial
x_j}\frac{\partial^2 \rho}{\partial x_j\partial x_k}\frac{\partial
w}{\partial x_k}+\mbox{div}\,(\vert \nabla
f_{N,\tau}(\varphi)\vert^2\nabla
\rho)\frac{\vert w\vert^2}{2})\right )dx\nonumber\\
+\int_{\partial\Omega}\vert\frac{\partial
w}{\partial\nu}\vert^2(\nabla\rho,\vec\nu) d\sigma-\frac
12\int_{\partial\Omega}(\nabla\rho,\vec \nu)\vert\frac{\partial
w}{\partial \nu} \vert^2d\sigma.
\end{eqnarray}
Here in order to obtain the last equality we used the equality
$\frac{\partial w}{\partial x_k}=\nu_k\frac{\partial
w}{\partial\nu}.$ The equality (\ref{verka}) and (\ref{gopnik})
imply
\begin{equation}\label{gopnik1}
\int_{\partial\Omega}\vert\frac{\partial w}{\partial \nu}
\vert^2d\sigma\le C(\Vert P(x,D,\tau)w\Vert^2_{L^2(\Omega)}+\Vert
w\Vert^2_{H^{1,\tau}(\Omega)}).
\end{equation}

From (\ref{gopnik1}) and (\ref{zoopark5}), the estimate (\ref{Lolo})
follows immediately.
 $\blacksquare$
\\
\vspace{0.2cm}
\\
{\bf Remark.} {\it Compare the Carleman estimate (\ref{Lolo}) with
(\ref{Lolo1}), we lose $\tau$ in front of the first term on the
right-hand side. On the other hand it can be shown that the inequality
(\ref{Lolo}) is sharp.}

Consider a boundary value problem
\begin{equation}\label{ad}
P(x,D)u=\Delta u+qu=f_\tau e^{\tau\varphi}\quad
\mbox{in}\,\,\Omega,\quad u\vert_{\partial\Omega_-}=a_\tau
e^{\tau\varphi}.
\end{equation}

For the problem (\ref{ad}) we can construct solutions with the
following properties:
\begin{proposition}
Let $b_j\in C^1(\overline\Omega), c\in L^\infty(\Omega),
f_\tau\in L^2(\Omega), a_\tau \in L^2(\partial\Omega_-)$
and a function $\varphi$ be weakly pseudoconvex with respect to the
principal symbol of the operator $P(x,D).$ Then there exists a
solution $u_\tau$ to problem (\ref{ad}) such that
\begin{equation}\label{Eb}
\Vert u_\tau e^{-\tau\varphi}\Vert_{L^2(\Omega)}\le C(\Vert
f_\tau\Vert_{L^2(\Omega)}/\tau+\Vert
a_\tau\Vert_{L^2(\partial\Omega_-)})\quad \forall \tau\ge\tau_0.
\end{equation}
If in addition
$a_\tau/\vert\frac{\partial\varphi}{\partial\nu}\vert^\frac 12 \in
L^2(\partial\Omega_-),$ then there exists a solution to problem
(\ref{ad}) such that
\begin{equation}\label{Eb1}
\Vert u_\tau e^{-\tau\varphi}\Vert_{L^2(\Omega)}\le C(\Vert
f_\tau\Vert_{L^2(\Omega)}/\tau+\Vert
a_\tau/\vert\frac{\partial\varphi}{\partial\nu}\vert^\frac 12
\Vert_{L^2(\partial\Omega_-)}/\root\of{\tau})\quad \forall
\tau\ge\tau_0.
\end{equation}
Here the constants $C$ and $\tau_0$ are independent of $\tau.$
\end{proposition}

{\bf Proof.} Let $\mathcal X=\{( (f,g)\in L^2(\Omega)\times
L^2(\partial\Omega\setminus\partial\Omega_-)\vert
P(x,D)^*w=f\quad\mbox{in}\,\,\Omega, \,\, w\vert_{\partial\Omega}=0,
\,\, \frac{\partial
w}{\partial\nu}\vert_{\partial\Omega\setminus\partial\Omega_-}=g\}$
be a linear subspace of the Hilbert space
$L^2_{e^{2\tau\varphi}}(\Omega)\times
L^2_{e^{2\tau\varphi}}(\partial\Omega\setminus\partial\Omega_-)$
which is equipped with the norm
\begin{equation}\label{EEb1}
\Vert (f,g)\Vert^2_{\mathcal X}=\Vert
fe^{\tau\varphi}\Vert^2_{L^2(\Omega)}+\tau\Vert
ge^{\tau\varphi}\Vert^2_{L^2(\partial\Omega\setminus\partial\Omega_-)}.
\end{equation}

By (\ref{Lolo}), the normed space  $\mathcal X$ is the closed
subspace of the Hilbert
space $L^2_{e^{2\tau\varphi}}(\Omega)\times
L^2_{e^{2\tau\varphi}}(\partial\Omega\setminus\partial\Omega_-).$
Hence $\mathcal X$ is a Hilbert space.
On $\mathcal X$, we consider the linear functional
\begin{equation}\label{EEb2}
\ell((f,g))=-\int_{\Omega} f_\tau
e^{\tau\varphi}wdx-\int_{\partial\Omega_-} a_\tau
ge^{\tau\varphi}d\sigma .
\end{equation}

In order to estimate the norm of the functional $\ell$, observe that
by Carleman estimate (\ref{Lolo}) we have
\begin{eqnarray}\label{zebra}
\Vert w e^{\tau\varphi}\Vert_{H^{1,\tau}(\Omega)}+\Vert
\frac{\partial w}{\partial\nu}
e^{\tau\varphi}\Vert_{L^2(\partial\Omega_-)}+\tau^\frac 12\Vert
\frac{\partial w}{\partial\nu}\vert\frac{\partial
\varphi}{\partial\nu}\vert^\frac 12
e^{\tau\varphi}\Vert_{L^2(\partial\Omega_-)}\nonumber\\\le C(\Vert
fe^{\tau\varphi}\Vert_{L^2(\Omega)}+\tau^\frac 12\Vert
g\vert\frac{\partial \varphi}{\partial\nu}\vert^\frac 12
e^{\tau\varphi}\Vert_{L^2(\partial\Omega\setminus\partial\Omega_-)}).
\end{eqnarray}
Then
\begin{equation}\label{AA1}
\Vert \ell\Vert=\mbox{sup}_{(f,g)\in \mathcal X\setminus
\{0\}}\frac{\vert \ell((f,g))\vert}{\Vert (f,g)\Vert_{\mathcal
X}}\le C( \Vert a_\tau\Vert_{L^2(\partial\Omega_-)}+\Vert
f_\tau\Vert_{L^2(\Omega)}/\tau)
\end{equation}
and if $a_\tau/\vert\frac{\partial\varphi}{\partial\nu}\vert^\frac
12 \in L^2(\partial\Omega_-)$, then
\begin{equation}\label{AA2}
\Vert \ell\Vert=\mbox{sup}_{(f,g)\in \mathcal X\setminus
\{0\}}\frac{\vert \ell((f,g))\vert}{\Vert (f,g)\Vert_{\mathcal
X}}\le C( \Vert
a_\tau/\vert\frac{\partial\varphi}{\partial\nu}\vert^\frac 12
\Vert_{L^2(\partial\Omega_-)}/\root\of{\tau}+\Vert
f_\tau\Vert_{L^2(\Omega)}/\tau),
\end{equation}
where the constant $C$ is independent of $\tau.$

This functional is
bounded on $\mathcal X$ and by the Banach theorem it can be extended on
the whole space $L^2_{e^{2\tau\varphi}}(\Omega)\times
L^2_{e^{2\tau\varphi}}(\partial\Omega\setminus \Omega_-)$ with
preservation of the norm. Hence, by the Riesz theorem,  there exists  a
pair $(w_\tau,\widetilde g)\in L^2_{e^{2\tau\varphi}}(\Omega)\times
L^2_{e^{2\tau\varphi}}(\partial\Omega\setminus\partial\Omega_-)$
such that
\begin{equation}\label{zarevo1}
\ell((f,g))=(\widetilde ge^{\tau\varphi},
ge^{\tau\varphi})_{L^2(\partial\Omega\setminus\partial\Omega_-)}+(
e^{\tau\varphi}w_\tau, fe^{\tau\varphi})_{L^2(\Omega)}
\end{equation}
and \begin{equation} \label{zarevo2}\Vert \ell\Vert=\Vert
(w_\tau,\widetilde g)\Vert_{ L^2_{e^{2\tau\varphi}}(\Omega)\times
L^2_{e^{2\tau\varphi}}(\partial\Omega\setminus\partial\Omega_-)}.
\end{equation}

By (\ref{EEb2}) and (\ref{zarevo1}), the function
$u_\tau=-e^{2\tau\varphi} w_\tau$ solves the problem (\ref{ad}).
From (\ref{zarevo2}) we obtain
$$
\Vert u_\tau e^{-\tau\varphi}\Vert_{L^2(\Omega)}=\Vert
w_\tau\Vert_{L^2_{e^{\tau\varphi}}(\Omega)}\le\Vert \ell\Vert.
$$
Hence from  this estimate  and (\ref{AA1}), (\ref{AA2}) imply (\ref{Eb})
and (\ref{Eb1}). $\blacksquare$

\begin{corollary}\label{ZZZ}
Let $b_j\in C^1(\overline\Omega), c\in L^\infty(\Omega),$ the families of
functions $f_\tau$ and $ a_\tau$ be uniformly bounded in
$L^2(\Omega)$ and $L^\infty(\partial\Omega_-)$  respectively and
a function $\varphi$ be weakly pseudoconvex with respect to the
principal symbol of the operator $P(x,D).$ Then there exist solutions
$u_\tau$, $\tau>0$, to the problem (\ref{ad}) such that
\begin{equation}\label{Eb6}
\Vert u_{\tau}
e^{-\tau\varphi}\Vert_{L^2(\Omega)}=o(1)\quad\mbox{as}\,\,\tau\rightarrow
\infty.
\end{equation}
\end{corollary}

{\bf Proof.} We set $\frak Y=\{x\in\partial\Omega\vert x\in
\partial(\partial \Omega_-)\}$ and
$\partial\Omega_{-,\epsilon}=\{x\in\partial\Omega_-\vert
dist(x,\frak Y)\ge \epsilon\}$ for any positive $\epsilon$.
Obviously
\begin{equation}\label{Eb3}
mes(\partial\Omega_{-,\epsilon}\setminus \Omega_-)\rightarrow
0\quad\mbox{as}\quad \epsilon\rightarrow +0.
\end{equation}

We set $g(\epsilon)=\Vert\frac{1}{
\frac{\partial\varphi}{\partial\nu}}
\Vert_{C^0(\overline{\Omega_{-,\epsilon}})}.$
Let $m(\tau)$ be a positive continuous function such that
\begin{equation}\label{lolol}
m(\tau)\rightarrow 0\quad\mbox{as}\quad\tau\rightarrow +\infty\quad
\mbox{and}\quad g(m(\tau))\le C\tau^\frac 14,
\end{equation}
where the constant $C$ is independent of $\tau.$
 We look for the function $u_\tau$ in the
form $u_\tau=u_{\tau,1}+u_{\tau,2}$ where
\begin{equation}\label{ad1}
P(x,D)u_{\tau,1}=f_\tau e^{\tau\varphi}\quad
\mbox{in}\,\,\Omega,\quad u_{\tau,1}\vert_{\partial\Omega_-}=0
\end{equation}
and
\begin{equation}\label{ad2}
P(x,D)u_{\tau,2}=0\quad \mbox{in}\,\,\Omega,\quad
u_{\tau,2}\vert_{\partial\Omega_-}=a_\tau e^{\tau\varphi}.
\end{equation}
By (\ref{Eb}) one can construct a solution to  problem (\ref{ad1})
such that
\begin{equation}\label{Eb10}
\Vert u_{\tau,1} e^{-\tau\varphi}\Vert_{L^2(\Omega)}=o(\frac
1\tau)\quad\mbox{as}\,\,\tau\rightarrow +\infty.
\end{equation}
By (\ref{Eb3}) and (\ref{lolol}), we have
\begin{equation}\label{Eb4}
\Vert \chi_{\partial\Omega_{-,m(\tau)}\setminus
\partial\Omega_-}a_\tau\Vert_{L^2(\partial\Omega)}\rightarrow 0\quad \mbox{as}\quad
\tau\rightarrow +0.
\end{equation}
Using (\ref{Eb}) we construct a solution $w_{\tau}$ to the boundary
value problem
\begin{equation}\label{ad3}
P(x,D) w_{\tau}=0\quad \mbox{in}\,\,\Omega,\quad
w_{\tau}\vert_{\partial\Omega_-}=\chi_{\partial\Omega_{-,m(\tau)}\setminus
\partial\Omega_-} a_\tau e^{\tau\varphi}
\end{equation}
such that
\begin{equation}\label{Eb4}
\Vert  w_{\tau}e^{-\tau\varphi} \Vert_{L^2(\Omega)}\rightarrow
0\quad \mbox{as}\quad \tau\rightarrow +\infty.
\end{equation}
On the other hand, we have
$(1-\chi_{\partial\Omega_{-,m(\tau)}\setminus \partial\Omega_-}
)a_\tau/\vert\frac{\partial\varphi}{\partial\nu}\vert^\frac 12\in
L^2(\partial\Omega_-)$ for all $\tau$ sufficiently large.

Applying (\ref{Eb1}) and (\ref{lolol}), we construct a solution $\widetilde
w_\tau$ to the boundary value problem
\begin{equation}\label{ad4}
P(x,D)\widetilde  w_{\tau}=0\quad \mbox{in}\,\,\Omega,\quad \widetilde
w_{\tau}\vert_{\partial\Omega_-}=(1-\chi_{\partial\Omega_{-,m(\tau)}\setminus
\partial\Omega_-} )a_\tau e^{\tau\varphi}
\end{equation}
such that  \begin{equation} \label{E5}\Vert \widetilde
w_{\tau}e^{-\tau\varphi} \Vert_{L^2(\Omega)}\le
\frac{C}{\root\of{\tau}}\Vert(1-\chi_{\partial\Omega_{-,m(\tau)}\setminus
\partial\Omega_-} )a_\tau
/\root\of{\vert\frac{\partial\varphi}{\partial\nu}\vert}\Vert_{L^2(\partial\Omega_-)}\le
g(m(\tau))\Vert a_\tau\Vert_{L^2(\partial\Omega_-)} =O(\frac
{1}{\tau^\frac 14})\quad \mbox{as}\quad \tau\rightarrow +\infty.
\end{equation}

Finally we set $u_{\tau, 2}= w_{\tau} +\widetilde w_{\tau} .$ By
(\ref{ad1}), (\ref{ad3}) and (\ref{ad4})  the function $u_{\tau,2}$
solves the problem (\ref{ad}) and by (\ref{Eb10}), (\ref{Eb4})  and
(\ref{E5}) the estimate  (\ref{Eb6}) holds true.
 $\blacksquare$

 In order to prove the uniqueness result in determining a potential
 of the  Schr\"odinger equation in dimension $n=2$, we need to further
 relax the notion of pseudoconvex function.
That is, as a solution of the Eikonal equation we should admit a holomorphic
function $\Phi$ which is degenerate at some points of domain $\Omega.$

More precisely, let $\Phi=\varphi+i\psi$ be a holomorphic function
in $\Omega$ such that $\varphi, \psi$ are real-valued and
\begin{equation}\label{1'} \Phi\in C^2(\overline\Omega),\quad
\mbox{Im}\, \Phi\vert_{\Gamma_0^*}=0, \quad \Gamma_0\subset\subset
\Gamma_0^*,
\end{equation}
where $\Gamma_0^*$ is some open set on $\partial\Omega.$ Denote by
$\mathcal H$ the set of the critical points of the function $\Phi.$
Assume that
\begin{equation}\label{22}
\mathcal H\ne \emptyset,\quad\partial^2_z\Phi(z)\ne 0\quad \forall
z\in \mathcal H,\quad \mathcal
H\cap\overline{\partial\Omega\setminus\Gamma_0}=\emptyset
\end{equation}
 and
\begin{equation}\label{kk}
\int_{\mathcal J}1d\sigma =0,\quad \mathcal J=\{x; \thinspace
\partial_{\vec \tau}\psi(x)=0,x\in \partial\Omega\setminus\Gamma_0^*\}.
\end{equation}

Then $\Phi$  has only a finite number of critical points and we can
set:
\begin{equation}\label{mona}
\mathcal H \setminus \Gamma_0= \{ \widetilde{x}_1, ...,
\widetilde{x}_{\ell} \},\quad \mathcal H \cap \Gamma_0=\{
\widetilde{x}_{\ell+1}, ..., \widetilde{x}_{\ell+\ell'} \}.
\end{equation}
The following proposition  was proved in \cite{IUY1}.

\begin{proposition}\label{Proposition -1}
Let $\widetilde x$ be an arbitrary point in $\Omega.$ There exists a
sequence of functions $\{\Phi_\epsilon\}_{\epsilon\in(0,1)}$
satisfying (\ref{1'})-(\ref{kk}) such that all the critical points
of $\Phi_\epsilon$ are nondegenerate and there exists a sequence
$\{\widetilde x_\epsilon\}, \epsilon\in (0,1)$ such that
$$
\widetilde x_\epsilon \in \mathcal H_\epsilon =
\{z\in\overline\Omega \vert \frac{\partial \Phi_\epsilon}{\partial
z}(z)=0 \},\quad \widetilde x_\epsilon\rightarrow \widetilde
x\,\,\mbox{ as}\,\, \epsilon\rightarrow +0.
$$
Moreover for any $j$ from $\{1,\dots,\mathcal N\}$ we have
$$
\mathcal H_\epsilon\cap\gamma_j=\emptyset\quad\mbox{if}\,\,
\gamma_j\cap (\partial\Omega\setminus\Gamma_0)\ne\emptyset,
$$
$$
\mathcal H_\epsilon\cap\gamma_j\subset \Gamma_0 \quad\mbox{if}\,\,
\gamma_j\cap (\partial\Omega\setminus\Gamma_0) = \emptyset,
$$
$$
\mbox{Im}\,\Phi_\epsilon(\widetilde x_\epsilon)\notin \{\mbox{Im}\,
\Phi_\epsilon(x)\vert x\in \mathcal H_\epsilon\setminus
\{\widetilde{x_\epsilon}\}\} \,\,\mbox{and}
\,\,\mbox{Im}\,\Phi_\epsilon(\widetilde x_\epsilon) \ne 0.
$$
\end{proposition}

Now we start the proof of the Carleman estimate for the
two-dimensional Schr\"odinger equation. The results of Theorem
\ref{guilty} cannot be applied directly to this case since the
weight function is allowed to have critical points. The proof of the
Carleman estimate is based on the decomposition of the Laplace
operator into $\partial_z$ and
$\partial_{\overline z}.$

First we establish a Carleman estimates for the operators
$\partial_z$ and $\partial_{\overline z}.$

\begin{proposition} \label{Proposition 2.1}
Let $\Phi$ satisfy (\ref{1'})-(\ref{kk}), $\tau \in \Bbb R^1$, and
the function $C=C_1+iC_2$ belong to $C^1(\overline\Omega)$ where
$C_1, C_2$ are real-valued. Let $\widetilde f\in L^2(\Omega)$, and
$\widetilde v\in H^1(\Omega)$ be a solution to
\begin{equation}\label{zina}
2 \frac{\partial}{\partial{z}}\widetilde v -\tau\frac{\partial\Phi}
{\partial{z}}\widetilde v+C \widetilde v =\widetilde f\quad \mbox{in
}\,\Omega
\end{equation}
or let $\widetilde v$ be a solution to
\begin{equation}\label{zina1}
2\frac{\partial}{\partial{\overline z}}\widetilde v - \tau
\frac{\partial\overline\Phi}{\partial{\overline z}} \widetilde v+
C\widetilde v =\widetilde f\quad\mbox{ in }\,\Omega.
\end{equation}
In the case (\ref{zina}) we have
\begin{eqnarray}\label{vika1}
\Vert\frac{\partial \widetilde{v}}{\partial {x_1}}
-i\mbox{Im}(\tau\frac{\partial\Phi}{\partial{z}}-C )\widetilde
v\Vert^2 _{L^2(\Omega)} -
\int_{\partial\Omega}\left(\tau\frac{\partial\varphi}{\partial\nu}
-(\nu_1C_1+\nu_2C_2)\right)\vert \widetilde v\vert^2d\sigma
- \int_\Omega \left(\frac{\partial C_1}{\partial x_1}+\frac{\partial C_2}
{\partial x_2}\right)\vert \widetilde v\vert^2dx\nonumber\\
+ \mbox{Re}\int_{\partial\Omega}i \frac{\partial \widetilde v}
{\partial\vec{\tau}}\overline{\widetilde v}d\sigma
+ \Vert -\frac 1i \frac{\partial \widetilde{v}}{\partial{x_2}} -
\mbox{Re}(\tau\frac{\partial\Phi} {\partial{z}}-C )\widetilde
v\Vert^2 _{L^2(\Omega)} = \Vert\widetilde f\Vert^2_{L^2(\Omega).}
\end{eqnarray}
In the case (\ref{zina1}) we have
\begin{eqnarray} \label{vika2}
\Vert \frac{\partial \widetilde{v}}{\partial{x_1}}
-i\mbox{Im}(\tau\frac{\partial\overline\Phi} {\partial{\overline z}}-C
)\widetilde v\Vert^2_{L^2(\Omega)} -
\int_{\partial\Omega}\left(\tau\frac{\partial \varphi}{\partial\nu}
-(\nu_1C_1-\nu_2C_2)\right)\vert \widetilde v\vert^2d\sigma -\int_\Omega
\left(\frac{\partial C_1}{\partial x_1} -\frac{\partial C_2}{\partial
x_2}\right)\vert
\widetilde v\vert^2dx\nonumber\\
- \mbox{Re}\int_{\partial\Omega}i \frac{\partial\widetilde
v}{\partial \vec \tau} \overline{\widetilde
v}d\sigma
+ \Vert\frac 1i\frac{\partial \widetilde{v}}{\partial{x_2}}
-\mbox{Re}(\tau\frac{\partial\overline\Phi} {\partial{\overline z}}-C
)\widetilde v \Vert^2 _{L^2(\Omega)} = \Vert\widetilde
f\Vert^2_{L^2(\Omega)}.
\end{eqnarray}
\end{proposition}

{\bf Proof.} We prove the equality (\ref{vika1}). The proof of
equality (\ref{vika2}) is the same. Denote $L_-(x,D,\tau){\widetilde
v}=\frac{\partial \widetilde{v}}{\partial {x_1}}
-i\mbox{Im}(\tau\frac{\partial\Phi}{\partial{z}}-C )\widetilde v$
and $L_+(x,D,\tau){\widetilde v}=\frac 1i \frac{\partial
\widetilde{v}}{\partial{x_2}} - \mbox{Re}(\tau\frac{\partial\Phi}
{\partial{z}}-C )\widetilde v.$ In the new notations we rewrite
equation (\ref{zina}) as

\begin{equation}\label{zanoza}
L_-(x,D,\tau){\widetilde
v}+L_+(x,D,\tau){\widetilde v}=\widetilde f
\quad\mbox{in}\quad\Omega.
\end{equation}
Taking the $L^2$- norm of the left- and right-hand sides of
(\ref{zanoza}), we obtain
\begin{equation}\label{zanoza1}
\Vert L_+ (x,D,\tau)\widetilde
 v\Vert^2_{L^2(\Omega)}+2\mbox{Re}(L_+(x,D,\tau)\widetilde  v,L_-(x,D,\tau)\widetilde  v)_{L^2(\Omega)}+\Vert L_-(x,D,\tau)
\widetilde  v\Vert^2_{L^2(\Omega)}=\Vert \widetilde
f\Vert^2_{L^2(\Omega)}.
\end{equation}
Integrating by parts the second term of (\ref{zanoza1}), we obtain
\begin{eqnarray}\label{zanoza2}
2\mbox{Re}(L_+(x,D,\tau)\widetilde  v,L_-(x,D,\tau)\widetilde v)_{L^2(\Omega)}=
\mbox{Re}\{([L_+,L_-]\widetilde
 v, \widetilde  v)_{L^2(\Omega)}\nonumber\\+\int_{\partial\Omega}((L_-(x,D,\tau)\widetilde
v)i\nu_2\overline{\widetilde v}+L_+(x,D,\tau)\widetilde
v\nu_1\overline{\widetilde v})d\sigma\}.
\end{eqnarray}

The Cauchy-Riemann equations yield
\begin{equation}
[L_+,L_-]=-\left(\frac{\partial C_1}{\partial x_1}+\frac{\partial C_2}
{\partial x_2}\right).
\end{equation}
Using the Cauchy-Riemann equations again, we observe
$$
\frac{\partial\Phi}{\partial z}=\frac{\partial\varphi}{\partial x_1}
-i\frac{\partial\varphi}{\partial x_2}.
$$
Therefore
\begin{eqnarray}
\label{nano}\mbox{Re}\,\int_{\partial\Omega}((L_-(x,D,\tau)\widetilde
v)i\nu_2\overline{\widetilde v}+L_+(x,D,\tau)\widetilde
v\nu_1\overline{\widetilde
v})d\sigma
= \mbox{Re}\,\int_{\partial\Omega} \left(\frac{\partial
\widetilde{v}}{\partial {x_1}}+i\left(\tau
\frac{\partial\varphi}{\partial x_2}+C_2\right)\widetilde v\right)
i\nu_2\overline{\widetilde v}\nonumber\\
+ \left(\frac 1i \frac{\partial
\widetilde{v}}{\partial{x_2}} -\tau\frac{\partial \varphi}{\partial
x_1}\widetilde{v}+C_1\widetilde{v}\right)\nu_1\overline{\widetilde
v}d\sigma
= -\int_{\partial\Omega}\left(\tau\frac{\partial\varphi}{\partial\nu}
-(\nu_1C_1+\nu_2C_2)\right)\vert \widetilde v\vert^2d\sigma+
\mbox{Re}\int_{\partial\Omega}i \frac{\partial \widetilde v}
{\partial\vec{\tau}}\overline{\widetilde v}d\sigma .
\end{eqnarray}

From (\ref{zanoza1})-(\ref{nano}) we obtain (\ref{vika1}).
$\blacksquare$

Consider a boundary value problem
$$
\mathcal{K}(x,{D})u = \left(4 \frac{\partial}{\partial{
z}}\frac{\partial}{\partial{\overline z}} + 2
{A}\frac{\partial}{\partial{z}} +2{B}
\frac{\partial}{\partial\overline z}\right)u = f
\quad \text{in} \quad \Omega, \quad
u \vert_{\partial \Omega} = 0.
$$

For this problem we have the following Carleman estimate with
boundary terms.

\begin{proposition}(\cite{IUY})\label{Theorem 2.1}
Suppose that $\Phi$ satisfies (\ref{1'})-(\ref{kk}), $u\in
H^1_0(\Omega)$ and $\Vert A\Vert_{L^\infty(\Omega)} +\Vert B\Vert
_{L^\infty(\Omega)}\le K$. Then there exist $\tau_0=\tau_0(K,\Phi)$
and $C=C(K,\Phi)$ independent of $u$ and $\tau$ such that for all
$\vert\tau\vert>\tau_0$

\begin{eqnarray}\label{suno4} \vert \tau\vert\Vert
ue^{\tau\varphi}\Vert^2_{L^2(\Omega)}+\Vert
ue^{\tau\varphi}\Vert^2_{H^1(\Omega)}+\Vert\frac{\partial
u}{\partial\nu}e^{\tau\varphi}\Vert^2_{L^2(\Gamma_0)}+
\tau^2\Vert\vert\frac{\partial\Phi}{\partial z} \vert
ue^{\tau\varphi}\Vert^2_{L^2(\Omega)}\nonumber \\
\le C\left(\Vert ({\mathcal K }(x,D)
u)e^{\tau\varphi}\Vert^2_{L^2(\Omega)}+\vert\tau\vert
\int_{\partial\Omega\setminus\Gamma^*_0}\vert \frac{\partial
u}{\partial\nu}\vert^2e^{2\tau\varphi}d\sigma\right).
\end{eqnarray}
\end{proposition}

{\bf Proof.} Denote $\widetilde v=ue^{\tau\varphi}$ and ${\mathcal
K}(x,D) u=f.$ Observe that
$\varphi(x_1,x_2)=\frac{1}{2}(\Phi(z)+\overline{\Phi(z)})$.
Therefore
\begin{equation}\nonumber
e^{\tau\varphi}\Delta (e^{-\tau\varphi} \widetilde v)
= \left(2\frac{\partial}{\partial z} -{\tau}\frac{\partial\Phi}
{\partial z}\right)
\left(2\frac{\partial}{\partial\overline z}
-{\tau}\frac{\partial\overline\Phi} {\partial \overline
z}\right)\widetilde v
=\widetilde f
= \left(f-2B\frac{\partial u}{\partial \overline
z} -2A\frac{\partial u}{\partial z}\right)e^{\tau\varphi}.
\end{equation}
Assume now that $u$ is a real-valued function. Denote $\widetilde
w=( 2\frac{\partial}{\partial\overline z}
-{\tau}\frac{\partial\overline\Phi} {\partial \overline
z})\widetilde v.
$

Thanks to the zero Dirichlet boundary condition for $u$, we have
$$
\widetilde w\vert_{\partial\Omega}= 2\frac{\partial\widetilde
v}{\partial \overline z}\vert_{\partial\Omega}=(\nu_1+i\nu_2)
\frac{\partial \widetilde v}{\partial
\nu}\vert_{\partial\Omega}.\,\,
$$
Let $\mathcal C$ be some smooth real-valued vector function in
$\Omega$ such that
$$
2\frac{\partial \mathcal C}{\partial z} =C(x)=C_1(x)+iC_2(x)\quad
\mbox{in}\,\,\Omega, \quad \mbox{Im} \,{\mathcal C}=0 \quad \mbox
{on}\,\,\Gamma_0,
$$
where $\vec C=(C_1,C_2)$ is a smooth function in $\Omega$ such that
\begin{equation}\label{zopaW}
div\, \vec C=1\quad \mbox{in}\,\,\Omega,\quad (\nu,\vec C)=-1 \quad
\mbox{on}\,\,\Gamma_0.
\end{equation}

By Proposition \ref{Proposition 2.1} we have the following integral
equality:
\begin{eqnarray}\label{ipolit}
\Vert \frac{\partial (\widetilde we^{N \mathcal C})}{\partial{x_1}}
-i\mbox{Im}\left(\tau\frac{\partial\overline\Phi} {\partial{\overline z}}
+NC \right)
(\widetilde we^{N \mathcal C})\Vert^2_{L^2(\Omega)}
- \int_{\partial\Omega}\left(\tau\frac{\partial \varphi}{\partial\nu} +
N(\nu_1C_1+\nu_2C_2)\right)\vert \frac{\partial\widetilde
v}{\partial\nu}e^{N\mathcal
C}\vert^2d\sigma\nonumber\\
+ N\int_\Omega\vert \widetilde we^{N \mathcal C}\vert^2dx +
\mbox{Re}\int_{\partial\Omega} i \frac{\partial}{\partial\vec\tau}
(\widetilde{w}e^{N \mathcal{C}})\overline{\widetilde
we^{N\mathcal{C}}}d\sigma\\
+\Vert -\frac 1i \frac{\partial (\widetilde {w} e^{N
\mathcal{C}})}{\partial{x_2}}-\mbox{Re}(\tau\frac{\partial\Phi}
{\partial{z}}+NC )(\widetilde we^{N \mathcal C})\Vert^2
_{L^2(\Omega)}=\Vert \widetilde fe^{\tau\varphi+N \mathcal
C}\Vert^2_{L^2(\Omega)}.\nonumber
\end{eqnarray}

We now simplify the integral $\mbox{Re}\,i\int_{\partial\Omega}
\frac{\partial}{\partial\vec\tau}(\widetilde we^{N\mathcal C})
\overline{\widetilde w_1e^{N\mathcal C}}d\sigma.$ We recall that
$\widetilde v=ue^{\tau \varphi}$ in $\Omega$ and $\widetilde
w=(\nu_1+i\nu_2) \frac{\partial \widetilde v}{\partial \nu} =
(\nu_1+i\nu_2)\frac{\partial u}{\partial \nu} e^{\tau \varphi}$ on
$\partial\Omega$. Denote $(\nu_1+i\nu_2)e^{N i\mbox{Im}\mathcal C} =
R + iP$ where $R, P$ are real-valued.  Therefore
\begin{eqnarray} \label{leonid}
\mbox{Re}\int_{\partial\Omega}
i\frac{\partial}{\partial\vec\tau}(\widetilde we^{ N\mathcal C})
\overline{\widetilde we^{ N\mathcal
C}}d\sigma\\
= \mbox{Re}\int_{\partial\Omega}i
\frac{\partial}{\partial\vec\tau}\left((R+iP) \frac{\partial
u}{\partial\nu}e^{\tau \varphi+N\mbox{Re} \,{\mathcal
C}}\right)(R-iP) \frac{\partial u}{\partial\nu}
e^{\tau\varphi+N\mbox{Re}\,{\mathcal C}}d\sigma\nonumber\\
= \mbox{Re}\int_{\partial\Omega}i
\frac{\partial}{\partial\vec\tau}(R+iP) \left\vert\frac{\partial
(\widetilde v e^{N{\mathcal C}})}{\partial\nu}\right\vert^2
(R-iP)d\sigma.\nonumber
\end{eqnarray}

Using the above formula in (\ref{ipolit}), we obtain
\begin{eqnarray} \label{suno2}
 \Vert \frac{\partial (\widetilde
we^{N \mathcal C})}{\partial{x_1}}-i\mbox{Im}(\tau\frac{\partial\overline\Phi}
{\partial{\overline z}}+NC ) (\widetilde we^{N \mathcal
C})\Vert^2_{L^2(\Omega)}
- \int_{\partial\Omega}\left(\tau\frac{\partial
\varphi}{\partial\nu} + N(\nu_1C_1+\nu_2C_2)\right)\vert
\frac{\partial\widetilde v}{\partial\nu}e^{N\mathcal
C}\vert^2d\sigma\nonumber\\
+N\int_\Omega\vert \widetilde we^{N \mathcal C}\vert^2dx +
\mbox{Re}\int_{\partial\Omega} i \frac{\partial}{\partial\vec\tau}
(R+iP) \left\vert\frac{\partial (\widetilde v
e^{N\mbox{Re}\,{\mathcal C}})}{\partial\nu}
\right\vert^2 (R-iP)d\sigma\nonumber\\
+\Vert -\frac 1i \frac{\partial (\widetilde we^{N \mathcal
C})}{\partial{x_2}}-\mbox{Re}(\tau\frac{\partial\Phi}
{\partial{z}}+NC )(\widetilde w e^{N \mathcal C})\Vert^2
_{L^2(\Omega)}=\Vert \widetilde fe^{\tau\varphi+N \mathcal
C}\Vert^2_{L^2(\Omega)}.
\end{eqnarray}

Taking a sufficiently large positive parameter $N$ and taking into
account that the function $R+iP$ is independent of $N$ on
$\Gamma^*_0$, we conclude from (\ref{suno2}), (\ref{zopaW})
\begin{eqnarray} \label{suno5}
- \int_{\partial\Omega}\left(\tau\frac{\partial\varphi}{\partial\nu} +
\frac{N}{2}(\nu_1C_1+\nu_2C_2)\right)\vert \frac{\partial\widetilde
v}{\partial\nu}e^{N\mathcal C}\vert^2d\sigma + N\int_\Omega\vert
\widetilde
we^{N \mathcal C}\vert^2dx\\
\le \Vert \widetilde fe^{\tau\varphi +N \mathcal C}\Vert^2_{L^2(\Omega)}
+ C(N)\int_{\partial\Omega\setminus\Gamma_0}\vert
\frac{\partial\widetilde v}{\partial\nu}e^{N\mathcal
C}\vert^2d\sigma. \nonumber
\end{eqnarray}

Simple computations give
\begin{eqnarray}4\Vert \frac{\partial
(\widetilde v e^{N\mbox{Re}\,\mathcal C})}{\partial \overline z}
\Vert^2_{L^2(\Omega)} +
\tau^2\Vert\overline{\frac{\partial\Phi}{\partial z}} (\widetilde v
e^{N\mbox{Re}\,\mathcal C})\Vert^2_{L^2(\Omega)} = \Vert
2\frac{\partial (\widetilde v e^{N\mbox{Re}\,\mathcal C})} {\partial
\overline z} -\tau\overline{\frac{\partial\Phi}{\partial z}} (\widetilde
v e^{N\mbox{Re}\,\mathcal C})\Vert^2_{L^2(\Omega)}
                                 \nonumber\\
= \Vert e^{N\mbox{Re}\,\mathcal C}(2\frac{\partial \widetilde v }
{\partial \overline z} -(\tau\overline{\frac{\partial\Phi}{\partial z}}
+2\frac{\partial N\mbox{Re}\,\mathcal C}{\partial \overline z} )
\widetilde v)\Vert^2_{L^2(\Omega)}                                                    \nonumber\\
\le 2 \Vert \widetilde w e^{N \mathcal C}\Vert^2_{L^2(\Omega)} +C(N)
\Vert u e^{\tau\varphi}\Vert^2_{L^2(\Omega)}.
\end{eqnarray}

Since the function $\Phi$ has zeros of at most second order by
assumption (\ref{22}), there exists a constant $C>0$ independent of
$\tau$ such that
\begin{equation}\label{(2.23)}
\tau\Vert \widetilde v  e^{N\mbox{Re}\,\mathcal
C}\Vert^2_{L^2(\Omega)}
\le C\left(\Vert \widetilde v
e^{N\mbox{Re}\,\mathcal C}\Vert^2 _{H^1(\Omega)} +
\tau^2\Vert\vert\frac{\partial\Phi}{\partial z} \vert \widetilde v
e^{N\mbox{Re}\,\mathcal C}\Vert^2_{L^2(\Omega)}\right).
\end{equation}
Therefore by (\ref{suno5})-(\ref{(2.23)}) there exists $N_0>0$ such
that for any $N>N_0$ there exists $\tau_0(N)$ such that
\begin{eqnarray}\label{xxx}
-\int_{\partial\Omega}\left(\tau\frac{\partial\varphi}{\partial\nu} +
\frac{N}{2}(\nu_1C_1+\nu_2C_2)\right)\vert \frac{\partial\widetilde
v}{\partial\nu}e^{N\mathcal C}\vert^2d\sigma +
\frac{N}{2}\int_\Omega\vert \widetilde
we^{N \mathcal C}\vert^2dx\nonumber\\
+\tau\Vert \widetilde v  e^{N\mbox{Re}\,\mathcal
C}\Vert^2_{L^2(\Omega)} +\Vert \widetilde v  e^{N\mbox{Re}\,\mathcal
C}\Vert^2
_{H^1(\Omega)}+\tau^2\Vert\vert\frac{\partial\Phi}{\partial z} \vert
\widetilde v e^{N\mbox{Re}\,\mathcal C}\Vert^2_{L^2(\Omega)}
\nonumber\\
\le \Vert \widetilde fe^{\tau\varphi+N \mathcal C}\Vert^2 _{L^2(\Omega)}
+ C(N)\int_{\partial\Omega\setminus\Gamma_0^*}\vert
\frac{\partial\widetilde v}{\partial\nu}e^{N\mathcal C}
\vert^2d\sigma
\end{eqnarray}
for all $\tau > \tau_0(N)$.

In order to remove the assumption that $u$ is real-valued,
we obtain (\ref{xxx}) separately for the real and
imaginary parts of $u$ and combine them. This concludes the proof of
the proposition. $\blacksquare$

As a corollary we derive a Carleman inequality for the function $u$
which satisfies the integral equality
\begin{equation}\label{nos}
(u, \mathcal K(x,D)^*w)_{L^2(\Omega)}+(f,w)_{H^{1,\tau}(\Omega)} +
(g e^{\tau\varphi},e^{-\tau\varphi} w)_{H^{\frac
12,\tau}(\partial\Omega\setminus\Gamma_0)}=0
\end{equation}
for all $w\in \mathcal X=\{w\in H^1(\Omega)\vert
w\vert_{\Gamma_0}=0, \quad\mathcal K(x,D)^*w\in L^2(\Omega)\}.$ We
have
\begin{corollary}\label{mymy}
Suppose that $\Phi$ satisfies (\ref{1'})-(\ref{kk}), $f\in
H^1(\Omega), g\in H^\frac 12 (\partial\Omega\setminus\Gamma_0),$
$u\in L^2(\Omega)$ and the coefficients $A,B$ of $\mathcal K(x,D)$
belong to $\{ C\in C^1(\overline\Omega)\vert \Vert
C\Vert_{C^1(\overline\Omega)}\le K\}.$ Then there exist
$\tau_0=\tau_0(K,\Phi)$ and $C=C(K,\Phi)$, independent of $u$ and
$\tau$, such that
\begin{eqnarray}\label{suno444}
\Vert ue^{\tau\varphi}\Vert^2_{L^2(\Omega)} \le C\vert\tau\vert
(\Vert fe^{\tau\varphi} \Vert^2_{H^{1,\tau}(\Omega)}+ \Vert g
e^{\tau\varphi}\Vert^2_{H^{\frac
12,\tau}(\partial\Omega\setminus\Gamma_0)}) \quad
\forall\vert\tau\vert\ge \tau_0
\end{eqnarray}
for solutions of (\ref{nos}).
\end{corollary}

{\bf Proof.} Let $\epsilon$ be some positive number and $d(x)$ be a
smooth positive function on $\partial\Omega\setminus\Gamma_0$ which
blows up like $\frac{1}{\vert x- y\vert^8}$ for any
$y\in\partial(\partial\Omega\setminus\Gamma_0).$ Consider an
extremal problem
\begin{equation}\label{1}
J_\epsilon(w)=\frac{1}{2}\Vert
we^{-\tau\varphi}\Vert^2_{L^2(\Omega)} +\frac{1}{2\epsilon}\Vert
\mathcal K(x,D)^*w-ue^{2\tau\varphi}\Vert^2
_{L^2(\Omega)}+\frac{1}{2\vert\tau\vert}\Vert w
e^{-\tau\varphi}\Vert^2
_{L^2_d(\partial\Omega\setminus\Gamma_0)}\rightarrow \inf
\end{equation}
for
\begin{equation}\label{2}
w\in \widehat{\mathcal X}=\{w\in H^\frac 12 (\Omega)\vert \mathcal
K(x,D)^*w\in L^2(\Omega), w\vert_{\Gamma_0}=0\}.
\end{equation}
There exists a unique solution to (\ref{1}), (\ref{2})  which we
denote by $\widehat w_\epsilon.$ By Fermat's theorem, we have
\begin{equation}\label{++}
J_\epsilon'(\widehat w_\epsilon)[\delta]=0\quad \forall \delta\in
\widehat{\mathcal X}. \end{equation}
 Using the notation
$p_\epsilon=\frac 1\epsilon (\mathcal K(x,D)^* \widehat
w_\epsilon-ue^{2\tau\varphi})$, we see that
\begin{equation}\label{3}
\mathcal K(x,D)p_\epsilon+\widehat w_\epsilon e^{-2\tau\varphi}=0
\quad\mbox{in}\,\,\Omega, \quad p_\epsilon\vert_{\partial\Omega}=0,
\quad \frac{\partial
p_\epsilon}{\partial\nu}\vert_{\partial\Omega\setminus\Gamma_0}
=d\frac{\widehat w_\epsilon}{\vert\tau\vert} e^{-2\tau\varphi}.
\end{equation}
By Proposition \ref{Theorem 2.1} we have
\begin{eqnarray}\label{suno4445A} \vert \tau\vert\Vert
p_\epsilon e^{\tau\varphi}\Vert^2_{L^2(\Omega)}+\Vert p_\epsilon
e^{\tau\varphi}\Vert^2_{H^1(\Omega)}+\Vert\frac{\partial
p_\epsilon}{\partial\nu}e^{\tau\varphi}\Vert^2_{L^2(\Gamma_0)}+
\tau^2\Vert\vert\frac{\partial\Phi}{\partial z} \vert p_\epsilon
e^{\tau\varphi}\Vert^2_{L^2(\Omega)}\nonumber\\
\le C\left(\Vert \widehat w_\epsilon
e^{-\tau\varphi}\Vert^2_{L^2(\Omega)} +\frac{1}{\vert\tau\vert}
\int_{\partial\Omega\setminus\Gamma_0}\vert\widehat
w_\epsilon\vert^2e^{-2\tau\varphi}d\sigma\right)
\le 2CJ_\epsilon(\widehat w_\epsilon).
\end{eqnarray}

Substituting $\delta = \widehat w_{\epsilon}$ in (\ref{++}),
we obtain
$$
2J_\epsilon(\widehat w_\epsilon)+\mbox{Re}(u  e^{2\tau\varphi},
p_\epsilon) _{L^2(\Omega)}=0.
$$
Applying estimate (\ref{suno4445A}) to the second term of the above equality,
we have
$$
{\vert\tau\vert}J_\epsilon(\widehat w_\epsilon) \le C\Vert
ue^{\tau\varphi}\Vert^2_{L^2(\Omega)}.
$$
Using this estimate, we pass to the limit in (\ref{3}) as $\epsilon$
goes to zero. We obtain
\begin{equation}\label{13'}
\mathcal K(x,D)p+\widehat we^{-2\tau\varphi}=0\quad\mbox{in}
\,\,\Omega,\quad  p\vert_{\partial\Omega}=0, \quad \frac{\partial
p}{\partial\nu}\vert_{\partial\Omega\setminus\Gamma_0} =
d\frac{\widehat w}{\vert \tau\vert}e^{-2\tau\varphi},
\end{equation}
\begin{equation}\label{003'}
\mathcal K(x,D)^*\widehat w- ue^{2\tau\varphi}=0\quad\mbox{in}
\,\,\Omega,\quad \widehat w\vert_{\Gamma_0}=0
\end{equation}
and
\begin{equation}\label{omo}
\vert\tau\vert\Vert \widehat w e^{-\tau\varphi}\Vert^2_{L^2(\Omega)}
+\Vert \widehat w
e^{-\tau\varphi}\Vert^2_{L^2(\partial\Omega\setminus\Gamma_0)} \le
C\Vert ue^{\tau\varphi}\Vert^2_{L^2(\Omega)}.
\end{equation}
Since $\widehat w\in L^2(\Omega)$, we have $p\in H^2(\Omega)$ and
$\frac{\partial p}{\partial \nu}\in H^\frac
12(\partial\Omega)$ by the trace theorem.
The relation (\ref{13'}) implies $\widehat w\in
H^\frac 12(\partial\Omega\setminus\Gamma_0)$.  Since $\widehat w\in
L^2_d(\partial\Omega\setminus\Gamma_0)$ and $\widehat
w\vert_{\Gamma_0}=0$, we have $\widehat w\in H^\frac
12(\partial\Omega).$ By (\ref{suno4445A})-(\ref{omo}), we obtain
\begin{equation}\label{elkao}
\Vert \widehat w e^{-\tau\varphi}\Vert_{H^{\frac 12, \tau}
(\partial\Omega)} \le  C \vert\tau\vert^\frac 12\Vert
ue^{\tau\varphi} \Vert_{L^2(\Omega)}.
\end{equation}
Taking the scalar product of (\ref{003'}) with $\widehat w
e^{-2\tau\varphi}$ and using estimates (\ref{elkao}) and
(\ref{omo}), we obtain
\begin{equation}\label{omm}
\frac{1}{\vert \tau\vert}\Vert \nabla \widehat w e^{-\tau\varphi}
\Vert^2_{L^2(\Omega)}+\vert\tau\vert\Vert \widehat w
e^{-\tau\varphi} \Vert^2_{L^2(\Omega)}+\frac{1}{\vert\tau\vert}
\Vert \widehat w e^{-\tau\varphi}\Vert^2 _{H^{\frac
12,\tau}(\partial\Omega\setminus\Gamma_0)}\le C\Vert
ue^{\tau\varphi} \Vert^2_{L^2(\Omega)}.
\end{equation}
From this estimate and a standard duality argument, the statement of
Corollary \ref{mymy} follows immediately. $\blacksquare$

Consider the following problem
\begin{equation}\label{(2.26)}
L_q(x,D)u = \Delta u  + qu = f e^{\tau\varphi}\quad
\mbox{in}\,\,\Omega,\quad u\vert_{\Gamma_0}=g e^{\tau\varphi}.
\end{equation}

We have
\begin{proposition}\label{Proposition 2.3}
Let $q\in L^\infty(\Omega),$ $\Phi$ satisfy (\ref{1'})-(\ref{kk}),
$f\in L^2(\Omega), g\in H^\frac 12(\partial\Omega). $ There exists
$\tau_0>0$ such that for all $\vert\tau\vert>\tau_0$ there exists a
solution to the boundary value problem (\ref{(2.26)}) such that
\begin{equation} \label{(2.27)} \frac{1}{\root\of{\vert
\tau\vert}}\Vert \nabla ue^{-\tau\varphi}
\Vert_{L^2(\Omega)}+\root\of{\vert \tau\vert}\Vert
ue^{-\tau\varphi}\Vert_{L^2(\Omega)}\le C(\Vert
f\Vert_{L^2(\Omega)}+\Vert g\Vert_{H^{\frac 12,
\tau}(\partial\Omega)}).
\end{equation}

\end{proposition}

{\bf Proof.} First we reduce the problem (\ref{(2.26)})  to the case
$g=0.$ Let $r(z)$ be a holomorphic function and $\widetilde
r(\overline z) $ be an antiholomorphic function  such that
$(r+\widetilde r)\vert_{\Gamma_0}=g$ and
$$
\Vert r\Vert_{H^1(\Omega)}+\Vert\widetilde r\Vert_{H^1(\Omega)} \le
C\Vert g\Vert_{H^\frac 12(\partial\Omega)}.
$$

 The existence of such
functions $r,\widetilde r$ follows from the Fredholm theorem
combined with the possibility of an arbitrary choice of the
Dirichlet data on the part of the boundary.

We look for a solution $u$ in the form
$$
u=(e^{\tau\Phi}r +e^{\tau\overline\Phi}\widetilde r)+\widetilde u,
$$
where
\begin{equation}
L_q(x,D)\widetilde u=\widetilde f e^{\tau\varphi}\quad \mbox{in}
\,\,\Omega,\quad \widetilde u\vert_{\Gamma_0}=0
\end{equation}
and $\widetilde f= f-qre^{i\tau\psi}-q\widetilde re^{-i\tau\psi}.$

In order to prove (\ref{(2.27)}), we consider the following extremal
problem:
\begin{equation}\label{00001'}
\widetilde I_\epsilon(u)=\frac{1}{2}\Vert ue^{-\tau\varphi}\Vert^2
_{H^{1,\tau}(\Omega)} + \frac{1}{2\epsilon}\Vert
L_q(x,D)u-\widetilde fe^{\tau\varphi}
\Vert^2_{L^2(\Omega)}+\frac{1}{2}\Vert u e^{-\tau\varphi}
\Vert^2_{H^{\frac
12,\tau}(\partial\Omega\setminus\Gamma_0)}\rightarrow \inf
\end{equation}
for
\begin{equation}\label{02'}
u\in {\mathcal Y}=\{w\in H^1(\Omega)\vert L_q(x,D)w\in
L^2(\Omega),\quad w\vert_{\Gamma_0}=0\}.
\end{equation}
There exists a unique solution to problem (\ref{00001'}),
(\ref{02'}) which we denote by $\widehat u_\epsilon.$ By Fermat's
theorem, we have
\begin{equation}\label{NANA}
\widetilde I_\epsilon'(\widehat u_\epsilon)[\delta]=0\quad \forall
\delta\in \mathcal Y.
\end{equation}

Let $p_\epsilon=\frac{1}{\epsilon}(L_q(x,D)\widehat u_\epsilon
-\widetilde fe^{\tau\varphi}).$ Applying  Corollary \ref{mymy} we
obtain from (\ref{NANA})
\begin{equation}\label{suno444}\frac{1}{ \vert \tau\vert}\Vert
p_\epsilon e^{\tau\varphi}\Vert^2_{L^2(\Omega)} \le C(\Vert \widehat
u_\epsilon e^{-\tau\varphi}\Vert^2 _{H^{1,\tau}(\Omega)}+\Vert
\widehat u_\epsilon e^{-\tau\varphi}\Vert^2 _{H^{\frac 12,\tau}
(\partial\Omega\setminus\Gamma_0)}) \le 2C\widetilde
I_\epsilon(\widehat u_\epsilon).
\end{equation}

Substituting in (\ref{NANA}) with $\delta=\widehat u_\epsilon$, we
obtain
$$
2\widetilde I_\epsilon(\widehat u_\epsilon)+\mbox{Re}(p_{\epsilon},
\widetilde f  e^{\tau\varphi})_{L^2(\Omega)}=0.
$$
Applying estimate (\ref{suno444}) to this equality, we have
$$
\widetilde I_\epsilon(\widehat u_\epsilon)\le C{\vert\tau\vert}
\Vert \widetilde f\Vert^2_{L^2(\Omega)}.
$$
Using this estimate, we pass to the limit as $\epsilon \rightarrow
+0$. We obtain
\begin{equation}\label{03'}
L_q(x,D) u- \widetilde
fe^{\tau\varphi}=0\quad\mbox{in}\,\,\Omega,\quad u\vert_{\Gamma_0}=0
\end{equation}
and
\begin{equation}\label{om63}
\Vert ue^{-\tau\varphi}\Vert^2_{H^{1,\tau}(\Omega)}+\Vert
ue^{-\tau\varphi} \Vert^2_{L^2(\partial\Omega\setminus\Gamma_0)}\le
C\vert \tau\vert \Vert\widetilde f\Vert^2_{L^2(\Omega)}.
\end{equation}
Since $\Vert\widetilde f\Vert_{L^2(\Omega)}\le C(\Vert
f\Vert_{L^2(\Omega)}+\Vert g\Vert _{H^\frac 12(\partial\Omega)})$,
inequality (\ref{om63}) implies (\ref{(2.27)}).

This finishes the proof of the proposition. $\blacksquare$

\section{Uniqueness in the three dimensional case by
Dirichlet-to-Neumann map on subboundaries}

On the basis of Carleman estimates in Section 2, we show uniqueness
results in three dimensions.
We recall that $\Gamma_\pm$ are some subsets of $\partial \Omega$
and for the
Schr\"odinger operator with potential $q$ we consider the
Dirichlet-to-Neumann map $\Lambda (q, \Gamma_-, \Gamma_+)$
on subboundaries $\Gamma_-$ and $\Gamma_+$:
$$
\Lambda (q, \Gamma_-, \Gamma_+) (f)= \frac{\partial u}{\partial
\nu}\vert_{\partial\Omega\setminus\Gamma_+},
$$
where
$$ L_{q}(x,D)u=\Delta u+qu=0\quad\mbox{in}\,\,\Omega, \quad
u\vert_{\Gamma_-}=0,\quad u\vert_{\partial\Omega\setminus
\Gamma_-}=f.
$$

Consider two particular cases of the subboundaries $\Gamma_\pm.$
Let $\vec v$ be a unit vector in $\Bbb R^3.$ We introduce two
subsets of the boundary $\partial\Omega$:
\begin{equation}\label{igrok}
\Gamma_+(\vec v)=\{ x\in \partial\Omega\vert (\vec v,\vec
\nu)>0\},\quad \Gamma_-(\vec v)=\{ x\in \partial\Omega\vert (\vec
v,\vec \nu)<0\}.
\end{equation}

We will show three uniqueness results and the first two are concerned with
determination of potentials.

\begin{theorem}\label{t1} Let $n=3,$ $q_1,q_2\in L^\infty(\Omega),$ $0$
be not an eigenvalue of the Schr\"odinger operators
$L_{q_1}(x,D), L_{q_2}(x,D)$
and $\Lambda (q_1, \Gamma_-(\vec v), \Gamma_+(\vec
v))=\Lambda (q_2, \Gamma_-(\vec v), \Gamma_+(\vec v))$ for some
unit vector $\vec v.$ Then $q_1=q_2$ in $\Omega$.
\end{theorem}

Let $x^0$ be a point in $\Bbb R^3$ such that this point and domain
$\Omega$ are separated by some plane.

We introduce the following subsets of $\partial\Omega$:
$$
\Gamma_+(x^0)=\{ x\in \partial\Omega\vert (x-x^0,\vec \nu)>0\},
\quad \mbox{and}\quad \Gamma_-(x^0)=\{ x\in
\partial\Omega\vert (x-x^0,\vec \nu)<0\}.
$$
We have
\begin{theorem}\label{t2}
Let $n=3,$ $q_1,q_2\in L^2(\Omega)$, $0$ be not an
 eigenvalue of the Schr\"odinger operators $L_{q_1}(x,D),L_{q_2}(x,D)$ and
$\Lambda (q_1, \Gamma_-(x^0),
\Gamma_+(x^0 ))=\Lambda (q_2, \Gamma_-(x^0), \Gamma_+(x^0))$ for
some $x^0$ which can be separated from $\overline \Omega$ by a plane.
 Then $q_1=q_2$ in $\Omega$.
\end{theorem}

{\bf Remark 1.} {\it Theorem \ref{t2} improved the result of
Theorem 1.1 of \cite{KSU}.  Unlike \cite{KSU}, we do not need consider the
 neighborhoods of the sets  $F(x_0)$ and $B(x_0)\cup
\{x\in\partial\Omega\vert (x-x_0,\vec\nu)=0\}$
(here we are using notations of \cite{KSU}),
but precisely these sets are sufficient as subboundaries.
This in turn reduces the amount of the information which is used
  for the determination of a potential of the Schr\"odinger equation.}
\\
{\bf Remark 2.}  {\it The assumptions of Theorems \ref{t1}, \ref{t2}
and the corresponding theorems from \cite{Bu}, \cite{KSU} require
the access to the whole boundary $\partial\Omega$, that is, to any point of
the boundary we have to either apply the voltage or measure current. The
Calder\'on's problem was motivated by the search of the oil fields
which are located underground, but voltage and current
should be measured only on the surface.
It is the most interesting and practically
important that we need not apply voltage and not measure the current
on the sufficiently large part of the boundary.
In the three dimensional case, there are a few results results in
this direction.  The paper \cite{I} treats the case when roughly
speaking $\Omega $ is a half of the plane or sphere.  During the
preparation of this manuscript two more articles appeared:
\cite{IY5} and \cite{KM}.  The paper \cite{IY5} established the
uniqueness in the case of cylindrical domain.}

{\bf Proof of Theorem 4.}
Without loss of generality, performing a rotation around the origin
if necessary, we can assume that $\vec v=\vec e_3=(0,0,1)$ and set
\begin{equation}\label{region}
\Gamma_+=\{ x\in \partial\Omega\vert (\vec e_3,\vec \nu)>0\},\quad
\Gamma_-=\{ x\in \partial\Omega\vert (\vec e_3,\vec \nu)<0\}.
\end{equation}

Let $b(s)\in C^2(\Bbb R^1)$ be an arbitrary function and $z\in \Bbb C^1, \theta\in [0,2\pi]$
are some parameters and $g(\theta,x')=b(sin(\theta)x_1-cos(\theta)x_2).$
 We construct a complex geometric optics solution  for
the Schr\"odinger equation in the form
\begin{equation}\label{zopaA}
u_1(x)=e^{(\tau+i z) \Phi}g(\theta,x')+e^{\tau
x_3}o_{L^2(\Omega)}(1)\quad \mbox{as}\,\,\tau\rightarrow +\infty,
\end{equation}
where $ \Phi(x)=x_3+i(cos(\theta)x_1+sin(\theta)x_2). $

Indeed
$$
\Delta (e^{(\tau+i z)
\Phi}g(\theta,x'))=((\tau+iz)^2(\nabla\Phi,\nabla\Phi)+2(\tau+i
z)(\nabla\Phi,\nabla g)+\Delta g)e^{(\tau+i z)\Phi}.
$$
Observing that $(\nabla\Phi,\nabla\Phi)=(\nabla\Phi,\nabla g)=0$,
we obtain
\begin{equation}\label{e11}
\Delta (e^{(\tau+i z) \Phi}g(\theta,x'))=(\Delta g)e^{(\tau+i
z)\Phi}.
\end{equation}

Then using Corollary \ref{ZZZ}  we construct the function
$w_\tau(z,\cdot)$ which solves the boundary value problem
\begin{equation}\label{e12}
L_{q_1}(x,D)w_\tau=-e^{(\tau+i z)\Phi} L_{q_1}(x,D)g\quad\mbox{in}
\,\,\Omega,\quad w_\tau\vert_{\Gamma_-}=-ge^{(\tau+i z)\Phi}
\end{equation}and satisfies the estimate
\begin{equation}\label{-0i}
\Vert e^{-\tau
x_3}w_\tau\Vert_{L^2(\Omega)}=o(1)\quad\mbox{as}\quad\tau\rightarrow
+\infty.
\end{equation}
 By (\ref{e11}) and (\ref{e12}), the function
$$
u_1=ge^{(\tau+i z)\Phi}+w_\tau
$$
solves the Schr\"odinger equation
\begin{equation}\label{eeq1}
L_{q_1}(x,D)u_1=0\quad\mbox{in}\,\,\Omega, \quad
u_1\vert_{\Gamma_-}=0
\end{equation}
and admits the asymptotic expansion (\ref{zopaA}) by (\ref{-0i}).

Since $\Lambda (q_1, \Gamma_-, \Gamma_+)=\Lambda (q_2, \Gamma_-,
\Gamma_+)$, there exists $u_2$ such that
\begin{equation}\label{eq2}
L_{q_2}(x,D)u_2=0\,\,\mbox{in}\,\,\Omega, \quad
(u_1-u_2)\vert_{\partial\Omega}=\frac{\partial
(u_1-u_2)}{\partial\nu}\vert_{\partial\Omega\setminus\Gamma_+}=0.
\end{equation}

Next, in a way similar to the construction of $u_1$, we construct the
complex geometric optics solution $v$ to the Schr\"odinger operator
with potential $q_2$
\begin{equation}\label{eq4}
L_{q_2}(x,D)v=0 \quad \mbox{in }\,\,\Omega, \quad
v\vert_{\Gamma_+}=0
\end{equation}
in the form
\begin{equation}\label{zopa1}
v(x)=e^{-\tau\Phi} +e^{-\tau x_3} o_{L^2(\Omega)}( 1)\quad
\mbox{as}\,\,\tau\rightarrow +\infty.
\end{equation}
Setting $u=u_1-u_2,$  by (\ref{eeq1}) and (\ref{eq2}), we have
\begin{equation}\label{eq3}
L_{q_2}(x,D)u = -(q_1-q_2)u_1\quad \mbox{in }\,\,\Omega, \,\,
u\vert_{\partial\Omega}=0,\,\, \frac{\partial u}{\partial
\nu}\vert_{\partial\Omega\setminus\Gamma_+}=0.
\end{equation}

Then using (\ref{eq4}) and (\ref{eq3}), we obtain
\begin{eqnarray}
-\int_\Omega (q_1-q_2)u_1vdx
= (L_{q_2}(x,D)u,v)_{L^2(\Omega)}=(u,L_{q_2}(x,D)v)_{L^2(\Omega)}
+(\frac{\partial u}{\partial\nu},v)_{L^2(\partial\Omega)}-(u,\frac{\partial
v}{\partial\nu})_{L^2(\partial\Omega)}   \nonumber \\
=(v,\frac{\partial u}{\partial\nu})_{L^2(\partial\Omega\setminus\Gamma_+)}
+(v,\frac{\partial u}{\partial\nu})_{L^2(\Gamma_+)} = 0.
\end{eqnarray}
Hence
$$
\int_\Omega (q_1-q_2)e^{i z \Phi}g(\theta,x')dx=0.
$$

Let $\Pi=G\times [-K,K]$ be such a cylinder that $\Omega\subset
\Pi.$ We extend the function $(q_1-q_2)$ by zero on
$\Pi\setminus\Omega$ and set
$$
r_z(x')=\int_{-K}^K(q_1-q_2)e^{i zx_3}dx_3, \quad
r_{z,k}(x')=\int_{-K}^K(q_1-q_2)e^{i zx_3}(ix_3)^kdx_3.
$$
Therefore we have
$$
\int_Gr_z(x') e^{- z(cos(\theta)x_1+sin(\theta)x_2)}g(\theta,x')dx'=0.
$$
Then for any $\omega\in \Bbb S^1$ and any $p\in \Bbb R^1$
\begin{equation}\label{e14!}
\frak P(z,\omega,p)=\int_{<\omega,x'>=p}r_z e^{ z (\omega^\bot
,x')}ds=0.
\end{equation}
For any fixed $(\omega,p)\in \Bbb S^1\times \Bbb R^1$ the function $
\frak P(z,\omega,p)$ is holomorphic in the variable $z$. Therefore,
by (\ref{e14!}) we obtain
\begin{equation}\label{eeq8} \frac{\partial^\ell \frak P}{\partial
z^\ell}(0,\omega,p)=\int_{<\omega,x'>=p}\int_{-K}^K(q_1-q_2)(ix_3+(\omega^\bot
,x'))^\ell ds=0 \quad \forall \ell \in\Bbb N_+.
\end{equation}

We claim that \begin{equation}\label{ee7} r_{0,k}\equiv 0 \quad
\forall k\in \Bbb N_+.
\end{equation}
From (\ref{eeq8}) there exist constants $C_{k,\ell}$ such that
\begin{equation}\label{eq9}
\int_{<\omega,x'>=p}r_{0,k}
ds=\sum_{\ell=0}^{k-1}C_{k,\ell}\int_{<\omega,x'>=p}(\omega^\bot
,x')^{k-\ell}r_{0,\ell} ds.
\end{equation}

The function $\frak P(0,\omega,p)$ is the Radon transform of the
function $r_0.$ By the classical uniqueness result for the Radon
transform  (see e.g. Theorem 5.5, p.30 in Helgason \cite{He}),
we obtain
\begin{equation}
r_0=r_{0,0}\equiv 0. \end{equation} Suppose that the equalities
(\ref{ee7}) are already proved for  all $\ell$ less than $k.$ Then
equality (\ref{eq9}) immediately implies (\ref{ee7}) for $\ell=k.$

The function $r_z$ is holomorphic in the variable $z$ for any fixed
$x'$. Equality (\ref{ee7}) implies that derivatives of any orders
with respect to $z$ of this function are equal to zero.
Hence
$$
r_z=0\quad\forall z\in \Bbb C^1\,\,\mbox{and}\,\, x'\in G.
$$
Hence $q_1=q_2.$  The proof of the Theorem \ref{t1} is complete.
$\blacksquare$

{\bf Proof of Theorem 5.}
Without loss of generality we can assume that
$x^0=0$ and set $\Gamma_\pm=\Gamma_\pm (0).$ In the spherical
coordinates the Laplace operator has the form:
\begin{equation}\label{eqq1}
\Delta u=\frac{1}{r^2}\frac{\partial}{\partial
r}(r^2\frac{\partial u}{\partial r})+ \frac
{1}{r^2sin(\theta)}\frac{\partial}{\partial\theta}(sin(\theta)\frac{\partial
u}{\partial\theta})+ \frac{1}{r^2 sin^2(\theta)}\frac{\partial^2
u}{\partial\varphi^2}=\frac{\partial^2 u}{\partial r^2}+
\frac{1}{r^2}\frac{\partial^2 u}{\partial^2\theta}+\frac{1}{r^2
sin^2(\theta)}\frac{\partial^2 u}{\partial\varphi^2}+
\frac{2}{r}\frac{\partial u}{\partial
r}+\frac{cot(\theta)}{r^2}\frac{\partial u}{\partial\theta}.
\end{equation}
The function $\Phi(x)=\varphi+i\psi=\ln r\pm i\theta$ satisfies the
Eikonal equation.  Short computations and formula (\ref{eqq1})
imply
$$
\Delta \Phi=\frac{1}{r^2}(1\pm i cot(\theta)).
$$
Let a function $a$ satisfy the transport equation:
$$
2(\nabla a,\nabla \Phi)+\Delta\Phi a=0.
$$
In the spherical coordinates, the transport equation has the form
\begin{equation}\label{e13}
\frac{2}{r}\frac{\partial a}{\partial r}\pm i
\frac{2}{r^2}\frac{\partial a}{\partial\theta}+\frac{1}{r^2}(1\pm i
cot(\theta)) a=0.
\end{equation}
This equation admits the following solution
$$
a(r,\theta,\varphi)=\frac{1}{\root\of{r}}e^{-\frac 12 ln (sin
(\theta))} a_0(\varphi),
$$
where $a_0$ is some function from $C^2_0[0,2\pi].$

Then short computations imply
$$
\Delta (ae^{(\tau+i z)\Phi}) = \Biggl((\tau+i z)^2
a\left(\left(\frac{\partial \Phi}{\partial r}\right
)^2+\frac{1}{r^2}\left(\frac{\partial \Phi}{\partial
\theta}\right)^2\right) + (\tau+i z)\left(2\frac{\partial
\Phi}{\partial r}\frac{\partial a}{\partial r} +
\frac{2}{r^2}\frac{\partial \Phi}{\partial \theta}\frac{\partial
a}{\partial \theta}+ \left(\frac{2}{r}\frac{\partial \Phi}{\partial
r}+\frac{cot(\theta)}{r^2}\frac{\partial \Phi}{\partial\theta}
\right)a\right)
$$
$$
+\Delta a \Biggr)e^{(\tau+i z)\Phi}=(\Delta a)e^{(\tau+i z)\Phi}.
$$
Then using Corollary \ref{ZZZ}, we construct the function
$w_\tau(z,\cdot)$ which solves the boundary value problem
$$
L_{q_1}(x,D)w_\tau=-e^{(\tau+i z)\Phi}L_{q_1}(x,D)a\quad\mbox{in}
\,\,\Omega,\quad w_\tau\vert_{\Gamma_-}=-ae^{(\tau+i z)\Phi}.
$$ and satisfies the estimate
\begin{equation}\label{-0}
\Vert
e^{-\tau\varphi}w_\tau\Vert_{L^2(\Omega)}=o(1)\quad\mbox{as}\quad
\tau\rightarrow +\infty.
\end{equation}
The  function
$$u_1=ae^{(\tau+i z)\Phi}+w_\tau
$$ solves the Schr\"odinger equation
$$
L_{q_1}(x,D)u_1=0\quad\mbox{in}\,\,\Omega, \quad
u_1\vert_{\Gamma_-}=0
$$
and admits the asymptotic expansion
\begin{equation}\label{0001'}
u_1(x)=ae^{\tau\Phi} +e^{\tau \ln (r)} o_{L^2(\Omega)}(1)\quad
\mbox{as}\,\,\tau\rightarrow +\infty
\end{equation}
by (\ref{-0}).
Similarly we construct complex geometric optics solutions $v$ for
the Schr\"odinger equation with the potential $q_2$
$$
L_{q_2}(x,D)v=0 \quad \mbox{in }\,\,\Omega,\quad v\vert_{\Gamma_+}=0
$$
in the form
\begin{equation}\label{0002}
v(x)=ae^{-\tau\Phi} +e^{-\tau \ln (r)} o_{L^2(\Omega)}(1)\quad
\mbox{as}\,\,\tau\rightarrow +\infty.
\end{equation}

Since the Dirichlet-to-Neumann maps are the same, there exists a
function $u_2$ which solves the Schr\"odinger equation with
potential $q_2$ in $\Omega$ and satisfies the following equations
$$
L_{q_2}(x,D)u_2=0\,\,\mbox{in}\,\,\Omega, \quad
(u_1-u_2)\vert_{\partial\Omega}=\frac{\partial
(u_1-u_2)}{\partial\nu}\vert_{\partial\Omega\setminus\Gamma_+}=0 .
$$

Setting $u=u_1-u_2$ we have
\begin{equation}\label{0001}
L_{q_1}(x,D)u=-(q_1-q_2)u_1\quad \mbox{in }\,\,\Omega, \,\,
u\vert_{\partial\Omega}=0,\,\, \frac{\partial u}{\partial
\nu}\vert_{\partial\Omega\setminus\Gamma_+}=0. \end{equation}

Taking the scalar product in $L^2(\Omega)$ of equation (\ref{0001})
and the function $v,$ after integration by parts, we have
$$\int_\Omega (q_1-q_2)u_1vdx=0.$$ Using the asymptotic formulae
(\ref{0001'}) and (\ref{0002}) for the functions $u_1$ and $v$, we
obtain
$$
\int_R (q_1-q_2)a^2e^{i
z\Phi}r^2sin(\theta)drd\varphi d\theta =0.
$$
Here $R$ denotes the image of the domain $\Omega$ after change of
coordinates from the Cartesian to the spherical one.
Taking a sequence of functions $a_0$ converging to
$\delta(\varphi-\varphi_0)$, we obtain
\begin{equation}\label{e14}\int_{R\cap\{\varphi=\varphi_0\}}
(q_1-q_2)e^{-ln (sin (\theta))}sin(\theta)e^{i z\Phi}r dr d\theta =
\int_{R\cap\{\varphi=\varphi_0\}} (q_1-q_2) e^{i z\Phi}r dr d\theta.
\end{equation}

We introduce the functions $r_z, m_k: \Bbb S^2\rightarrow \Bbb
R^1$ as follows : for each  point on the sphere we choose the ray
$\ell$ starting from the origin and passing through this point. Then
we set $r_z=\int_\ell (q_1-q_2)r e^{i z \ln r}dr$ and
$m_k=\int_\ell (q_1-q_2)r (i\ln r)^kdr$ where $k\in \Bbb N_+.$

Then from (\ref{e14}) we obtain
\begin{equation}\label{ZA11}
 \int_0^\pi r_z e^{\pm z\theta}d\theta=0, \quad \forall
z\in \Bbb C^1 \quad\mbox{and}\quad \varphi=\varphi_0\in [0,2\pi].
\end{equation}
There exists  a hemisphere such that for each $z$, the support of the
function $r_z$ is included in this hemisphere. Let $\Xi$ be the set of
"big circles" on $\Bbb S^2.$  By "big circle" we mean any
intersection of sphere $\Bbb S^2$ and a plane which passes through the
origin.  The function $\varphi(x)=\ln r$ is invariant under
rotations around the origin.  Consequently (\ref{ZA11}) implies that

\begin{equation}\label{ZA1}
 \frak H(z,\xi)=\int_\xi r_z e^{\pm z\cdot}d\sigma=0\quad \forall
z\in \Bbb C^1\quad\mbox{and}\quad \forall \xi\in \Xi.
\end{equation}

If $z=0$, then after proper rotation of the rectangular
coordinate system around the origin, the above formula implies

\begin{equation}\label{ZA}
\int_\xi r_0d\sigma =0 \quad\forall \xi\in \Xi.
\end{equation}
The equality (\ref{ZA}) can be reformulated in the following way:
the Minkowski-Funk transform of the function $
r_0$ is identically equal to zero. Then the classical  Minkowski's
result implies  $r_0=0 $ on $\Bbb S^2$ (see e.g. \cite{P2}).

Then (\ref{ZA1}) implies that for any $\ell \in \Bbb N_+$ there
exist constants $C_{k,\ell}$ such that
\begin{equation}\label{vikont1}
\frac{\partial^\ell\frak H}{\partial z^\ell}(0,\xi)=\int_\xi m_\ell
d\sigma+\sum_{k=0}^{\ell-1}C_{k,\ell}\int_\xi\theta^{\ell-k} m_k d\sigma=0
\quad\forall \xi\in \Xi.
\end{equation}
From the above formula, the induction argument yields
\begin{equation}\label{vikont}
m_\ell=0\quad \forall \ell\in \Bbb N_+.
\end{equation}
Indeed, since $m_0=r_0$ we have (\ref{vikont}) for $\ell=0.$ Suppose
that (\ref{vikont}) is established for $\ell<k.$ Then formula
(\ref{vikont1}) implies
$$
\int_\xi m_k
d\sigma=-\sum_{\ell=0}^{k-1}C_{j,k}\int_\xi\theta^{k-\ell} m_\ell
d\sigma=0 \quad\forall \xi\in \Xi.
$$
Hence the Minkowski-Funk transform of the function $ m_\ell, \ell<k$ is
identically equal to zero and applying the  Minkowski's result again we
have $m_k\equiv 0.$

 On the other hand the function $r_z(y)$ is
holomorphic in variable $z$ for any fixed $y\in \Bbb S^2$. Since
$$
\frac{\partial^\ell r_z}{\partial z^\ell}\vert_{z=0}=m_\ell, \quad
\forall \ell\in \Bbb N_+\quad \mbox{and}\,\,y\in \Bbb S^2,
$$
we obtain
$$
r_z(y)=0\quad \mbox{on} \,\, \Bbb C^1\times \Bbb S^1.
$$
Using the definition of the function $r_z$ we obtain
immediately that
$$
q_1=q_2 \quad \mbox{in $\Omega$}.
$$
The proof of Theorem \ref{t2} is complete. $\blacksquare$
\\
\vspace{0.2cm}

Next we consider the Schr\"odinger equation with the first order terms:
$$
L_{q,A}(x,D)u=\Delta u+(A,\nabla u)+qu=0\quad \mbox{in}\,\,\Omega,
$$
where $A=(A_1,A_2,A_3)$ is a regular real-valued  vector field.
We consider the problem of determination of the potential $q$ and the vector field $A$ from the following Dirichlet-to-Neumann map:
 $$
\Lambda (q,A, \Gamma_-, \Gamma_+) (f)= \frac{\partial u}{\partial
\nu}\vert_{\partial\Omega\setminus\Gamma_+},
$$
where
\begin{equation}\label{igla} L_{q,A}(x,D)u=0\quad\mbox{in}\,\,\Omega, \quad
u\vert_{\Gamma_-}=0,\quad u\vert_{\partial\Omega\setminus
\Gamma_-}=f.
\end{equation}

We can see that a vector field $A$ and a potential $q$ cannot be
determined simultaneously from the Dirichlet-to-Neumann map.
More precisely we have the following proposition.
\begin{proposition}\label{volja}
Let $\eta\in C^2(\overline\Omega)$ be a function such that $\eta\vert_{\partial\Omega\setminus\Gamma_+\cup\partial\Omega\setminus
\Gamma_-}=0$ and $\frac{\partial\eta}{\partial
\nu}\vert_{(\partial\Omega\setminus\Gamma_+)\setminus \Gamma_-}=0.$
Then the operators $ L_{q,A}(x,D)$ and $e^{-\eta}
L_{q,A}(x,D)e^{\eta}$ generate the same Dirichlet-to-Neumann map
on $\Gamma_-$ and $\Gamma_+$.
\end{proposition}

{\bf Proof.} Denote $\widetilde q=q+\vert\nabla \eta\vert^2+\Delta
\eta+(A,\nabla \eta).$ If $u$ is the solution to equation
(\ref{igla}), then $w=u e^{-\eta}$ solves the boundary value problem
\begin{equation}\label{igla1} e^{-\eta} L_{q,A}(x,D)e^{\eta}w=L_{\widetilde q,A+\nabla\eta}(x,D)w=0\quad\mbox{in}\,\,\Omega, \quad
w\vert_{\Gamma_-}=0,\quad w\vert_{\partial\Omega\setminus
\Gamma_-}=f.
\end{equation}
Obviously
$$\frac{\partial w}{\partial
\nu}\vert_{\partial\Omega\setminus\Gamma_+\cap\Gamma_-}=(\frac{\partial u}{\partial
\nu}e^{-\eta})\vert_{\partial\Omega\setminus\Gamma_+\cap\Gamma_-} -(u e^{-\eta}\frac{\partial\eta}{\partial\nu})\vert_{\partial\Omega\setminus\Gamma_+\cap\Gamma_-}=\frac{\partial u}{\partial
\nu}\vert_{\partial\Omega\setminus\Gamma_+\cap\Gamma_-}
$$
and
$$
\frac{\partial w}{\partial
\nu}\vert_{\partial\Omega\setminus(\Gamma_+\cup\Gamma_-)}=(\frac{\partial
u}{\partial
\nu}e^{-\eta})\vert_{\partial\Omega\setminus(\Gamma_+\cup\Gamma_-)}
-(u
e^{-\eta}\frac{\partial\eta}{\partial\nu})\vert_{\partial\Omega\setminus(\Gamma_+\cup\Gamma_-)}=\frac{\partial
u}{\partial
\nu}\vert_{\partial\Omega\setminus(\Gamma_+\cup\Gamma_-)}.
$$
The proof of the proposition is finished. $\blacksquare$

We have

\begin{theorem}\label{t3}
Let $\Omega\subset\Bbb R^3$ be a bounded strictly convex domain with
smooth boundary,  $q_1,q_2\in L^\infty(\Omega), A\in C^2(\overline\Omega),$
$\vec v\ne 0$ be an arbitrary vector from $\Bbb R^3,$
the sets $ \Gamma_\pm(\vec v)$ given by $(\ref{igrok})$ , and
$\Lambda (q_1,A_1, \Gamma_-(\vec v), \Gamma_+(\vec v))=\Lambda
(q_2,A_2, \Gamma_-(\vec v), \Gamma_+(\vec v)).$  Then $\mbox{rot}\,
A_1=\mbox{ rot}\, A_2$ in $\Omega$.
\end{theorem}

{\bf Proof.} Without loss of generality, performing a rotation
around the origin if necessary, we can assume that $\vec
v=\vec e_3=(0,0,1)$ and sets $\Gamma_\pm$ are defined in
(\ref{region}).  We set $z_\theta=x_3+i(\vec\omega, x')$ and $\vec
\omega=(cos(\theta),sin(\theta)).$

Let $ \Phi(x)=x_3-i(cos(\theta)x_1+sin(\theta)x_2) $ where
$\theta\in [0,2\pi).$ We define the function $\mathcal A_1(\theta)$
as a solution to differential equation $$
4\partial_{z_\theta}\mathcal A_1(\theta)+(A_1, ( i\vec
\omega,1))=0\quad\mbox{in}\,\,\Omega, \quad \mbox{Im}\, \mathcal
A_1(\theta)\vert_{\partial\Omega}=0
$$
and set $a=\widetilde b({\overline z}_\theta)\widetilde{g}(\theta,x')
e^{\mathcal A_1(\theta)}$ where $\widetilde{g}(\theta,x')=\widetilde
b(sin(\theta)x_1-cos(\theta)x_2)$ and  $\widetilde b(s)\in C^2(\Bbb
R^1)$ be an arbitrary function. Then the function $a$ solves the
differential equation:

\begin{equation}\label{xz1}
2(\nabla\Phi,\nabla a)+(A_1,\nabla\Phi)a=0\quad\mbox{in}\,\,\Omega.
\end{equation}

Let $a_{-1}$ satisfy
\begin{equation}\label{xz2}
2(\nabla\Phi,\nabla a_{-1})+((A_1,\nabla\Phi))a_{-1}=-(\Delta+(A_1,\nabla)+q_1) a\quad\mbox{in}\,\,\Omega.
\end{equation}
 We construct the complex geometric optics solution to the Schr\"odinger
equation in the form
\begin{equation}\label{zopaA01}
u_1(x)=e^{\tau \Phi}(a+\frac{a_{-1}}{\tau})+e^{\tau
x_3}o_{H^{1,\tau}(\Omega)}(1)\quad \mbox{as}\,\,\tau\rightarrow
+\infty.
\end{equation}

Indeed
$$
(\Delta+(A_1,\nabla)+q_1) (e^{\tau
\Phi}(a+\frac{a_{-1}}{\tau}))=(\tau^2(\nabla\Phi,\nabla\Phi)+2\tau(\nabla\Phi,\nabla
(a+\frac{a_{-1}}{\tau}))+\Delta a+\frac{\Delta
a_{-1}}{\tau}+q_1(a+\frac{a_{-1}}{\tau}))e^{\tau\Phi}
$$
$$
+\tau(A_1,\nabla
\Phi)(a+\frac{a_{-1}}{\tau})e^{\tau\Phi}+(A_1,\nabla
(a+\frac{a_{-1}}{\tau}))e^{\tau\Phi}
$$
Observing that $(\nabla\Phi,\nabla\Phi)=0$ and using (\ref{xz1}),
(\ref{xz2}), we obtain
\begin{equation}\label{e110}
(\Delta+(A_1,\nabla)+q_1)(e^{\tau
\Phi}(a+\frac{a_{-1}}{\tau}))=[(\Delta+(A_1,\nabla)+q_1)
a_{-1}]e^{\tau\Phi}.
\end{equation}

Then using Corollary \ref{ZZZ}, we construct functions
$w_\tau(z,\cdot)$ which solve the boundary value problem
\begin{equation}\label{e121}
L_{q_1,A_1}(x,D)w_\tau=-\frac{e^{\tau\Phi}}{\tau}(\Delta+(A_1,\nabla)+q_1)a_{-1}\quad\mbox{in}
\,\,\Omega,\quad
w_\tau\vert_{\Gamma_-}=-(a+\frac{a_{-1}}{\tau})e^{\tau\Phi}
\end{equation}
and satisfy the estimate
\begin{equation}\label{-01}
\Vert e^{-\tau
x_3}w_\tau\Vert_{H^{1,\tau}(\Omega)}=o(1)\quad\mbox{as}\quad\tau\rightarrow
+\infty.
\end{equation}
By (\ref{e110}) and (\ref{e121}), the function
\begin{equation}\label{yoyo0}
u_1=(a+\frac{a_{-1}}{\tau})e^{\tau\Phi}+w_\tau
\end{equation}
solves the Schr\"odinger equation
\begin{equation}\label{eq111}
L_{q_1,A_1}(x,D)u_1=0\quad\mbox{in}\,\,\Omega, \quad
u_1\vert_{\Gamma_-}=0
\end{equation}
and admits the asymptotic expansion (\ref{zopaA01}) by (\ref{-01}).

Since $\Lambda (q_1, \Gamma_-, \Gamma_+)=\Lambda (q_2, \Gamma_-,
\Gamma_+)$, there exists $u_2$ such that
\begin{equation}\label{eq22}
L_{q_2,A_2}(x,D)u_2=0\,\,\mbox{in}\,\,\Omega, \quad
(u_1-u_2)\vert_{\partial\Omega}=\frac{\partial
(u_1-u_2)}{\partial\nu}\vert_{\partial\Omega\setminus\Gamma_+}=0.
\end{equation}

Next, in a way similar to the construction of $u_1$, we construct the
complex geometric optics solution $v$ to the Schr\"odinger operator
with potential $L_{q_2,A_2}(x,D)^*:$
\begin{equation}\label{eq44}
L_{q_2,A_2}(x,D)^*v=\Delta v-(A_2,\nabla)v-(\nabla ,A_2)v+q_2v=0
\quad \mbox{in }\,\,\Omega, \quad v\vert_{\Gamma_+}=0
\end{equation}
in the form
\begin{equation}\label{zopa11}
v(x)=(\widetilde a+\frac{\widetilde a_{-1}}{\tau})e^{-\tau\Phi} +e^{-\tau
x_3} o_{H^{1,\tau}(\Omega)}( 1)\quad \mbox{as}\,\,\tau\rightarrow
+\infty.
\end{equation}
Let a function $\mathcal A_2(\theta)$ solve the differential equation
$$
4\partial_{z_\theta}\mathcal A_2(\theta)-(A_2, ( i\vec
\omega,1))=0\quad\mbox{in}\,\,\Omega,  \quad \mbox{Im}\, \mathcal
A_2(\theta)\vert_{\partial\Omega}=0.
$$
Then $\widetilde a=e^{\mathcal A_2(\theta)}$
solves the ordinary differential equations
\begin{equation}\label{49}
2(\nabla\Phi,\nabla \widetilde a)-(A_2,\nabla\Phi)\widetilde
a=0\quad\mbox{in}\,\,\Omega.
\end{equation}

A function $\widetilde a_{-1}$ solves the differential equation:
\begin{equation}
2(\nabla\Phi,\nabla {\widetilde a}_{-1})-(A_2,\nabla\Phi){\widetilde
a}_{-1}=(\Delta-(A_2,\nabla)-(\nabla ,A_2)+q_2) \widetilde
a\quad\mbox{in}\,\,\Omega.
\end{equation}
Setting $u=u_1-u_2,$  by (\ref{eq111}) and (\ref{eq22}), we have
\begin{equation}\label{eq33}
L_{q_2,A_2}(x,D)u=-(A_1-A_2,\nabla)u_1-(q_1-q_2)u_1\quad \mbox{in
}\,\,\Omega, \,\, u\vert_{\partial\Omega}=0,\,\, \frac{\partial
u}{\partial \nu}\vert_{\partial\Omega\setminus\Gamma_+}=0.
\end{equation}

Then using (\ref{eq44}) and (\ref{eq33}), we obtain
\begin{eqnarray}\label{kjb}
(L_{q_2,A_2}(x,D)u,v)_{L^2(\Omega)}=(u,L_{q_2,A_2}(x,D)^*v)_{L^2(\Omega)}
+(\frac{\partial
u}{\partial\nu},v)_{L^2(\partial\Omega)}-(u,\frac{\partial
v}{\partial\nu})_{L^2(\partial\Omega)}\nonumber\\+\int_\Omega
((A_1-A_2,\nabla)u_1v+ (q_1-q_2)u_1v)dx\nonumber
\\=(v,\frac{\partial
u}{\partial\nu})_{L^2(\partial\Omega\setminus\Gamma_+)}+(v,\frac{\partial
u}{\partial\nu})_{L^2(\Gamma_+)}+\int_\Omega((A_1-A_2,\nabla u_1)v+
(q_1-q_2)u_1v)dx\nonumber \\
= \int_\Omega((A_1-A_2,\nabla u_1)v+ (q_1-q_2)u_1v)dx=0.
\end{eqnarray}

We are interested in the asymptotic expansion of the right-hand
side of (\ref{kjb}). By (\ref{zopaA01}) and (\ref{zopa11}), we have
\begin{equation}
\int_\Omega((A_1-A_2,\nabla u_1)v+ (q_1-q_2)u_1v)dx=\tau{\frak
S}_1+{\frak S}_0+o(1)\quad\mbox{as}\quad \tau\rightarrow +\infty.
\end{equation}
Since $\frak S_1$ is independent of $\tau$, the above asymptotic
formula implies $\frak S_1=0.$ Integrating by parts and using the
equalities (\ref{xz1}) and (\ref{49}), we obtain
\begin{eqnarray}\label{inos}
0={\frak S}_1=\int_\Omega(A_1-A_2,\nabla\Phi)a\widetilde a
dx=\int_\Omega (-(2\nabla\Phi,\nabla a) \widetilde
a-(2\nabla\Phi,\nabla \widetilde a) a) dx
\nonumber\\
= \int_\Omega (-(2\nabla\Phi,\nabla a)\widetilde a- a(2\nabla\Phi,\nabla
\widetilde a))dx =2\int_{\partial\Omega} \widetilde b(\overline
z_\theta)e^{\mathcal A_1(\theta)+\mathcal A_2(\theta)}\widetilde
g(\theta,x')\frac{\partial\Phi}{\partial\nu}d\sigma.
\end{eqnarray}

Let $\frak L$ be the set of all planes in $\Bbb R^3$ orthogonal to
the plane $x_3=0.$

From (\ref{inos}) we have
$$
\int_{\partial\Omega\cap P}\widetilde b(\overline z_\theta)e^{\mathcal
A_1(\theta)+\mathcal
A_2(\theta)}\frac{\partial\Phi}{\partial\nu}d\sigma=0\quad \forall
P\in\frak L.
$$

By Proposition \ref{govno}, there exists the antiholomorphic function $\Theta(\overline
z_\theta)$ such that $$e^{\mathcal A_1(\theta)+\mathcal
A_2(\theta)}= \Theta\quad\mbox{ on}\quad
\partial \Omega\cap P.$$
Observe that the function $\Theta$ does not have any zeros in
$\Omega$. Indeed, since the domain $\Omega$ is assumed to be convex,
the two dimensional domain $\Omega\cap P$ is symply connected. Then by
the well-know formula the number of zeros $N$ of the function
$\Theta$ is given by formula
$$
N=\frac{1}{2\pi i}\int_{\partial \Omega\cap
P}\frac{\overline\Theta'}{\overline\Theta}dz_\theta=\frac{1}{2\pi}\Delta_{\partial
\Omega\cap P} \,\mbox{arg}\,
\overline\Theta=\frac{1}{2\pi}\Delta_{\partial \Omega\cap P}
\,\mbox{arg}\, e^{\overline{\mathcal A_1(\theta)+\mathcal
A_2(\theta)}}=\frac{1}{2\pi}\Delta_{\partial \Omega\cap P}
\,\mbox{arg}\, e^{\mbox{Re}\,(\mathcal A_1(\theta)+\mathcal
A_2(\theta))}=0.
$$

Consider the form $\alpha=d\overline\Theta /\overline\Theta$. This form is
closed and since $\Omega\cap P$ is symply connected, the
differential form $\alpha$ is exact. Hence there exists a function
$a(x)$ such that $\alpha=da.$ Then
$\partial_{ z_\theta}a=\partial_{z_\theta}\overline \Theta/\overline\Theta.$ Consider
this equality as a first-order differential equation. The general
solution to this differential equation is written as $\overline\Theta=
c(\overline z_\theta)e^{a}.$ On the other hand $\partial_{\overline
z_\theta}\Theta=0$ . Hence $c(\overline z_\theta)=const$ and since
the function $a$ defined up to a constant, we have
$$
\overline\Theta= e^{a}.
$$
Then $a$ is a holomorphic function and we set $\ln
\Theta =\overline a.$

The function $\mathcal A_1(\theta)+\mathcal A_2(\theta)$ satisfies
the equation
\begin{equation}\label{gnil}
4\partial_{z_\theta}(\mathcal A_1(\theta)+\mathcal
A_2(\theta))+(A_1-A_2, (i\vec \omega,1))=0\quad\mbox{in}\quad \Omega
\end{equation}
and
$$
\mathcal A_1(\theta)+\mathcal A_2(\theta)=\ln \Theta\quad
\mbox{on}\quad
\partial \Omega\cap P.
$$

Integrating the equation (\ref{gnil}) over $\Omega\cap P$, we have
 $$
 \int_{\Omega\cap P} (A_1-A_2, \nabla \Phi)dx=- \int_{\Omega\cap P} 4\partial_{z_\theta}(\mathcal A_1(\theta)+\mathcal
A_2(\theta))dx=\int_{\partial\Omega\cap P}(\nu_3+i(\vec
\omega,\vec\nu'))(\mathcal A_1(\theta)+\mathcal
A_2(\theta))d\sigma
$$
$$
= \int_{\partial\Omega\cap P}(\nu_3+i(\vec \omega,\vec\nu')) \ln
\Theta d\sigma=0.
$$
Since $A_1$ and $A_2$ are real-valued vector fields, from the above
equality we obtain

\begin{equation}\label{0X}
 \int_{\Omega\cap P} (A_{1,3}-A_{2,3})dx=0, \quad\forall P\in \frak L
\end{equation}
and
\begin{equation}\label{01X}
 \int_{\Omega\cap P} (A_{1}-A_{2}, (\vec \omega, 0))dx=0,
\quad\forall P\in \frak L.
\end{equation}
We extend the vector fields $A_j$ by zero outside of domain
$\Omega.$ From (\ref{0X}) applying the uniqueness result for
the Radon transform, we obtain that
\begin{equation}\label{2X}
\int_{-K}^K(A_{1,3}-A_{2,3})dx_3=0, \quad \forall x'\in\Bbb R^2.
\end{equation}

By (\ref{2X}) there exists a function $\Psi(x)$ such that
\begin{equation}\label{3X}
\frac{\partial \Psi}{\partial
x_3}=(A_{1,3}-A_{2,3})\quad\mbox{in}\,\,\Omega, \quad
\Psi\vert_{\partial\Omega}=0\quad\mbox{in}\,\,\Omega.
\end{equation}

By Proposition \ref{volja} and the assumption of strict convexity of the
domain $\Omega$, the operators $L_{q_2,A_2}(x,D)$ and
$e^{-\Psi}L_{q_2,A_2}(x,D)e^{\Psi}$ generate the same Dirichlet-to-Neumann
map. The convection terms in the operator $e^{-\Psi}L_{q_2,A_2}(x,D)e^{\Psi}$
have the form
$$
(A_2+\nabla \Psi,\nabla).
$$
Hence by  (\ref{3X}) without loss of generality we can assume that
\begin{equation}\label{4X}
A_{1,3}=A_{2,3}\quad\mbox{in}\,\,\Omega.
\end{equation}
Then from (\ref{01X}) we have
\begin{equation}\label{1X}
 \int_{<\omega, x'>=p}
\left(\int_{-K}^Kg(\overline z_\theta) (A_{1,1}-A_{2,1})dx_3\right)
dx_1
+ \left(\int_{-K}^Kg(\overline z_\theta) (A_{1,2}-A_{2,2})dx_3\right)dx_2
=0,
\end{equation}
where ${<\omega, x'>=p}$ is an arbitrary line from $\Bbb R^2$.
We claim that
\begin{equation}\label{5X}
\left(\int_{-K}^Kx_3^k (A_{1,1}-A_{2,1})dx_3,\int_{-K}^K x_3^k
(A_{1,2}-A_{2,2})dx_3\right)
=(0,0), \quad\forall k\in \Bbb N_+\quad\mbox{and}\quad\forall x'\in \Bbb R^2.
\end{equation}

Our proof is by the induction method. Setting in (\ref{1X}) the
function $g=1$, we obtain (see e.g. \cite{P1}, p. 78) that there
exists a function $f$ with compact support such that
\begin{equation}
\nabla_{x'}f
= \left(\int_{-K}^K(A_{1,1}-A_{2,1})dx_3,\int_{-K}^K(A_{1,2}-A_{2,2})dx_3
\right), \quad\forall x'\in \Bbb R^2.
\end{equation}

Setting $g(\overline z_\theta)=\overline z_\theta$ in (\ref{1X}), we obtain
\begin{equation}\int_{< \omega, x'>=p}(\vec\omega, x')\frac{\partial f}{\partial x_1} dx_1+(\vec\omega,
x')\frac{\partial f}{\partial x_2}dx_2=0, \quad\forall p\in \Bbb
R^1\quad\mbox{and}\quad\forall \omega\in \Bbb S^1.
\end{equation}
Integrating by parts in this equation we obtain
\begin{equation}
\int_{<\omega, x'>=p}fds=0, \quad\forall p\in \Bbb
R^1\quad\mbox{and}\quad\forall \omega\in \Bbb S^1.
\end{equation}
By the uniqueness theorem for the Radon transform, we obtain $f\equiv
0.$ Hence the beginning step of the induction  method is established.
Suppose that (\ref{5X}) is already proved for all $k< \widehat k.$

Setting $g(\overline z_\theta)=(x_3-i(\vec\omega,x'))^{\widehat k}$
in (\ref{1X}), we obtain
\begin{eqnarray}
 \int_{<\omega, x'>=p}\left(\int_{-K}^K(x_3-i(\vec\omega,x'))^{\widehat k}
(A_{1,1}-A_{2,1})dx_3\right)dx_1
+ \left(\int_{-K}^K(x_3-i(\vec\omega,x'))^{\widehat k}
 (A_{1,2}-A_{2,2})dx_3\right)dx_2\nonumber\\
= \int_{<\omega, x'>=p}\left(\int_{-K}^Kx_3^{\widehat k} (A_{1,1}-A_{2,1})
dx_3\right)dx_1
+ \left(\int_{-K}^Kx_3^{\widehat k}
 (A_{1,2}-A_{2,2})dx_3\right)dx_2=0.
\end{eqnarray}
Hence there exists a function $f_{\widehat k}$  with compact support such that
\begin{equation}
\nabla_{x'}f_{\widehat k}
= \left(\int_{-K}^Kx_3^{\widehat k}(A_{1,1}-A_{2,1})dx_3,\int_{-K}^K
x_3^{\widehat k}(A_{1,2}-A_{2,2})dx_3\right), \quad\forall x'\in \Bbb R^2.
\end{equation}
Setting $g(\overline z_\theta)=(x_3-i(\vec\omega,x'))^{{\widehat k}+1}$
in (\ref{1X}), we obtain
\begin{equation}
\int_{<\omega, x'>=p}(\vec\omega, x')\frac{\partial f_{\widehat k}}
{\partial x_1} dx_1
+ (\vec\omega,x')\frac{\partial f_{\widehat k}}{\partial x_2}dx_2
=0, \quad\forall p\in
\Bbb R^1\quad\mbox{and}\quad\forall \omega\in \Bbb S^1.
\end{equation}
Integrating by parts in this equation we obtain
\begin{equation}
\int_{<\omega, x'>=p}f_{\widehat k} ds=0,\quad\forall p\in \Bbb
R^1\quad\mbox{and}\quad\forall \omega\in \Bbb S^1.
\end{equation}
By the uniqueness theorem for the Radon transform, we obtain
$f_{\widehat k}\equiv 0$ and (\ref{5X}) is proved. On the other hand the equality $(\ref{5X})$ implies that
$$
A_{1,1}-A_{2,1}=0\quad\mbox{and}\quad
A_{1,2}-A_{2,2}=0\quad\mbox{in}\,\,\Omega.
$$
The proof of Theorem \ref{t3} is complete.
$\blacksquare$

For more results on recovery of coefficients of the Schr\"odinger
equation with the first terms, see \cite{DKSU} where
the function $\Phi=\ln r+i\theta$ was used
for construction of the complex geometric optics solution.
In the proof of Theorem \ref{t3} we used some ideas from \cite{DKSU}.
\\
\vspace{0.3cm}

We conclude this section with
\begin{proposition}\label{pzpz}
We assume that $\Lambda(q_1,\emptyset, \emptyset) =
\Lambda(q_2,\emptyset,\emptyset )$ with $q_1, q_2$ in some
admissible set implies $q_1=q_2$ in $\Omega$. If $q_1 = q_2$ near
$\partial\Omega$ and $\Lambda(q_1,\Gamma_-,\Gamma_+) =
\Lambda(q_2,\Gamma_-,\Gamma_+)$ with arbitrarily subboundaries
$\Gamma_-, \Gamma_+$, then $q_1=q_2$ in $\Omega$.
\end{proposition}

Thus if we can assume that the coefficients are equal near
$\partial\Omega$, then the uniqueness by Dirichlet-to-Neumann map
on subboundaries is trivial from the uniqueness by the
Dirichlet-to-Neumann map on the whole boundary.

{\bf Proof.}
We can choose an open neighborhood $\widetilde{\omega}$
of $\partial\Omega$ such that $q:= q_1 = q_2$ in $\omega:=
\widetilde{\omega} \cap \Omega$.  Let $u_j$, $j=1,2$ satisfy
$$
L_{q_j}(x,D)u_j=\Delta u_j + q_ju_j = 0 \quad \mbox{in $\Omega$},
\quad u_j\vert_{\partial\Omega} = f.
$$
First we prove $\frac{\partial u_1}{\partial \nu} =
\frac{\partial u_2}{\partial \nu}$ on $\partial\Omega$ if
$f=0$ on $\Gamma_-$.
In fact, setting $u=u_1-u_2$, we have
$$
L_{q}(x,D)u=\Delta u + qu = 0 \quad \mbox{in $\omega$}, \quad
u\vert_{\partial\Omega} = 0
$$
and
$$
\frac{\partial u}{\partial\nu}=0 \quad \mbox{on
$\partial\Omega\setminus \Gamma_+$}.
$$
Therefore the unique continuation for the Schr\"odinger equation
(e.g., H\"ormander \cite{Ho1}) yields $u=0$ in $\omega$, which
implies $\frac{\partial u_1}{\partial\nu}=
\frac{\partial u_2}{\partial\nu}$ on $\partial\Omega$.

Next let $f \in H^{\frac{1}{2}}(\partial\Omega)$ be arbitrary.
Then we will prove $\frac{\partial u_1}{\partial\nu}=
\frac{\partial u_2}{\partial\nu}$ on $\partial\Omega$.
Let $w_j$, $j=1,2$, satisfy
$$
L_{q_j}(x,D)w_j=\Delta w_j + q_jw_j = 0 \quad \mbox{in $\Omega$},
\quad w_j\vert_{\partial\Omega} = g,
$$
where $g=0$ on $\Gamma_-$.  For $j=1,2$, we have
\begin{eqnarray*}
&& 0 = \int_{\Omega} w_jL_{q_j}(x,D)u_j dx = \int_{\Omega}
u_jL_{q_j}(x,D)w_j dx + \int_{\partial\Omega}\left(
w_j\frac{\partial u_j}{\partial\nu}
-  u_j\frac{\partial w_j}{\partial\nu}\right) d\sigma\\
=&& \int_{\partial\Omega}g\frac{\partial u_j}{\partial\nu} d\sigma
- \int_{\partial\Omega} f\frac{\partial w_j}{\partial\nu} d\sigma,
\end{eqnarray*}
that is,
$$
\int_{\partial\Omega\setminus\Gamma_-}g\frac{\partial u_j}{\partial\nu} d\sigma
= \int_{\partial\Omega} f\frac{\partial w_j}{\partial\nu} d\sigma,
\quad j=1,2.
$$
By $\Lambda(q_1,\Gamma_-,\Gamma_+) = \Lambda(q_2,\Gamma_-,\Gamma_+)$ and
the fact proved above, we see that $\frac{\partial w_1}{\partial\nu}
= \frac{\partial w_2}{\partial\nu}$ on $\partial\Omega$.
Therefore
$$
\int_{\partial\Omega\setminus\Gamma_-}g\frac{\partial u_1}{\partial\nu} d\sigma
= \int_{\partial\Omega\setminus\Gamma_-}g\frac{\partial u_2}{\partial\nu}
d\sigma.
$$
Since we can choose $g$ arbitrarily, for example, any $g \in
C^{\infty}_0(\partial\Omega\setminus \Gamma_-)$, we obtain
$\frac{\partial u_1}{\partial\nu} = \frac{\partial
u_2}{\partial\nu}$ on $\partial\Omega\setminus\Gamma_-$. Again
setting $u=u_1-u_2$, we have $\Delta u + qu = 0$ in $\omega$, $u=0$
on $\partial\Omega$ and $\frac{\partial u}{\partial\nu} = 0$ on
$\partial\Omega \setminus \Gamma_-$.  The unique continuation yields
$u=0$ in $\omega$. Hence $\frac{\partial u_1}{\partial\nu} =
\frac{\partial u_2}{\partial\nu}$ on $\partial\Omega$.  Hence we
prove $\Lambda(q_1,\emptyset,\emptyset) =
\Lambda(q_2,\emptyset,\emptyset)$. Thus the proof of the proposition
is completed. $\blacksquare$
%
%
%
%
%
%
%
%
%
%
\section{2-D Calder\'on's problem.}

Let $\Omega$ be a bounded domain in $\Bbb R^2$ with smooth
boundary such that $\partial\Omega=\cup_{k=1}^{\mathcal N}
\gamma_k$, where $\gamma_k$, $1 \le k \le {\mathcal N}$, are smooth
closed contours, and $\gamma_{\mathcal N}$ is the external contour.
Let $\Gamma_0$ be an arbitrarily chosen relatively open subset
of $\partial \Omega.$

For the Schr\"odinger operator with potential $q$ we consider the following
Dirichlet-to-Neumann map $\Lambda (q, \Gamma_0, \Gamma_0):$
$$
\Lambda (q, \Gamma_0,\Gamma_0) (f)= \frac{\partial u}{\partial
\nu}\vert_{\partial\Omega\setminus\Gamma_0},
$$
where
$$ L_{q}(x,D)u=\Delta u+qu=0\quad\mbox{in}\,\,\Omega, \quad
u\vert_{\Gamma_0}=0,\quad u\vert_{\partial\Omega\setminus
\Gamma_0}=f.
$$
Henceforth we write $\Lambda(q,\Gamma_0) = \Lambda(q,\Gamma_0,
\Gamma_0)$.

We have
\begin{theorem}\label{balalaika}(\cite{IY})
Let $q_1,q_2\in
W^1_p(\Omega)$ for some $p>2$ and $\Lambda (q_1,
\Gamma_0)=\Lambda(q_2,\Gamma_0).$ Then $q_1=q_2$
in $\Omega$.
\end{theorem}

We modify the argument in \cite{IY} and
describe the proof.
Before starting the proof of the theorem we recall the classical
results for the properties of the operators
$\partial_{z}^{-1}$ and $\partial_{\overline z}^{-1}$ which are given by
$$
\partial_{\overline z}^{-1}g=-\frac1\pi\int_\Omega
\frac{g(\zeta,\overline\zeta)}{\zeta-z} d\xi_2d\xi_1,\quad
\partial_{ z}^{-1}g
= \overline{\partial^{-1}_{\overline{z}}\overline{g}}.
$$

The following is proved in \cite{VE} (p.47, 56, 72):
\begin{proposition}\label{Proposition 3.00}
{\bf A)} Let $m\ge 0$ be an integer number and $\alpha\in (0,1).$
Then $\partial_{\overline z}^{-1},\partial_{ z}^{-1}\in \mathcal
L(C^{m+\alpha}(\overline{\Omega}),C^{m+\alpha+1}
(\overline{\Omega})).$
\newline
{\bf B}) Let $1\le p\le 2$ and $ 1<\gamma<\frac{2p}{2-p}.$ Then
 $\partial_{\overline z}^{-1},\partial_{ z}^{-1}\in
\mathcal L(L^p( \Omega),L^\gamma(\Omega)).$
\newline
{\bf C})Let $1< p<\infty.$ Then  $\partial_{\overline z}^{-1},
\partial_{ z}^{-1}\in \mathcal L(L^p( \Omega),W^1_p(\Omega)).$
\end{proposition}
\vspace{0.3cm}

{\bf Proof of Theorem 7.} We define two other operators:
\begin{equation}\label{anna}
\mathcal R_{\tau}g = \frac 12e^{\tau(\Phi - \overline{\Phi})}
\partial_{\overline z}^{-1}(g
e^{\tau(\overline{\Phi}-\Phi)}),\,\, \widetilde {\mathcal R}_{\tau}g
= \frac 12 e^{\tau(\overline {\Phi}-{\Phi})}
\partial_{ z}^{-1}(ge^{\tau( {\Phi}
-\overline {\Phi})}),
\end{equation}
where $\Phi\in C^2(\overline\Omega)$ is a holomorphic function which
satisfies (\ref{1'})-(\ref{kk}). Observe that
\begin{equation}\label{begemot1} 2\frac{\partial}{\partial
z}(e^{\tau\Phi}\widetilde {\mathcal R}_{\tau}g)=ge^{\tau\Phi},\quad
2\frac{\partial}{\partial \overline z}(e^{\tau\overline\Phi}
{\mathcal R}_{\tau}g)=ge^{\tau\overline\Phi} \quad \forall g\in
L^2(\Omega).
\end{equation}
Let $a\in C^6(\overline\Omega)$ be some holomorphic function, not
identically equal to a constant  on $\Omega,$ such that
\begin{equation}\label{LLM}
\mbox{Re}\, a\vert_{\Gamma^*_0}=0 ,\quad \lim_{z\rightarrow \widehat
z}a(z)/\vert z-\widehat z\vert^{100}=0, \quad \forall \widehat z\in \mathcal
H\cap\Gamma^*_0.
\end{equation}
We recall that $\mathcal H=\{\widetilde z\in \Omega\vert
\partial_z\Phi(\widetilde z)=0\}$  is the set of critical points of the function $\Phi.$  Moreover, for some $\widetilde x\in
\mathcal H$, we assume that
\begin{equation}\label{begemot}
a(\widetilde x) \ne 0.
\end{equation}
The existence of such a function is proved in Proposition
\ref{nikita} in Section 7.
 Let polynomials $M_{1}(z)$ and $M_{3}(\overline z)$
satisfy
\begin{equation}\label{begemot2}
(\partial^{-1}_{\overline z}q_{1} -M_{1})(\widetilde x)=0, \quad
\quad (\partial^{-1}_{z}q_{1} - M_{3})(\widetilde x)= 0.
\end{equation}

We define the function $U_1$ by
\begin{eqnarray}\label{mozilaal}
U_1(x)=e^{\tau{\Phi}}{(a+a_{1}/\tau)} +e^{\tau\overline{\Phi}}
{(\overline a+b_{1}/\tau)}
- \frac 12e^{\tau\Phi}\widetilde {\mathcal
R}_\tau\{{a(\partial^{-1}_{\overline z} q_{1}-M_{1})}\} - \frac 12
 e^{\tau\overline\Phi}{\mathcal
R}_\tau\{\overline{a}(\partial^{-1}_{z} q_{1}-M_{3})\},
\end{eqnarray}
where $a_1$ is some holomorphic function and $b_1$ some
antiholomorphic function.  We set
$$
g_\tau=q_1(e^{i\tau\psi}{a_{1}/\tau} +e^{-i\tau\psi} {b_{1}/\tau}-
\frac{e^{i\tau\psi}}{2}\widetilde {\mathcal
R}_\tau\{{a(\partial^{-1}_{\overline z} q_{1}-M_{1})}\} -
 \frac{e^{-i\tau\psi}}{2}{\mathcal
R}_\tau\{\overline{a}(\partial^{-1}_{z} q_{1}-M_{3})\}).
$$
After short computations,  using (\ref{mozilaal}), (\ref{begemot1})
and the factorization of the Laplace operator in the form
$\Delta=4\partial_{\overline z}\partial_z$ we reach the
following equation
\begin{equation}\label{lob}
L_{q_1}(x,D)U_1=e^{\tau\varphi}g_\tau\quad \mbox{in}\,\,\Omega.
\end{equation}
We make a choice of the functions $a_1, b_1$ in such a way that

  \begin{equation}
\label{inka}\Vert g_\tau\Vert_{L^2(\Omega)}=O(\frac 1\tau
)\quad\mbox{as}\,\,\tau\rightarrow +\infty
\end{equation}
and
\begin{equation}\label{lob1}
\quad U_1 \vert_{\Gamma_0}=e^{\tau\varphi}O_{H^\frac
12(\overline{\Gamma^*_0})}(\frac
1\tau)\quad\mbox{as}\,\,\tau\rightarrow +\infty.
\end{equation}

The holomorphic function $a_1$ and the antiholomorphic function
$b_1$ are defined by $a_1(z)=a_{1,1}(z)+a_{1,2}(z)$ and
$b_1(\overline z)=b_{1,1}(\overline z)+b_{1,2}(\overline z) $ where
the functions $a_{1,1}, b_{1,1}\in C^1(\overline\Omega)$ satisfy
$$
a_{1,1}(z)+b_{1,1}(\overline z)=\bigg( \frac{a(\partial^{-1}_{\overline
z} q_{1}-M_{1})}{4\partial_z\Phi} +
 \frac{\overline{a}(\partial^{-1}_{z}
q_{1}-M_{3})}{4\overline{\partial_z\Phi}} \bigg
)\quad\mbox{on}\,\,\Gamma_0,
$$
and the functions $a_{1,2}(z,\tau),b_{1,2}(\overline z,\tau)\in
C^1(\overline \Omega)$ for each $\tau$ are holomorphic and
antiholomorphic function  such that
$$
a_{1,2}(z,\tau)=-\frac{1}{8\pi}\int_{\partial\Omega}
\frac{(\nu_1+i\nu_2)\overline{a}(\partial_\zeta^{-1}q_{1}-M_{3})
e^{\tau(\overline\Phi-\Phi)}}{(\zeta-z)\partial
_{\overline\zeta}\overline\Phi} d\sigma
$$
and
$$
b_{1,2}(\overline z,\tau)=-\frac{1}{8\pi}\int_{\partial\Omega}
\frac{(\nu_1-i\nu_2) a(\partial^{-1}_{\overline
\zeta}q_{1}-M_{1})e^{\tau(\Phi-\overline\Phi)}}{(\overline\zeta-\overline
z)\partial_\zeta\Phi} d\sigma .
$$

Here the denominators of the integrands vanish in ${\cal H} \cap
\Gamma^*_0$, but thanks to the second condition in (\ref{LLM}),
the integrability is guaranteed. We represent the functions
$a_{1,2}(z,\tau),b_{1,2}(\overline z,\tau)$ in the form
$$
a_{1,2}(z,\tau)=a_{1,2,1}(z)+a_{1,2,2}(z,\tau),\quad
b_{1,2}(\overline z,\tau)=b_{1,2,1}(\overline z) +
b_{1,2,2}(\overline z,\tau),
$$
where
$$
a_{1,2,1}(z)=-\frac{1}{8\pi}\int_{\Gamma_0^*}
\frac{(\nu_1+i\nu_2)\overline{a}(\partial_\zeta^{-1}q_{1}-M_{3})
}{(\zeta-z)\partial _{\overline\zeta}\overline\Phi} d\sigma,\quad
b_{1,2,1}(\overline z)=-\frac{1}{8\pi}\int_{\Gamma_0^*}
\frac{(\nu_1-i\nu_2) a(\partial^{-1}_{\overline
\zeta}q_{1}-M_{1})}{(\overline\zeta-\overline z)\partial_\zeta\Phi}
d\sigma.
$$

By (\ref{LLM}), the functions $a_{1,2,1}, b_{1,2,1}$ belong to
$C^1(\overline \Omega).$ By (\ref{kk}) and Proposition \ref{granata}
in Section 7, we have
\begin{equation}\label{yoyo!}
\Vert b_{1,2,2}(\cdot,\tau)\Vert_{L^2(\Omega)}+\Vert
a_{1,2,2}(\cdot,\tau)\Vert_{L^2(\Omega)}\rightarrow 0\quad\mbox{as}
\,\,\tau\rightarrow +\infty .
\end{equation}

In order to establish (\ref{lob1}), we use the following proposition:
\begin{proposition} The following asymptotic formula is true
\begin{eqnarray}\label{zopa}
\left\Vert \int_{\Omega}
\partial_\zeta\left (\frac{ a(\partial^{-1}_{\overline
\zeta}q_{1}-M_{1})}{\partial_\zeta\Phi}\right
)\frac{e^{\tau(\Phi-\overline\Phi)}}{\overline\zeta-\overline z}
d\xi_2d\xi_1\right\Vert_{H^{\frac 12}(\Gamma_0^*)} \nonumber\\
+ \left\Vert \int_{\Omega}
\partial_{\overline\zeta}\left (\frac{\overline{a}(\partial^{-1}_{\overline
\zeta}q_{1}-M_{3})}{\partial_{\overline\zeta}\overline\Phi}\right)
\frac{e^{\tau(\overline\Phi-\Phi)}}{\zeta-z}
d\xi_2d\xi_1\right\Vert_{H^{\frac 12}(\Gamma_0^*)}
=o(1)\quad\mbox{as}\,\,\tau\rightarrow +\infty.
\end{eqnarray}
\end{proposition}

{\bf Proof.} In order to prove (\ref{zopa}), consider a function
$e\in C^\infty_0(\Omega)$ such that
\begin{equation}\label{eblo}
e\equiv 1\quad\mbox{in some neighborhood of the set}\quad \mathcal
H\setminus\Gamma_0^*. \end{equation}
 The family of functions
$\int_{\Omega} e
\partial_\zeta\left (\frac{ a(\partial^{-1}_{\overline
\zeta}q_{1}-M_{1})}{\partial_\zeta\Phi}\right
)\frac{e^{\tau(\Phi-\overline\Phi)}}{\overline\zeta-\overline z}
d\xi_2d\xi_1\in C^\infty(\partial\Omega),$  are uniformly bounded in
$\tau$  in $C^2(\partial\Omega)$ and by Proposition \ref{gandonnal}
in Section 7, this function converges pointwisely to zero. Therefore
\begin{equation}\label{zopa1}
\left\Vert\int_{\Omega}e
\partial_\zeta\left (\frac{ a(\partial^{-1}_{\overline
\zeta}q_{1}-M_{1})}{\partial_\zeta\Phi}\right
)\frac{e^{\tau(\Phi-\overline\Phi)}}{\overline\zeta-\overline z}
d\xi_2d\xi_1\right\Vert_{H^1(\partial\Omega)}
=o(1)\quad\mbox{as}\,\,\tau\rightarrow +\infty.
\end{equation}

Integrating by parts we obtain
$$
\int_{\Omega}(1-e)
\partial_\zeta\left (\frac{ a(\partial^{-1}_{\overline
\zeta}q_{1}-M_{1})}{\partial_\zeta\Phi}\right
)\frac{e^{\tau(\Phi-\overline\Phi)}}{\overline\zeta-\overline z}
d\xi_2d\xi_1 = \frac{(1-e)}{\partial_z\Phi}
\partial_z\left (\frac{ a(\partial^{-1}_{\overline z}q_{1}-M_{1})}
{\tau\partial_z\Phi}\right)e^{\tau(\Phi-\overline\Phi)}
$$
$$
-\frac{1}{\tau}\int_{\Omega}\partial_{\zeta}\left(\frac{(1-e)}
{\partial_\zeta\Phi}\partial_\zeta\left (\frac{
a(\partial^{-1}_{\overline
\zeta}q_{1}-M_{1})}{\partial_\zeta\Phi}\right )\right)
\frac{e^{\tau(\Phi-\overline\Phi)}} {\overline\zeta-\overline
z}d\xi_2d\xi_1 .
$$

Thanks to (\ref{22}) and (\ref{LLM}), we have
\begin{equation}\label{zopa2}
\left\Vert \frac{1-e}{\partial_z\Phi}
\partial_z\left (\frac{ a(\partial^{-1}_{\overline z}q_{1}-M_{1})}
{\tau\partial_z\Phi}\right
)e^{\tau(\Phi-\overline\Phi)}\right\Vert_{H^{\frac
12}(\Gamma_0^*)}=o(1)\quad\mbox{as}\,\,\tau\rightarrow +\infty.
\end{equation}

By (\ref{eblo}) and  Proposition \ref{Proposition 3.00}, the
functions $\partial_{\zeta}\left(\frac{1-e}{\partial_\zeta\Phi}
\partial_\zeta\left (\frac{ a(\partial^{-1}_{\overline
\zeta}q_{1}-M_{1})}{\partial_\zeta\Phi}\right
)\right)e^{\tau(\Phi-\overline\Phi)}$ are bounded in $L^p(\Omega)$
uniformly in $\tau.$  Therefore by Proposition \ref{Proposition
3.00}, the functions
$\int_{\Omega}\partial_{\zeta}\left(\frac{1-e}{\partial_\zeta\Phi}
\partial_\zeta\left (\frac{ a(\partial^{-1}_{\overline
\zeta}q_{1}-M_{1})}{\partial_\zeta\Phi}\right
)\right)\frac{e^{\tau(\Phi-\overline\Phi)}}{\overline\zeta-\overline
z} d\xi_2d\xi_1$ are uniformly bounded in $W^1_p(\Omega).$
The trace theorem yields
\begin{equation}\label{zopa3}
\left\Vert\frac{1}{\tau}\int_{\Omega}\partial_{\zeta}\left(
\frac{1-e}{\partial_\zeta\Phi}
\partial_\zeta\left (\frac{ a(\partial^{-1}_{\overline
\zeta}q_{1}-M_{1})}{\partial_\zeta\Phi}\right
)\right)\frac{e^{\tau(\Phi-\overline\Phi)}} {\overline\zeta-\overline
z} d\xi_2d\xi_1\right\Vert _{H^{\frac
12}(\Gamma_0^*)}=o(1)\quad \mbox{as}\,\,\tau\rightarrow +\infty.
\end{equation}
By (\ref{zopa1})-(\ref{zopa3}) we obtain (\ref{zopa}).
$\blacksquare$
\\
\vspace{0.2cm}

We note that $\frac{a}{\partial_z\Phi}\in C^2(\partial\Omega)$
by (\ref{LLM}).
Integrating by parts, we obtain the following:
\begin{equation}\label{01'}
e^{\tau\Phi}\widetilde {\mathcal R}_\tau
\{{a(\partial^{-1}_{\overline z} q_{1}-M_{1})}\} =
\frac{1}{\tau}\Biggl(2b_{1,2}e^{\tau\overline\Phi} +
\frac{e^{\tau\Phi}a(\partial^{-1}_{\overline z}
q_{1}-M_{1})}{2\partial_z\Phi}
\end{equation}
$$
+ \frac{e^{\tau\overline\Phi}}{2\pi}\int_{\Omega}
\partial_\zeta\left (\frac{ a(\partial^{-1}_{\overline
\zeta}q_{1}-M_{1})}{\partial_\zeta\Phi}\right
)\frac{e^{\tau(\Phi-\overline\Phi)}}{\overline\zeta-\overline z}
d\xi_2d\xi_1\Biggr)
$$
and
\begin{equation}\label{022}
e^{\tau\overline\Phi}{\mathcal
R}_\tau\{\overline{a}(\partial^{-1}_{z} q_{1}-M_{3})\} =
\frac{1}{\tau}\Biggl(2a_{1,2}e^{\tau\Phi}+\frac
{e^{\tau\overline\Phi}\overline{a}(\partial^{-1}_{z}
q_{1}-M_{3})}{2\partial_{\overline z}\overline\Phi}
\end{equation}
$$
+ \frac{e^{\tau\Phi}}{2\pi}\int_{\Omega}
\partial_{\overline\zeta}\left (\frac{\overline{a}(\partial^{-1}_{\overline
\zeta}q_{1}-M_{3})}{\partial_{\overline\zeta}\overline\Phi}\right)
\frac{e^{\tau(\overline\Phi-\Phi)}}{\zeta-z} d\xi_2d\xi_1\Biggr).
$$

We have
\begin{proposition}
The following asymptotic formula is true:
\begin{eqnarray}\label{PPPP}
\left\Vert\frac{e^{-i\tau\psi}}{2\pi}\int_{\Omega}
\partial_\zeta\left (\frac{ a(\partial^{-1}_{\overline
\zeta}q_{1}-M_{1})}{\partial_\zeta\Phi}\right
)\frac{e^{\tau(\Phi-\overline\Phi)}}{\overline\zeta-\overline z}
d\xi_2d\xi_1\right\Vert_{L^2(\Omega)}\nonumber\\
+ \left\Vert \frac{e^{i\tau\psi}}{2\pi}\int_{\Omega}
\partial_{\overline\zeta}\left (\frac{\overline{a}(\partial^{-1}_{\overline
\zeta}q_{1}-M_{3})}{\partial_{\overline\zeta}\overline\Phi}
\right)\frac{e^{\tau(\overline\Phi-\Phi)}}{\zeta-z}
d\xi_2d\xi_1\right\Vert_{L^2(\Omega)} \rightarrow 0\quad
\mbox{as}\,\,\tau\rightarrow + \infty.
\end{eqnarray}
\end{proposition}

{\bf Proof.} We prove the asymptotic behavior of the
first term in (\ref{PPPP}).  The proof for the second term is the
same. Denote $r_\tau(\xi)=\partial_\zeta\left (\frac{
a(\partial^{-1}_{\overline
\zeta}q_{1}-M_{1})}{\partial_\zeta\Phi}\right
)e^{\tau(\Phi-\overline\Phi)}.$  By (\ref{kk}), (\ref{begemot}) and
(\ref{begemot2}), the family of these functions is bounded in
$L^p(\Omega)$ for any $p<2.$ Hence by Proposition \ref{Proposition
3.00} there exists a constant $C$ independent of $\tau$ such that
\begin{equation}\label{mk}
\left\Vert\frac{e^{-i\tau\psi}}{2\pi}\int_{\Omega}
\partial_\zeta\left (\frac{ a(\partial^{-1}_{\overline
\zeta}q_{1}-M_{1})}{\partial_\zeta\Phi}\right
)\frac{e^{\tau(\Phi-\overline\Phi)}}{\overline\zeta-\overline z}
d\xi_2d\xi_1\right\Vert_{L^4(\Omega)}\le C.
\end{equation}

By (\ref{22}), (\ref{begemot}) and (\ref{begemot2}), for any $z\ne
\widetilde x_1+i\widetilde x_2$, the function
$r_\tau(\xi)/(\overline\zeta-\overline z)$ belongs to $L^1(\Omega).$
Therefore by Proposition \ref{gandonnal}
\begin{equation}\label{mk1}
\frac{e^{-i\tau\psi}}{2\pi}\int_{\Omega}
\partial_\zeta\left (\frac{ a(\partial^{-1}_{\overline
\zeta}q_{1}-M_{1})}{\partial_\zeta\Phi}\right
)\frac{e^{\tau(\Phi-\overline\Phi)}}{\overline\zeta-\overline z}
d\xi_2d\xi_1\rightarrow 0\quad \mbox{a.e.  in }\quad \Omega.
\end{equation}

By (\ref{mk}), (\ref{mk1}) and Egorov's theorem, the asymptotic
behavior of the first term in (\ref{PPPP}) follows immediately.
$\blacksquare$
\\
\vspace{0.2cm}

The asymptotic formula (\ref{inka}) follows from
(\ref{yoyo!}), (\ref{PPPP}), (\ref{01'}) and (\ref{022}).

In order to prove (\ref{lob1}), we set $U_1=I_1+I_2$, where
\begin{equation}\label{luba}
I_1=((a+a_{1,1}/\tau)e^{\tau\Phi}+(\overline
a+b_{1,1}/\tau)e^{\tau\overline\Phi})
 = \left(
\frac{a(\partial^{-1}_{\overline z}
q_{1}-M_{1})}{4\partial_z\Phi}+\frac {\overline{a}(\partial^{-1}_{z}
q_{1}-M_{3})}{4\partial_{\overline
z}\overline\Phi}\right)e^{\tau\varphi}
\end{equation}
and
\begin{eqnarray}\label{luba1}
I_2 = (a_{1,2}e^{\tau\Phi} +b_{1,2}e^{\tau\overline\Phi}) - \frac 12
(e^{\tau\Phi}\widetilde {\mathcal
R}_\tau\{{(a(\partial^{-1}_{\overline z}
q_{1}-M_{1})}\} \nonumber\\
+ e^{\tau\overline\Phi}{\mathcal
R}_\tau\{\overline{a}(\partial^{-1}_{z} q_{1}-M_{3})\})\nonumber\\
= -\frac 12\Biggl( \frac{e^{\tau\Phi}a(\partial^{-1}_{\overline z}
q_{1}-M_{1})}{2\partial_z\Phi}+\frac{e^{\tau\overline\Phi}}{2\pi}
\int_{\Omega}
\partial_\zeta\left (\frac{ a(\partial^{-1}_{\overline
\zeta}q_{1}-M_{1})}{\partial_\zeta\Phi}\right
)\frac{e^{\tau(\Phi-\overline\Phi)}}{\overline\zeta-\overline z}
d\xi_2d\xi_1                         \nonumber\\
+ \frac{e^{\tau\overline\Phi}\overline{a}(\partial^{-1}_{z}
q_{1}-M_{3})}{2\partial_{\overline
z}\overline\Phi}+\frac{e^{\tau\Phi}}{2\pi}\int_{\Omega}
\partial_{\overline\zeta}\left (\frac{\overline{a}(\partial^{-1}_{\overline
\zeta}q_{1}-M_{3})}{\partial_{\overline\zeta}\overline\Phi}\right)
\frac{e^{\tau(\overline\Phi-\Phi)}}{\zeta-z} d\xi_2d\xi_1\Biggr)
                                                \nonumber\\
= -\left(\frac {{a}(\partial^{-1}_{\overline z}
q_{1}-M_{1})}{4\partial_{ z}\Phi}+\frac
{e^{\tau\overline\Phi}\overline{a}(\partial^{-1}_{z}
q_{1}-M_{3})}{4\partial_{\overline
z}\overline\Phi}\right)e^{\tau\varphi} +e^{\tau\varphi} O_{H^{\frac
12}(\Gamma^*_0)}(\frac 1\tau).
\end{eqnarray}
Here in order to obtain the last equality, we used (\ref{zopa}).

From (\ref{luba}) and (\ref{luba1}), we obtain (\ref{lob1}).

Finally we construct the last term of the complex geometric optics
solution $e^{\tau\varphi}w_\tau.$ Consider the boundary value
problem
\begin{equation}\label{lena}
L_{q_1}(x,D)(w_\tau e^{\tau\varphi})=-g_\tau
e^{\tau\varphi}\quad\mbox{in}\,\,\Omega,\quad (w_\tau
e^{\tau\varphi})\vert_{\Gamma_0}=-U_1.
\end{equation}

By (\ref{inka}) and Proposition \ref{Proposition 2.3}, there exists
a solution to problem (\ref{lena}) such that
\begin{equation}\label{ioio}
\Vert w_\tau\Vert_{L^2(\Omega)}=o(\frac 1\tau) \quad
\mbox{as}\,\tau\rightarrow +\infty.
\end{equation}
Finally we set
\begin{equation}\label{ioioio}
u_1=U_1+e^{\tau\varphi} w_\tau.
\end{equation}
By (\ref{ioio}), (\ref{ioioio}), (\ref{PPPP}), (\ref{01'}) and
(\ref{022})  we can represent the complex geometric optics solution
$u_1$ in the form
\begin{eqnarray}\label{mozilaall}
u_1(x)=e^{\tau{\Phi}}{(a+(a_{1,1}+a_{1,2,1})/\tau)}
+e^{\tau\overline{\Phi}} {(\overline a+(b_{1,1}+b_{1,2,1})/\tau)}\nonumber\\
- \bigg( e^{\tau\Phi}\frac{a(\partial^{-1}_{\overline z}
q_{1}-M_{1})}{4\tau\partial_z\Phi} +
 e^{\tau\overline\Phi}\frac{\overline{a}(\partial^{-1}_{z}
q_{1}-M_{3})}{4\tau\overline{\partial_z\Phi}} \bigg
)+e^{\tau\varphi}o_{L^2(\Omega)}(\frac 1\tau)\quad
\mbox{as}\,\tau\rightarrow +\infty.
\end{eqnarray}

Since the Dirichlet-to-Neumann maps for the potentials $ q_1$ and
$q_2$ are equal, there exists a solution $u_2$  to the Schr\"odinger
equation with potential $q_2$ such that $\frac{\partial
u_1}{\partial \nu}=\frac{\partial u_2}{\partial\nu}$ on
$\partial\Omega\setminus\Gamma_0$ and $ u_1= u_2$ on
$\partial\Omega\setminus\Gamma_0$. Setting $u=u_1-u_2$, we obtain
\begin{equation}\label{pp}
(\Delta+q_2)u=(q_2-q_1)u_1\quad \mbox{in}\,\,\Omega, \quad
u\vert_{\partial\Omega\setminus\Gamma_0}=\frac{\partial u}{\partial
\nu}\vert_{\partial\Omega\setminus\Gamma_0}=0.
\end{equation}

In a similar way to the construction of $u_1$, we construct a
complex geometric optics solution $v$ for the Schr\"odinger
equation with potential $q_2.$ The construction of $v$ repeats the
corresponding steps of the construction of $u_1.$ The only
difference is that instead of $q_{1}$ and $\tau$, we use $q_{2}$ and
$-\tau,$ respectively. We skip the details of the construction and
point out that similarly to (\ref{mozilaall}) it can be represented
in the form
\begin{eqnarray}\label{mozilaa}
v(x)=e^{-\tau{\Phi}}{(a+(\widetilde a_{1,1}+\widetilde
a_{1,2,1})/\tau)} +e^{-\tau\overline{\Phi}} {(\overline
a+(\widetilde b_{1,1}
+\widetilde b_{1,2,1})/\tau)}\nonumber\\
+ \left (e^{-\tau\Phi}\frac{a(\partial^{-1} _{\overline z}
q_{2}-M_{2})}{4\tau\partial_z\Phi} +e^{-\tau\overline\Phi}
\frac{\overline{a}(\partial^{-1}_{z} q_{2}-M_{4})}
{4\tau\overline{\partial_z\Phi}}\right
)+e^{-\tau\varphi}o_{L^2(\Omega)}(\frac 1\tau) \quad
\mbox{as}\,\tau\rightarrow +\infty,\quad  v\vert_{\Gamma_0}=0,
\end{eqnarray}
where $M_{2}(z)$ and $M_{4}(\overline z)$ satisfy
$$
(\partial^{-1}_{\overline z}q_{2} -M_{2})(\widetilde x)=0, \quad
\quad (\partial^{-1}_{z} q_{2} - M_{4})(\widetilde x)= 0.
$$
The functions $\widetilde a_1(z)=\widetilde a_{1,1}(z)+\widetilde
a_{1,2}(z)$ and $\widetilde b_1(z)=\widetilde b_{1,1}(z)+\widetilde
b_{1,2}(z)$ are given by
$$
\widetilde a_{1,1}(z) +\widetilde b_{1,1}(\overline
z)=\frac{a(\partial^{-1} _{\overline z}
q_{2}-M_{2})}{4\tau\partial_z\Phi} +
\frac{\overline{a}(\partial^{-1}_{z} q_{2}-M_{4})}
{4\tau\overline{\partial_z\Phi}} \quad\mbox{on}\,\,\Gamma_0,
$$
\begin{equation}
\quad \widetilde a_{1,1}, \widetilde b_{1,1}\in C^1(\overline\Omega)
\end{equation}
and $\widetilde a_{1,2,1}(z), \widetilde b_{1,2,1}(\overline z)\in
C^1(\overline \Omega)$ are a holomorphic function and an antiholomorphic
function respectively such that
$$
\widetilde a_{1,2,1}(z)=\frac{1}{8\pi}\int_{\Gamma_0^*}
\frac{(\nu_1+i\nu_2)\overline{a}(\partial^{-1}_\zeta
q_{2}-M_{4})e^{\tau(\overline\Phi-\Phi)}}{(\zeta-z)\partial_{\overline\zeta}
\overline\Phi} d\sigma
$$
and
$$
\widetilde b_{1,2,1}(\overline z)=\frac{1}{8\pi}\int_{\Gamma_0^*}
\frac{(\nu_1-i\nu_2) a(\partial^{-1}_{\overline
\zeta}q_{2}-M_{2})e^{\tau(\Phi-\overline\Phi)}}{(\overline\zeta-\overline
z)\partial_\zeta\Phi} d\sigma .
$$

Denote $q=q_1-q_2.$ Taking the scalar product of equation (\ref{pp})
with the function $v$, we have:
\begin{equation}
\int_\Omega q u_1vdx=0.
\end{equation}
From formulae (\ref{mozilaall}) and (\ref{mozilaa}) in the
construction of complex geometric optics solutions, we have
\begin{eqnarray}\label{lala}
0=\int_\Omega q u_1vdx = \int_\Omega q(a^2+\overline
a^2)dx\nonumber\\
+ \frac 1\tau\int_\Omega q(
a(a_{1,1}+a_{1,2,1}+b_{1,1}+b_{1,2,1})+\overline{a}(\widetilde
a_{1,1}+\widetilde a_{1,2,1}+\widetilde b_{1,1}+\widetilde
b_{1,2,1}))dx                  \nonumber\\
+ \int_\Omega q(a\overline a e^{2\tau i\psi}+ a\overline a e^{-2\tau
i\psi})dx
                                       \nonumber\\
+ \frac{1}{4\tau}\int_{\Omega} \left( qa^2 \frac{\partial_{\overline
z} ^{-1}q_{2}-M_{2}} {\partial_z\Phi} + q\overline{a}^2
\frac{\partial_{z}^{-1}q_{2}
-{M_{4}}}{\overline{\partial_z\Phi}}\right)dx           \nonumber\\
- \frac{1}{4\tau}\int_\Omega\left( qa^2\frac{\partial_{\overline
z}^{-1} q_{1}-M_{1}}{\partial_z\Phi} +q\overline
a^2\frac{\partial_{z}^{-1}q_{1}-{
M_{3}}}{\overline{\partial_z\Phi}}\right)dx\nonumber\\
+ o(\frac{1}{\tau})=0\quad\mbox{as}\,\,\tau \rightarrow +\infty.
\end{eqnarray}

Since the potentials $q_j$ are not necessarily from
$C^{\infty}_0(\overline\Omega)$, we can not directly use the stationary phase
argument (see Proposition {\ref{jabloko} in Section 7).
If the function $q$ is not identically equal to zero on $\Omega$, then for
some positive $\alpha'$ we set $\frak X=\{x\in \Omega\vert
\vert q(x)\vert>\alpha'\}.$
Since the holomorphic function $a$ is not identically equal to the
constant, this function is not equal to zero on open dense set $\mathcal
V.$  The set ${\frak X}\cap V$ has positive measure. Let a point
$\widetilde x_*\in \Omega$ be some point from $ {\frak X}\cap V.$
Proposition \ref{Proposition -1} states that there exists a
holomorphic function $\Phi$ such that (\ref{1'})-(\ref{kk}) are
satisfied and a point $\widetilde x\in \mathcal H$ can be chosen
arbitrarily close to any given point in $\Omega $. Therefore such a
point $\widetilde x$ can be chosen close to $\widetilde x_*$.
Since $q_j \in W^1_p(\Omega)$ with $p>2$, the function $q$ is
continuous on $\overline\Omega.$  Therefore for the
point $\widetilde x\in \mathcal H$ we have
\begin{equation}\label{zvezda}
q(\widetilde x)\ne 0\quad \mbox{and}\quad a(\widetilde x)\ne 0.
\end{equation}

Let $\widehat q\in C^\infty_0(\Omega)$ satisfy $\widehat q(\widetilde
x)=q(\widetilde x).$ We have
\begin{equation}\label{rono}
\int_\Omega q\mbox{Re}\,(a\overline a e^{2\tau i\psi})dx
=\int_\Omega \widehat q\mbox{Re}\,(a\overline a e^{2\tau
i\psi})dx+\int_\Omega (q-\widehat q)\mbox{Re}\,(a\overline a e^{2\tau
i\psi})dx.
\end{equation}
Using Proposition \ref{gandonnal1} and (\ref{begemot}) we obtain
\begin{equation}\label{masa}
\int_\Omega \widehat q(a\overline a e^{2\tau i\psi} + a\overline a
e^{-2\tau i\psi})dx=\frac{2\pi (q\vert a\vert^2)(\widetilde
x)\mbox{Re}\,e^{2{\tau} i\psi(\widetilde x)}} {{\tau}
\vert(\mbox{det}\, H_\psi)(\widetilde x)\vert^\frac
12}+o\left(\frac{1}{{\tau}}\right)\quad\mbox{as}\,\,\tau \rightarrow
+\infty.
\end{equation}
The second term on the right-hand side of (\ref{rono}) after
integration by parts is written as
$$
\int_\Omega (q-\widehat q)(a\overline a e^{2\tau i\psi}+ a\overline a
e^{-2\tau i\psi})dx=\int_\Omega (q-\widehat q)\left (a\overline a \frac
{(\nabla\psi,\nabla)e^{2\tau i\psi}}{2\tau i\vert
\nabla\psi\vert^2}-a\overline a  \frac {(\nabla\psi,\nabla)e^{-2\tau
i\psi}}{2\tau i\vert \nabla\psi\vert^2}\right )dx
$$
$$
= \int_{\partial\Omega} q\left (a\overline a \frac
{(\nabla\psi,\nu)e^{2\tau i\psi}}{2\tau i\vert
\nabla\psi\vert^2}-a\overline a  \frac {(\nabla\psi,\nu)e^{-2\tau
i\psi}}{2\tau i\vert \nabla\psi\vert^2}\right )d\sigma
$$
\begin{equation}\label{opl}
-\frac{1}{2\tau i}
\int_\Omega\left \{ e^{2\tau i\psi}\mbox{div}\,\left ((q-\widehat
q)a\overline a \frac {\nabla\psi}{\vert \nabla\psi\vert^2}\right ) -
e^{-2\tau i\psi}\mbox{div}\,\left ( (q-\widehat q )a\overline a
\frac{\nabla\psi}{\vert \nabla\psi\vert^2}\right )\right\}dx.
\end{equation}
Since $\psi\vert_{\Gamma_0}=0$, we have
$$
\int_{\partial\Omega} qa\overline a \left (\frac
{(\nabla\psi,\nu)e^{2\tau i\psi}}{2\tau i\vert \nabla\psi\vert^2}-
\frac {(\nabla\psi,\nu)e^{-2\tau i\psi}}{2\tau i\vert
\nabla\psi\vert^2}\right )d\sigma=\int_{\partial\Omega\setminus
\Gamma_0^*} \frac{qa\overline a}{2\tau i\vert \nabla\psi\vert^2}
(\nabla\psi,\nu)(e^{2\tau i\psi}- e^{-2\tau i\psi})d\sigma.
$$
By (\ref{1'}), (\ref{kk}) and Proposition \ref{granata} we conclude
that
\begin{equation}\nonumber
\int_{\partial\Omega} qa\overline a \left (\frac
{(\nabla\psi,\nu)e^{2\tau i\psi}}{2\tau i\vert \nabla\psi\vert^2}-
\frac {(\nabla\psi,\nu)e^{-2\tau i\psi}}{2\tau i\vert
\nabla\psi\vert^2}\right )d\sigma=o(\frac 1\tau)\quad
\mbox{as}\,\,\tau\rightarrow+\infty.
\end{equation}
The last integral over $\Omega$ in formula (\ref{opl})
is $o(\frac{1}{\tau})$ by Proposition \ref{gandonnal} and therefore
\begin{equation}\label{-3}
\int_\Omega(q-\widehat q)(a\overline a e^{2\tau i\psi} + a\overline a
e^{-2\tau i\psi})dx=o(\frac 1\tau)\quad\mbox{as}\,\,\tau \rightarrow
+\infty.
\end{equation}
Taking into account that $\psi(\widetilde x)\ne 0$ and using
(\ref{masa}), (\ref{-3}), we have from (\ref{lala}) that
\begin{equation}
\frac{2\pi (q\vert a\vert^2)(\widetilde x)} { \vert(\mbox{det}\,
H_\psi)(\widetilde x)\vert^\frac 12}=0.
\end{equation}
Hence $q(\widetilde x)=0$, and we have a contradiction with
(\ref{zvezda}).
The proof of the theorem is completed. $\blacksquare$

In the case $\Gamma_0=\emptyset$, the uniqueness in determining a
potential $q$ in the two dimensional case was proved for the
conductivity equation by Nachman in \cite{N} within $C^2$
conductivities, and later in \cite {AP} within $L^\infty$
conductivities.  The case of the Schr\"odinger equation
was solved by Bukhegim \cite {Bu} and for the improvement of regularity assumption of potential for Bukhgeim's uniqueness result, see \cite{IY4}.
Theorem \ref{balalaika} was originally proved in \cite{IUY} for
$C^{2+\alpha}(\overline\Omega)$ potentials, and in \cite {IY}, the
regularity assumption on potentials was improved to up to
$W^1_p(\Omega)$ with $p>2.$   The case of general second-order
elliptic equation was studied in the papers \cite{IUY2} and
\cite{IUY1}.  See also \cite{EB}, \cite {BrU}.
The results of \cite{IUY} were extended to a Riemannian
surface in \cite{GT}.   Conditional stability estimates in
determining a potential are obtained in \cite{Nov}.
As for reconstruction, see e.g., \cite{Nov-San1}. An analog of the main
theorem of \cite{IUY} for the Neumann-to-Dirichlet map was proved
in \cite{IUY4}.

In \cite{IY1} the result of Theorem \ref{balalaika} was extended to
the weakly coupled systems of elliptic equations. More precisely,
consider the
following boundary value problem:
\begin{equation}\label{o-1}
L(x,D)u=\Delta u+2A\partial_z u+2B\partial_{\overline z} u+Qu=0\quad
\mbox{in}\,\,\Omega,\quad  u\vert_{\Gamma_0}=0.
\end{equation}
Here $u=(u_1,\dots, u_N)$ and $A(x),B(x),Q(x)$ be smooth complex-valued
$N\times N$ matrix-valued functions.

Consider the following Dirichlet-to-Neumann map
$\Lambda(A,B,Q,\Gamma_0)$:

\begin{equation}\label{vivi}
\Lambda (A,B,Q,\Gamma_0) (f)= \frac{\partial u}{\partial
\nu}\vert_{\partial\Omega\setminus\Gamma_0},
\end{equation}
where \begin{equation}\label{vivi1} L(x,D)u=\Delta u+2A\partial_z
u+2B\partial_{\overline z} u+Qu=0\quad\mbox{in}\,\,\Omega, \quad
u\vert_{\Gamma_0}=0,\quad u\vert_{\partial\Omega\setminus
\Gamma_0}=f.
\end{equation}

We have
\begin{theorem}(\cite{IY1})\label{vokal}
Let $A_j,B_j \in C^{5+\alpha}(\overline \Omega)$ and $Q_j\in
C^{4+\alpha}(\overline \Omega)$ for $j=1,2$ and some $\alpha\in
(0,1)$. Suppose that $\Lambda (A_1,B_1,Q_1,\Gamma_0)=\Lambda
(A_2,B_2,Q_2,\Gamma_0).$ Then
\begin{equation}\label{op!}
A_1=A_2\quad\mbox{and}\,\, B_1=B_2\quad \mbox{on}
\,\,\partial\Omega\setminus \Gamma_0,
\end{equation}
\begin{equation}\label{A1}
2\partial_z(A_1-A_2)+B_2(A_1-A_2)+(B_1-B_2)A_1-(Q_1-Q_2)=0\quad\mbox{in}
\,\,\Omega
\end{equation}
and
\begin{equation}\label{A2}
2\partial_{\overline
z}(B_1-B_2)+A_2(B_1-B_2)+(A_1-A_2)B_1-(Q_1-Q_2)=0\quad\mbox{in}\,\,\Omega.
\end{equation}
\end{theorem}

{\bf Remark 1.}{\it The proof of Theorem \ref{vokal} is based on the
construction of the complex geometric optics solutions, which is
performed in a way similar to  one presented in the proof of
Theorem \ref{balalaika}. Therefore it is critically important
that the principal parts of all the equations in (\ref{vivi1}) are
the Laplace operator for the construction of complex geometric
optics solutions. If the principal parts of the operators in (\ref{vivi1})
are different, then such a construction in general is impossible and
Calder\'on's problem for such a system is still open.  In Section 6, we
treat the Lam\'e system whose principal parts are different but a special
structure allows us to construct complex geometric optics solutions.}
\\
\vspace{0.3cm}

The simultaneous determination of all three matrices $A,B, Q$ from
the Dirichlet-to-Neumann map is impossible. Theorem \ref{vokal}
asserts that any two coefficient matrices among three are uniquely
determined by Dirichlet-to-Neumann map defined by (\ref{vivi}) and
(\ref{vivi1}) for the system of elliptic differential equations.
That is,
\begin{corollary}\label{coroAA} (\cite{IY1})
Let $(A_j,B_j,Q_j) \in C^{5+\alpha}(\overline\Omega) \times
C^{5+\alpha} (\overline\Omega)\times C^{4+\alpha}(\overline
\Omega)$, $j=1,2$ for some $\alpha\in (0,1)$ and be complex-valued.
We assume either $A_1\equiv A_2$ or  $B_1 \equiv B_2$ or $Q_1
\equiv Q_2$ in $\Omega$.  Then $\Lambda (A_1,B_1,Q_1,\Gamma_0)=
\Lambda (A_2,B_2,Q_2,\Gamma_0)$ implies $(A_1,B_1,Q_1) =
(A_2,B_2,Q_2)$ in $\Omega$.
\end{corollary}

Next we consider other form of elliptic systems:
\begin{equation}\label{OM}
\widetilde L(x,D)u=\Delta u+\mathcal A\partial_{x_1} u +\mathcal
B\partial_{x_2} u + Qu.
\end{equation}
Here $\mathcal A$, $\mathcal B$, $Q$ are complex-valued $N\times N$
matrices. Let us define the following Dirichlet-to-Neumann map
$\widetilde \Lambda(\mathcal A,\mathcal B,Q,\Gamma_0)$:
\begin{equation}\label{vova}
\widetilde \Lambda (\mathcal A,\mathcal B, Q,\Gamma_0) (f)=
\frac{\partial u}{\partial
\nu}\vert_{\partial\Omega\setminus\Gamma_0},
\end{equation}
where
\begin{equation}\label{vova1}
 \thinspace \widetilde L(x,D)u= \Delta u+\mathcal
A\partial_{x_1} u +\mathcal B\partial_{x_2} u +
Qu=0\,\,\mbox{in}\,\,\Omega, \, u\vert_{\Gamma_0}=0,\quad
u_{\partial\Omega\setminus\Gamma_0}=f,\quad  u\in H^1(\Omega).
\end{equation}
Then one can prove the following corollary.
\begin{corollary}(\cite{IY1})\label{coroB}
Let $Q_1,Q_2\in C^{4+\alpha}(\overline\Omega)$ and
$(\mathcal A_{1}, \mathcal B_{1}), (\mathcal A_{2}, \mathcal B_{2})
\in C^{5+\alpha}(\overline\Omega)\times C^{5+\alpha}(\overline \Omega)$
for some  $\alpha\in (0,1).$
We assume that $Q_1\equiv Q_2$ in $\Omega$ and $\widetilde
\Lambda(\mathcal A_{1},\mathcal B_{1},Q_1,\Gamma_0)=\widetilde
\Lambda(\mathcal A_{2},\mathcal B_{2},Q_1,\Gamma_0)$. Then
$(\mathcal A_{1}, \mathcal B_{1})\equiv (\mathcal A_{2}, \mathcal
B_{2})$ in $\Omega$.
\end{corollary}

{\bf Proof.}  Observe that $\widetilde L(x,D)=\Delta + A\partial_{z}
+B\partial_{\overline z}  + Q$ where $A=\mathcal A+i\mathcal B$ and
$B=\mathcal A-i\mathcal B.$ Therefore, applying Corollary
\ref{coroAA}, we complete the proof. $\blacksquare$

This corollary generalizes the result of \cite{ChengYama} where for
the scalar elliptic operator $ \Delta  + a\frac{\partial }{\partial
x_1} + b\frac{\partial }{\partial x_2} $ the uniqueness  of
determination of the coefficients $a,b$ was proved assuming that the
measurements are made on the whole boundary.
\\
{\bf Remark 2. } {\it Unlike Corollary \ref{coroAA}, in the two cases
of $\mathcal A_1\equiv\mathcal A_2$ and $\mathcal B_1\equiv\mathcal
B_2$, we can not, in general, claim that $(\mathcal A_1,\mathcal
B_1, Q_1)= (\mathcal A_2,\mathcal B_2, Q_2).$
We can prove only \\
(i) $\frac{\partial \mathcal B_1}{\partial x_1} = \frac{\partial
\mathcal B_2}{\partial x_1}$ in $\Omega$ if $\mathcal A_1 = \mathcal
A_2$ in $\Omega$.
\\
(ii) $\frac{\partial \mathcal A_1}{\partial x_2} = \frac{\partial
\mathcal A_2}{\partial x_2}$ in $\Omega$ if $\mathcal B_1 = \mathcal
B_2$ in $\Omega$.

Moreover consider the following example
$$
\Omega = (0,1)\times (0,1),
$$
$$
\partial\Omega\setminus \Gamma_0=\{(x_1,x_2); \thinspace x_2=0, \,\,
0<x_1<1\}\cup \{(x_1,x_2); \thinspace x_2=1, \,\,  0<x_1<1\},
$$
and let us choose $\eta(x_2)\in C^\infty_0(0,1)$. Then the operators
$\widetilde L(x,D)$ and $e^{s\eta}\widetilde L(x,D)e^{-s\eta}$
generate the same Dirichlet-to-Neumann map (\ref{vova}),
(\ref{vova1}), but the matrix coefficient matrices are not equal.}
\\
\vspace{0.3cm}

{\bf General second-order elliptic operator.} We consider  a general
second-order elliptic operator:
\begin{equation}\label{OMX}
L(x,D)u=\Delta_gu + 2A\frac{\partial u}{\partial z}
+ 2B\frac{\partial u}{\partial \overline z} + qu.
\end{equation}
Here $g = g(x) = \{ g_{jk}\}_{1\le j,k\le 2}$ is
a positive definite symmetric matrix in $\Omega$ and
$\Delta_g$ is the Laplace-Beltrami operator
associated to the Riemannian metric $g$:
$$
\Delta_g=  \frac{1}{\sqrt{\mbox{det} g}}
\sum_{j,k=1}^2\frac{\partial}{\partial x_k}
(\sqrt{\mbox{det} g}\,g^{jk}\frac{\partial }
{\partial x_j}),
$$
where we set $\{g^{jk}\} = g^{-1}$. Assume that $g\in
C^{7+\alpha}(\overline\Omega)$, $(A,B,q), (A_j,B_j,q_j) \in
C^{5+\alpha} (\overline\Omega)\times C^{5+\alpha}(\overline\Omega)
\times C^{4+\alpha}(\overline \Omega)$, $j=1,2$ for some $\alpha\in
(0,1)$, are complex-valued functions. We set
$$
L_k(x,D)=\Delta_{g_k} + 2A_{k}\frac{\partial}{\partial z}
+ 2B_{k}\frac{\partial}{\partial \overline z}
+ q_k.
$$

We define the Dirichlet-to-Neumann map by formula:
\begin{equation}\label{vova3}
\Lambda_{g,A,B,q,\Gamma_0}(f) = \frac{\partial u}{\partial
\nu_g}|_{\partial\Omega\setminus\Gamma_0}, \end{equation} where
\begin{equation}\label{vova4}
L(x,D)u = 0 \quad \mbox{in }\,\,\Omega, \quad \thinspace
u\vert_{\Gamma_0}=0,\quad
u\vert_{\partial\Omega\setminus\Gamma_0}=f,\quad u\in H^1(\Omega)
\end{equation}
and $\frac{\partial}{\partial\nu_g}=\root\of{\mbox{det}
g}\sum_{j,k=1}^2 g^{jk}\nu_k\frac{\partial}{\partial x_j}$ is the
conormal derivative with respect to the metric $g.$

Our goal is to determine the metric $g$ and coefficients $A,B,q$
from the Dirichlet-to-Neumann map ${ \Lambda}_{g,A,B,q,\Gamma_0}$
given by (\ref{vova3}) and (\ref{vova4}). In general, the uniqueness
is impossible. There are the following main invariance properties of the
Dirichlet-to-Neumann map in the problem.
\begin{itemize}
\item {\sl Conformal invariance}. Let $\beta\in C^{7+\alpha}
(\overline\Omega)$ be a strictly positive function.
Then
\begin{equation}
\Lambda_{g,A,B,q,\Gamma_0}= \Lambda_{\beta g, \frac{A}{\beta},
\frac{B}{\beta}, \frac{ q}{\beta},\Gamma_0}.
\end{equation}
This follows since the Laplace-Beltrami operator is conformal invariant
in two dimensions:
$$
\Delta_{\beta g} = \frac{1}{\beta} \Delta_g.
$$
\item {\sl Gauge transformation.}
It is easy to see that the Dirichlet-to-Neumann map for the
operators $e^{-\eta} L(x,D) e^{\eta}$ and $L(x,D)$ are the same
provided that $\eta$ is a smooth complex-valued function such that
\begin{equation}\label{victoryy}\eta\in C^{6+\alpha}(\overline \Omega),\quad \eta\vert_{\partial\Omega\setminus\Gamma_0}
=\frac{\partial\eta}{\partial\nu}\vert
_{\partial\Omega\setminus\Gamma_0}=0.
\end{equation}

\item {\sl Diffeomorphism invariance.}
Let $F=(F_1,F_2):\overline \Omega\rightarrow \overline \Omega$  be a
diffeomorphism such that
$F\vert_{\partial\Omega\setminus\Gamma_0}=Identity$. The pull back of a
Riemannian metric $g$ is given as composition of matrices by
\begin{equation}\label{zz}
F^*g = ((DF)\circ g\circ
(DF)^T )\circ F^{-1}
\end{equation}
and $DF$ denotes the differential of $F$, $(DF)^T$ its
transpose and $\circ$ denotes the matrix composition.

Moreover we introduce the functions: $A_F= \{(A+B)(\frac{\partial
F_1}{\partial x_1} -i\frac{\partial F_2}{\partial
x_1})+i(B-A)(\frac{\partial F_1}{\partial x_2}-i\frac{\partial
F_2}{\partial x_2})\}\circ F^{-1} \vert det \,D F^{-1}\vert, B_F=
\{(A+B)(\frac{\partial F_1}{\partial x_1}+i\frac{\partial
F_2}{\partial x_1}) +i(B-A)(\frac{\partial F_1}{\partial
x_2}+i\frac{\partial F_2}{\partial x_2})\}\circ F^{-1} \vert det \,D
F^{-1}\vert, q_F= \vert det \,D F^{-1}\vert (q\circ F^{-1}) .$
Then we can verify
\begin{equation}
\Lambda_{g,A,B,q,\Gamma_0}= \Lambda_{F^*g, A_F, B_F, q_F,\Gamma_0}.
\end{equation}
\end{itemize}

We can show the converse, that is, the above three kinds of the invariance
exhaust all the possibilities.  We have
\begin{theorem}(\cite{IUY1})\label{general}
Suppose that for some
$\alpha \in (0,1)$, there exists a positive function $\widetilde
\beta\in C^{7+\alpha} (\overline \Omega)$ such that
$(g_1-\widetilde\beta g_2)\vert_{\partial\Omega\setminus\Gamma_0}
=\frac{\partial(g_1-\widetilde\beta
g_2)}{\partial\nu}\vert_{\partial\Omega\setminus\Gamma_0}=
(A_1-\frac {A_2}{\widetilde\beta
})\vert_{\partial\Omega\setminus\Gamma_0}=(B_1-\frac{B_2}{\widetilde\beta}
)\vert_{\partial\Omega\setminus\Gamma_0}=0.$ Then
$\Lambda_{g_1,A_1,B_1,q_1,\Gamma_0} =
\Lambda_{g_2,A_2,B_2,q_2,\Gamma_0}$ if and only if there exist a
diffeomorphism $F\in C^{8+\alpha}(\overline\Omega), F: \overline
\Omega\rightarrow \overline \Omega$ satisfying
$F\vert_{\partial\Omega\setminus\Gamma_0}$=Id, a positive function
$\beta\in C^{7+\alpha} (\overline \Omega)$ and a complex valued
function $\eta$ satisfying (\ref{victoryy}) such that
$$
L_2(x,D) = e^{-\eta}K(x,D)e^{\eta},
$$
where
$$
K(x,D) = \Delta_{\beta F^*g_1}
+\frac{2}{\beta}({A_1}_F\frac{\partial}{\partial z}
+{B_1}_F\frac{\partial} {\partial \overline z}) + \frac{1}{\beta}
{q_1}_F.
$$
\end{theorem}
\bigskip

{\bf Calder\'on's problem for the matrix conductivity.}
The question proposed by
Calder\'on \cite{C} is whether one can uniquely determine the electrical
conductivity of a medium by making voltage and current measurements
at the boundary.

In the anisotropic case the conductivity depends on direction and is
represented by a positive definite symmetric matrix
$\{\sigma^{jk}\}$. The conductivity equation with voltage potential
$f$ on $\partial\Omega$ is given by
$$
\mathcal L(x,D)u=\sum_{j,k=1}^2\frac{\partial}{\partial x_j}
(\sigma^{jk}\frac{\partial u}{\partial x_k}) = 0
\quad\mbox{in}\,\,\Omega, \quad u\vert_{\partial\Omega}=f.
$$

The Dirichlet-to-Neumann map is defined  by
$$
\Lambda_{\sigma}(\Gamma_0)f=\sum_{i,j=1}^2\sigma^{ij}\nu_i\frac{\partial
u} {\partial x_j}\vert_{\Gamma_0},\quad \mathcal
L(x,D)u=0\quad\mbox{in}\,\,\Omega, \quad
u\vert_{\partial\Omega\setminus \Gamma_0}=f, \quad \quad
u\vert_{\Gamma_0}=0.
$$

It has been known for a long time (e.g., \cite{K-V}) that
$\Lambda_{\sigma}$ does not determine $\sigma$ uniquely in the anisotropic
case.  Let $F:\overline\Omega\rightarrow \overline\Omega$ be a diffeomorphism
such that $F(x)=x$ for $x$ on ${\partial\Omega\setminus\Gamma_0}.$
Then
$$
\Lambda_{F^*\sigma}=\Lambda_\sigma,
$$
where
\begin{equation}\label{star}F^*\sigma=\left(\frac{(DF)\circ\sigma\circ
(DF)^T}{\vert det DF\vert}\right)\circ F^{-1}.
\end{equation}

In the case of  $\Gamma_0= \emptyset$, the question whether one
can determine the conductivity up to the above obstruction has been
solved in two dimensions for $C^2$ conductivities in \cite{N} and
merely $L^\infty$ conductivities in \cite{APL}.  See also \cite{AP}.
The method of the proof in all these papers is based on the reduction to
the isotropic case performed using isothermal coordinates \cite{Ah}.

We can prove the uniqueness in Calder\'on's problem for the
anisotropic conductivity:

\begin{theorem}(\cite{IUY1}) Let  $\sigma_1, \sigma_2 \in C^{7+\alpha}
(\overline\Omega)$ with some
$\alpha\in (0,1) $ be positive definite symmetric matrices on
$\overline\Omega$ such that
$(\sigma_1-\sigma_2)\vert_{\partial\Omega\setminus\Gamma_0}
=\frac{\partial}{\partial\nu}(\sigma_1-\sigma_2)\vert
_{\partial\Omega\setminus\Gamma_0}=0.$
If $\Lambda_{\sigma_1}(\Gamma_0)= \Lambda_{\sigma_2} (\Gamma_0)$,
then there exists a diffeomorphism $F:\overline{\Omega} \rightarrow
\overline{\Omega}$ satisfying
$F\vert_{\partial\Omega\setminus\Gamma_0}=$ Identity and $F\in
C^{8+\alpha}(\overline\Omega)$ such that
$$
F^*\sigma_1 = \sigma_2.
$$
\end{theorem}

The uniqueness corresponding to the isotropic case was proven in
\cite{IUY} and in
fact follows from Theorem \ref{general} in the case where
$g=Identity$ and $A=B=0.$ We mention that \cite{GT} has proven a
similar result  for general Riemann surfaces in the case where $g$
is not the identity but fixed.
\\
\vspace{0.3cm}

{\bf General case where the principal part is the Laplacian.} Assume
that the principal parts of second-order elliptic operators under
consideration are the Laplacian: $g = I$. Then we can prove a bit sharper
result than Theorem \ref{general}:

\begin{theorem}(\cite{IUY1})\label{xyz}
The relation $\Lambda_{I,A_1,B_1,q_1,\Gamma_0} =
\Lambda_{I,A_2,B_2,q_2,\Gamma_0}$ holds true if and only if there
exists a function $\eta\in C^{6+\alpha}(\overline \Omega),$
$\eta\vert_{\partial\Omega\setminus\Gamma_0}
=\frac{\partial\eta}{\partial\nu}\vert_{\partial\Omega\setminus\Gamma_0}=0$
such that
\begin{equation}\label{opana}
L_1(x,D)=e^{-\eta}L_2(x,D)e^{\eta}.
\end{equation}
\end{theorem}
{\bf Proof.} For simplicity we consider only the case when domain
$\Omega$ is simply connected. The proof for the general domain is given
in \cite{IUY1}. By Theorem \ref{vokal} we have
\begin{equation}\label{bound}
A_1=A_2, \quad B_1=B_2 \quad \mbox{on}\quad
{\partial\Omega\setminus\Gamma_0},
\end{equation}
and in $\Omega$ we have
\begin{equation}\label{vovka}
-2\frac{\partial}{\partial z}({A}_1-{A}_2) - A_1{B}_1 +{A}_2B_2 +
(q_1-q_2)=0,
\end{equation}
\begin{equation}\label{vovka1}
-2\frac{\partial }{\partial \overline{z}} ({B}_1-{B}_2) -{A}_1B_1
+A_2{B}_2 +(q_1-q_2)=0.
\end{equation}
We only prove the
sufficiency since the necessity of the condition is easy to be
checked.
By (\ref{vovka}) and (\ref{vovka1}), we have
$\frac{\partial}{\partial z}({A}_1-{A}_2) =\frac{\partial}{\partial
\overline{z}}({B}_1-{B}_2)$. This equality is equivalent to
$$
\frac{\partial (A-B)}{\partial x_1} = i\frac{\partial
(B+A)}{\partial x_2}\,\quad\mbox{where} \quad
(A,B)=(A_1-A_2,B_1-B_2).
$$
Since the domain $\Omega$ simply connected, there exists a function
$\widetilde\eta$ such that:
\begin{equation}\label{zaika}
( i(B+A),(A-B))=\nabla \widetilde\eta.
\end{equation}
By (\ref{bound}) we have
$$
\widetilde\eta\vert_{\partial\Omega\setminus\Gamma_0}=\nabla
\widetilde\eta\vert_{\partial\Omega\setminus\Gamma_0}=0.
$$

Setting $2\eta=-i\widetilde \eta$ we have from (\ref{zaika})
$$
((B+A),i(B-A))=2\nabla \eta.
$$
Therefore (\ref{vovka}) yields
\begin{equation}\label{finish}
q_1=q_2+\Delta\eta +4\frac{\partial\eta}{\partial z}
\frac{\partial\eta}{\partial\overline z} +2\frac{\partial\eta}{\partial
z}A_2 +2\frac{\partial\eta}{\partial \overline z}B_2.
\end{equation}
The operator $L_1(x,D)$ given by the right-hand side of
$(\ref{opana})$ has the Laplace operator as the principal part, the
coefficients of $\frac{\partial}{\partial x_1}$ is
$A_2+B_2+2\frac{\partial\eta}{\partial x_2}$, the coefficient of
$\frac{\partial}{\partial x_2}$ is
$i(B_2-A_2)+2\frac{\partial\eta}{\partial  x_1}$, and the
coefficient of the zeroth order term is given by the right-hand side
of (\ref{finish}). The proof of the proposition is complete. $ \blacksquare$
\\
\vspace{0.3cm}

{\bf The magnetic Schr\"odinger equation.} Denote $\widetilde
A=(\widetilde A_1,\widetilde A_2)$, where $\widetilde A_j$, $j=1,2$,
are real-valued, $\widetilde {\mathcal A}=\widetilde A_1 -
i\widetilde A_2$, rot $\widetilde A=\frac{\partial \widetilde
A_2}{\partial x_1}-\frac{\partial \widetilde A_1} {\partial x_2}$.
The magnetic Schr\"odinger operator is defined by
$$
{\mathcal L}_{\widetilde A,\widetilde q}(x,D) = \sum_{k=1}^2( \frac
1i\frac{\partial}{\partial x_k} + \widetilde A_k)^2+\widetilde q.
$$
Let us define the following Dirichlet-to-Neumann map $$\widetilde
\Lambda_{\widetilde A,\widetilde q,\Gamma_0}(f)= \frac{\partial
u}{\partial \nu}\vert _{\partial\Omega\setminus\Gamma_0}, $$ where
$$ {\mathcal L}_{\widetilde A,
\widetilde q}(x,D)u=0\,\,\mbox{in}\,\,\Omega, \quad
u\vert_{\partial\Omega\setminus {\Gamma_0}}=f,\quad
u\vert_{{\Gamma_0}}=0, \thinspace u\in H^1(\Omega).
$$

Theorem  \ref{vokal} implies

\begin{corollary}(\cite{IUY1})\label{2.1}
{\it Let real-valued vector fields $\widetilde A^{(1)}, \widetilde
A^{(2)} \in C^{5+\alpha}(\overline \Omega)$ and complex-valued
potentials $\widetilde q^{(1)}, \widetilde q^{(2)}  \in
C^{4+\alpha}(\overline\Omega)$ with some $\alpha \in (0,1)$, satisfy
$\widetilde \Lambda_{\widetilde A^{(1)},\widetilde q^{(1)},\Gamma_0}
=\widetilde \Lambda_{\widetilde A^{(2)},\widetilde
q^{(2)},\Gamma_0}$. Then $\widetilde q^{(1)}=\widetilde q^{(2)}$ and
$\mbox{rot}\, \widetilde A^{(1)}=\mbox{rot}\, \widetilde
A^{(2)}$.}\end{corollary}

In the case of the Dirichet-to-Neumann map on the whole boundary,
see \cite{KU} and \cite{S}: \cite{S} proved a uniqueness result
provided that both electric and magnetic potentials are small, and
\cite{KU} proved a uniqueness result for a special case
of the magnetic Schr\"odinger equation, namely the Pauli
Hamiltonian.  See also \cite{S} and \cite{Sun1}.
\\
\vspace{0.2cm}

We conclude this section with the uniqueness in the case where
the subboundaries of Dirichlet input and measured Neumann data
are disjoint.

Let $\partial\Omega=\overline{\Gamma_1\cup\Gamma_2\cup \Gamma_0}$
where $\Gamma_1\cap\Gamma_2 =\Gamma_0\cap\Gamma_k=\emptyset$,
$k=1,2$. Then we consider the unique identifiability of the
conductivity by taking all pairs of Dirichlet data on the
subboundary $\Gamma_1$ and the corresponding Neumann data on the
subboundary $\Gamma_2$:
\begin{equation}\label{eq:0.2}
\mathcal A_\gamma(\Gamma_1,\Gamma_2)(f) =  \gamma\frac{\partial
u}{\partial \nu}\Big|_{\Gamma_2}, \quad \mbox{div}(\gamma\nabla u)
= 0 \mbox{ in }\Omega, \quad u\big|_{\Gamma_0\cup\Gamma_2}  =
0,\quad u\vert_{\Gamma_1}=f.
\end{equation}
We consider that the input is located on $\Gamma_1$, while the
output is measured on $\Gamma_2$. In the case where
$\Gamma_1=\Gamma_2$ and is an arbitrary open subset of the boundary,
the global uniqueness was shown in \cite{IUY} with $\gamma \in
C^{3+\alpha}(\overline\Omega)$, with some $\alpha\in (0,1)$.
See also Theorem \ref{balalaika}.

In order to state our main result, we need the following geometric
assumption on the position of the sets $\Gamma_1, \Gamma_2,
\Gamma_0$ on $\partial\Omega.$
\\
{\bf Assumption A.} {\it Let $\Gamma_1$, $\Gamma_2$, $\Gamma_0
\subset \partial\Omega$ be non-empty open subsets of
the boundary such that $\partial \Omega= \overline{\Gamma_1\cup
\Gamma_2\cup\Gamma_0}$, $\Gamma_1\cap \Gamma_2= \Gamma_k \cap
\Gamma_0 = \emptyset$, $\Gamma_k=\cup_{j=1}^2\Gamma_{k,j}$,
$\Gamma_0 = \cup_{\ell=1}^4 \Gamma_{0,\ell}$, where $\Gamma_{k,j}$,
$j,k=1,2$, $\Gamma_{0,\ell}$, $\ell=1,2,3,4$ are not empty open connected
subsets of $\partial\Omega$ and mutually disjoint. Then
$\partial\Omega$ is separated into
$$
\Gamma_{0,1}, \Gamma_{2,1}, \Gamma_{0,2}, \Gamma_{1,1},
\Gamma_{0,3}, \Gamma_{2,2}, \Gamma_{0,4}, \Gamma_{1,2}
$$
in the clockwise order.}

We note that $\Gamma_1, \Gamma_2$ can be arbitrarily small provided
that the above separation condition is satisfied.
%
%
%

Then
\begin{theorem}(\cite{IUY3})\label{main1}
We suppose Assumption A. Let $\gamma_j > 0$ on $\overline\Omega$ and
$\gamma_j \in C^{4+\alpha}(\overline \Omega)$, $j=1,2$ for some
$\alpha>0$. Assume $ \mathcal{A}_{\gamma_1}(\Gamma_1,\Gamma_2)=
\mathcal{A}_{\gamma_2}(\Gamma_1,\Gamma_2)$ and that $(\gamma_1 -
\gamma_2)\vert_{\Gamma_1 \cup \Gamma_2}=
\frac{\partial}{\partial\nu}(\gamma_1-\gamma_2)\vert_{\Gamma_*} =
0,$ where $\Gamma_*\subset \Gamma_1 \cup \Gamma_2$ is some open set.
Then $\gamma_1\equiv \gamma_2$ on $\overline{\Omega}$.
\end{theorem}

Next for the Schr\"odinger equation
$
L_q(x,D)u=\Delta u+qu=0\quad\mbox{in}\,\,\Omega,
$
we consider the problem of determining a complex-valued potential
$q$ by the following Dirichlet-to-Neumann map:
\begin{equation}\label{Z}
\Lambda_{q,\Gamma_1,\Gamma_2}(f)= \frac{\partial u}{\partial
\nu}\Big|_{\Gamma_2}, \,\,\mbox{ where}\,\, L_q(x,D)u=0 \quad
\mbox{in}\quad\Omega,\quad u\vert_{\Gamma_0\cup\Gamma_2}=0,\quad
u\vert_{\Gamma_1}=f,\quad u\in H^1(\Omega).
\end{equation}

Next we state the corresponding result for the Schr\"odinger equation.
\begin{theorem}(\cite{IUY3})\label{main}
We suppose Assumption A. Let $q_j \in C^{2+\alpha}(\overline
\Omega)$, $j=1,2$ for some $\alpha>0$ and let $q_j$ be
complex-valued.  If
$$
\Lambda_{q_1,\Gamma_1,\Gamma_2}= \Lambda_{q_2,\Gamma_1,\Gamma_2},
$$
then we have
$$
q_1\equiv q_2 \quad \mbox{in $\Omega$}.
$$
\end{theorem}

{\bf Proof of Theorem \ref{main1}.} If $u$ is some solution to the
conductivity equation then the function $u^* = u\sqrt{\gamma}$
solves in domain $\Omega$ the Schr\"odinger with the potential $ q =
-\frac{\Delta\sqrt{\gamma}}{\sqrt{\gamma}}.$ We claim that the
Dirichlet-to-Neumann maps  (\ref{eq:0.2}) for the Schr\"odinger
equations with potentials
$q_j=-\frac{\Delta\sqrt{\gamma_j}}{\sqrt{\gamma_j}}$  are the same,
provided that the Dirichlet-to-Neumann maps (\ref{eq:0.2})are the
same. Indeed let $f\in L^2(\partial\Omega), \mbox{supp} f\subset
\Gamma_1.$ Setting $\tilde f=f/\root\of{\gamma_j}$ we have that
$\mathcal A_{\gamma_1}(\Gamma_1,\Gamma_2)(\tilde f)=\mathcal
A_{\gamma_2}(\Gamma_1,\Gamma_2)(\tilde f).$ Denote by $\tilde u_j$
the corresponding solutions to the conductivity equation
(\ref{eq:0.2}) with the Dirichlet boundary condition $\tilde f.$
Then $u_j=\tilde u_j\sqrt{\gamma_j}$ is the solution to the
Shr\"odinger equation with the potential $q_j$ and the Dirichlet
boundary condition $f.$ Observe that
$$
\mathcal A_{\gamma_1}(\Gamma_1,\Gamma_2)(f)=\frac{\partial
u_1}{\partial\nu}\vert_{\Gamma_2}=\frac{\partial (\tilde
u_1\sqrt{\gamma_1})}{\partial\nu}\vert_{\Gamma_2}=\frac{\sqrt{\gamma_1}\partial
\tilde
u_1}{\partial\nu}\vert_{\Gamma_2}=\frac{\sqrt{\gamma_2}\partial
\tilde u_2}{\partial\nu}\vert_{\Gamma_2}=\frac{\partial (\tilde
u_2\sqrt{\gamma_2})}{\partial\nu}\vert_{\Gamma_2}=\frac{\partial
u_2}{\partial\nu}\vert_{\Gamma_2}=\mathcal
A_{\gamma_2}(\Gamma_1,\Gamma_2)(f).
$$
Applying the theorem \ref{main} we obtain that $q_1=q_2.$ Then the
function  $w =\sqrt{\gamma_1} - \sqrt{\gamma_2}$ verifies
$$
\Delta w - \frac{\Delta\sqrt{\gamma_2}}{\sqrt{\gamma_2}}w = 0
\quad\mbox{in $\Omega$}, \quad
w\vert_{\partial\Omega}=\frac{\partial
w}{\partial\nu}\vert_{\Gamma_*}=0.
$$
Applying to the above problem  the classical unique continuation for
the second order elliptic operator (see e.g. Corollary  2.9 Chapter
XIV of \cite{Taylor}), we obtain $\gamma_1\equiv\gamma_2.$
$\blacksquare.$

%
%
%
%
%
%
%
\section{ Calder\'on's problem for semilinear elliptic
equations}
In this section, we assume that $\Gamma_0 \subset \partial\Omega$ is an
arbitrarily fixed relatively open subset of $\partial\Omega$.

Consider the following boundary value problem:
\begin{equation}\label{o-1}
P(x,D)u = \Delta u + q(x)u - f(x,u) =0 \quad
\mbox{in}\,\,\Omega,\quad  u\vert_{\Gamma_0}=0,
\end{equation}

We introduce the Dirichlet-to-Neumann  map $\Lambda_{q,f}$:
$$
\Lambda_{q,f}(g) = \frac{\partial u}{\partial
\nu}\vert_{\partial\Omega\setminus \Gamma_0},\,\,\mbox{ where
}\,\,P(x,D)u=0\quad\mbox{in}\,\,\Omega,\quad
u\vert_{\Gamma_0}=0,\,\,u\vert_{\partial\Omega\setminus
\Gamma_0}=g,\quad u\in H^1(\Omega).
$$

This section is concerned with  the following inverse problem: {\it
Determine a coefficient $q$ and a nonlinear term $f$ from the
Dirichlet-to-Neumann map $\Lambda_{q,f}$. }

In this section, we always assume that
$
f, \frac{\partial f}{\partial y}, \frac{\partial^2 f}{\partial
y^2}\in C^0(\overline\Omega\times {\Bbb R^1}).
$
We state other conditions on semilinear terms $f$:
\begin{equation}\label{00}
f(x,0) = \frac{\partial f}{\partial y}(x,0) = 0, \quad
x \in \Bbb R^1
\end{equation}
and for some positive constants $p>1, C_1, C_2$, the following holds
true:
\begin{equation}\label{01}
f(x,y)y\ge C_1 \vert y\vert^{p+1}-C_2, \quad \forall
(x,y)\in\Omega\times {\Bbb R}^1.
\end{equation}
Moreover for some $p_1>0, p_2>0$, $C_3>0$ and $C_4>0$, the following
inequalities holds true:
\begin{equation}\label{02}
\left\vert \frac{\partial f}{\partial y} (x,y)\right\vert\le C_3(1+\vert
y\vert^{p_1}),\quad
\left\vert \frac{\partial^2 f}{\partial y^2} (x,y)\right\vert
\le C_4(1+\vert
y\vert^{p_2}),\quad \forall (x,y)\in\Omega\times {\Bbb R}^1.
\end{equation}

The first result is concerned with the uniqueness in determining a
linear part, that is, a potential $q$.
\begin{theorem}\label{liska} (\cite{IY3})
Let functions $f_1,f_2$ satisfy (\ref{00}), (\ref{01}), (\ref{02})
and $q_j\in C^{2+\alpha}(\overline \Omega)$, $j=1,2$, with some
$\alpha\in (0,1).$ Suppose that
$\Lambda_{q_1,f_1}=\Lambda_{q_2,f_2}$. Then $q_1=q_2$ in $\Omega$.
\end{theorem}

Theorem \ref{liska} is concerned with the determination of
potentials in spite of unknown nonlinear terms, and the proof is similar to
Theorem \ref{balalaika}.
\\
{\bf Remark 1.} {\it Since our assumptions on the
potential $q$ and nonlinear term $f$ in general do not imply the
uniqueness of a solution for the boundary value problem for the
elliptic operator $P(x,D)$, by the equality
$\Lambda_{q_1,f_1}=\Lambda_{q_2,f_2}$, we mean the following: for
any pair $(v_1,v_2)$ such that $$ P_1(x,D) w=\Delta w+q_1 w-f_1(x,
w)=0, \quad w\vert_{\Gamma_0}=0,\quad \frac{\partial w}{\partial
\nu}\vert_{\partial\Omega\setminus \Gamma_0}=v_2,\quad
w\vert_{\partial\Omega\setminus \Gamma_0}=v_1
$$
there exists a function $\widetilde w\in H^1(\Omega)$ such that
$P_2(x,D)\widetilde w=\Delta \widetilde w+q_2\widetilde w-f_2(x,\widetilde w)
=0, \widetilde w\vert_{\Gamma_0}=0, \widetilde w\vert_{\partial\Omega\setminus
\Gamma_0}=v_1$ and
$\frac{\partial \widetilde w}{\partial \nu}\vert
_{\partial\Omega\setminus \Gamma_0} = v_2$.}
\\
\vspace{0.3cm}
\\
{\bf Remark 2.} {\it Theorem \ref{vokal} is still true if condition
(\ref{01}) is replaced by following: there exists a continuous
function $G$ such that a solution to the boundary value problem
$$
P(x,D)u=0\quad \mbox{in}\,\,\Omega, \quad u\vert_{\partial\Omega}=g
$$
satisfies the estimate
$$
\Vert u\Vert_{H^1(\Omega)}\le G(\Vert g\Vert_{H^\frac
12(\partial\Omega)}).
$$
 }
\\
\vspace{0.4cm}
\\
For any $F(t)\in C([0,1];C^{2+\alpha}(\overline\Omega))$ with
$\alpha\in (0,1)$, we introduce the set
$$
\mathcal O_F=\bigcup_{0\le t\le 1, x\in \Omega}
\{(x,F(x,t))\}.
$$
Let
\begin{eqnarray*}
&&\mathcal U_j=\{F\in C([0,1];C^{2+\alpha}(\overline\Omega));
\thinspace
F(\cdot,0)=0 \quad \mbox{$u(\cdot,t):= F(t)$ satisfies}\\
&& \Delta u(x,t) + q_ju(x,t) - f_j(x, u(x,t)) = 0, \thinspace x \in
\Omega, \quad u(\cdot,t)\vert_{\Gamma_0}=0\}, \quad j=1,2.
\end{eqnarray*}
\\
The next theorem asserts the uniqueness for semilinear terms $f_k$,
$k=1,2$ in some range provided that the potential $q$ is known:
\\
\vspace{0.3cm}
\begin{theorem}\label{liska1}(\cite{IY3})
Let $q_1=q_2=q \in C^{2+\alpha}(\overline{\Omega})$ be arbitrarily
fixed. Let functions $f_1,f_2\in C^{3+\alpha}(\overline\Omega\times
\Bbb R^1)$ for some $\alpha\in (0,1)$, satisfy (\ref{01}),
(\ref{02}) and $f_1(\cdot,0) = f_2(\cdot,0)=0$. Suppose that
$\Lambda_{q,f_1}=\Lambda_{q,f_2}.$ Then
$$
f_1-f_2 = 0 \quad \mbox{in $\bigcup_{j\in\{ 1,2\}} \bigcup_{F\in
\mathcal U_j} \mathcal O_{F}$}.
$$
\end{theorem}
\begin{corollary}
{\it Let $q_1, q_2 \in
C^{2+\alpha}(\overline{\Omega})$ and let functions $f_1,f_2\in
C^{3+\alpha}(\overline\Omega\times \Bbb R^1)$ with some $\alpha\in
(0,1)$, satisfy (\ref{00}), (\ref{01}) and (\ref{02}). Suppose that
$\Lambda_{q_1,f_1}=\Lambda_{q_2,f_2}.$ Then $q_1 = q_2$ in $\Omega$
and
$$
f_1-f_2 = 0 \quad \mbox{in $\bigcup_{j\in\{ 1,2\}} \bigcup_{F\in
\mathcal U_j} \mathcal O_{F}$}.
$$
}\end{corollary}

\begin{corollary} {\it Let $q_1, q_2 \in
C^{2+\alpha}(\overline{\Omega})$ and let functions
$f_1,f_2 \in C^{3+\alpha}(\Bbb R^1)$ be independent of the variable $x$
with some $\alpha\in (0,1)$, and satisfy (\ref{00}),
(\ref{01}) and (\ref{02}). Suppose that
$\Lambda_{q_1,f_1}=\Lambda_{q_2,f_2}.$ Then $q_1=q_2$ and $f_1 = f_2$
in $\Omega$.}
\end{corollary}
In fact, since $f_1, f_2$ are independent of $x$, Theorems \ref{liska}
and \ref{liska1} yields the conclusion.
%
\\
\vspace{0.3cm}
{\bf Remark 3.} {\it Under the condition of Theorem
\ref{liska}, we can not completely recover the nonlinear term.
Indeed, if $\rho\in C^2(\overline\Omega)$,
$\rho\vert_{\partial\Omega}=0$, $\frac{\partial\rho}{\partial\nu}<0$
on $\partial\Omega$ and $\rho>0$ in $\Omega,$ under assumptions
(\ref{00}) and (\ref{01}), we have the following a priori estimate
proved in \cite{FI}:
$$
\int_\Omega \rho^{\kappa}(\vert \nabla u\vert^2+\vert
u\vert^{p+1})dx \le C
$$
for $u \in H^1(\Omega)$ satisfying $P(x,D)u=0$ in $\Omega$. Here a
constant $C$ is independent of $u$ and $\kappa$ depends on $p.$ Such
a estimate immediately implies that for any $\Omega_1\subset\subset
\Omega$, there exists a constant $C(\Omega_1)>0$ such that
$$
\Vert u\Vert_{C^0(\overline{\Omega_1})}\le C(\Omega_1).
$$
This estimate and (\ref{01}) imply that for any $x\in \Omega_1$  a
nonlinear term $f(x,y)$ in general can not be recovered for all
sufficiently large $y$. }
\vspace{0.3cm}

The uniqueness results for recovery of the nonlinear term in the
semilinear elliptic equation were  first obtained for the case
$\Gamma_0= \emptyset$ in three or higher dimensional cases  by
Isakov and Sylvester in \cite{IN} and in two dimensional case by
Isakov and Nachman in \cite{INa}. It should be mentioned that
their papers requires
the uniqueness of solution  for the Dirichlet boundary problem for
the operator $P(x,D)$. Later, by Isakov in \cite{IS2}, this result
was extended to the case of a system of semilinear elliptic
equations with Dirichlet-to-Neumann map on a certain subboundary.
Also see Kang and Nakamura \cite{KN} for determination of
coefficients of the linear and the quadratic nonlinear terms in the
principal part of a quasilinear elliptic equation.
As for the determination of quasilinear part, see Sun \cite{Sun}.
In a special case
where a nonlinear term is independent of $x$, the uniqueness was
proved in determining such a nonlinear term from partial Cauchy data
\cite{IS1}.  Moreover we note that in \cite{IS1} and \cite{IN}, the
monotonicity of $f(x,u)$ with respect to $u$ is assumed. In general,
if a nonlinear term depends on  $x$, $u$ and the gradient of $u$,
then it is impossible to prove the uniqueness even for the linear
case. This can be seen by \cite {IY1} if we consider the term
$-f(x,u,\nabla u) = A(x)\cdot \nabla u + q(x)u$.

Theorem \ref{liska1} is concerned with the determination of
nonlinear terms and the proof needs a different ingredient from
any previous arguments.  Thus for completeness, we describe the proof
from \cite{IY3} with modifications.

{\bf Proof of Theorem \ref{liska1}.}
 We set $P_k(x,D)u = \Delta u
+ q(x)u - f_k(x,u)$, $k=1,2$, and $u_{1,t}(x)=u(x,t)\in
C([0,1];C^{2+\alpha}(\overline\Omega))$ for some $\alpha\in (0,1)$ be a
function such that any $t$ function $u_{1,t}(x)$, $t\in [0,1]$ solves
the boundary value problem:
$$
P_1(x,D)u_{1,t}=0\quad\mbox{in}\,\,\Omega, \quad
u_{1,t}\vert_{\Gamma_0}=0.
$$
Let $u_{2,t}\in H^1(\Omega)$, $t\in [0,1]$ satisfy
$$
P_2(x,D)u_{2,t}=0\quad\mbox{in}\,\,\Omega, \quad u_{2,t} = u_{1,t}
\quad \mbox{on $\partial\Omega$}, \quad \forall t \in [0,1].\quad
u_{2,\tau}\in C([0,1];H^1(\Omega)).
$$
Then $\Lambda_{q,f_1} = \Lambda_{q,f_2}$ yields
$$
\left( \frac{\partial u_{1,t}}{\partial\nu} - \frac{\partial
u_{2,t}}{\partial\nu}\right) \vert_{\partial\Omega\setminus
\Gamma_0}=0, \quad \forall t\in [0,1].
$$
By (\ref{02}) and the Sobolev embedding theorem,
$f_2(\cdot,u_{2,t}(\cdot))\in L^{\kappa}(\Omega)$ for any
$\kappa>1.$ The standard solvability theory for the Dirichlet
boundary value problem for the Laplace operator in Sobolev spaces
implies $u_{2,t}\in H^2(\Omega).$ Hence
$f_2(\cdot,u_{2,t}(\cdot))\in C^{\widetilde\alpha}(\Omega)$ for any
$\widetilde \alpha\in (0,1).$ Then, since $u_{2,t}\in
C^{2+\alpha}(\partial\Omega)$, the solvability theory for the
Dirichlet boundary value problem for the Laplace operator in
H\"older spaces implies $u_{2,t}\in C^{2+\alpha}(\overline\Omega).$
By the assumption, there exists a constant $K>0$ such that
\begin{equation}\label{-10}
\sup_{t\in [0,1]}\Vert
u_{1,t}\Vert_{C^{2+\alpha}(\overline\Omega)}\le K.
\end{equation}
Next we show that \begin{equation}\label{zona} u_{2,t}\in
C([0,1];C^{2+\alpha}(\Omega)).
\end{equation}
Indeed, suppose that at some point $t_0\in [0,1]$ the function
$u_{2,\tau}$ is discontinuous. Then there exists a sequence
$t_j\rightarrow t_0$ such that
$$
\lim_{j\rightarrow +\infty}\Vert
u_{2,t_j}-u_{2,t_0}\Vert_{C^{2+\alpha}(\overline\Omega)}\ne 0.
$$
Without loss of generality, by (\ref{-10}) we can assume that there
exists a function $\widehat u\in H^\frac{9}{5}(\Omega),
u\vert_{\Gamma_0}=0$ such that
$$
u_{2,t_j}\rightarrow \widehat u\quad \mbox{in}\,\, H^\frac{9}{5}(\Omega)
\quad\mbox{as}\,\,t_j\rightarrow +\infty
$$
and
\begin{equation}\label{contradiction}
\widehat u\ne u_{2,t_0}.
\end{equation}

Obviously the function $\widehat u$ satisfies
$$
P_2(x,D)\widehat u=0\quad\mbox{in}\,\,\Omega, \quad\widehat
u\vert_{\Gamma_0}=0.
$$

In addition, since $(u_{2,t},\frac{\partial
u_{2,t}}{\partial\nu})=(u_{1,t},\frac{\partial
u_{1,t}}{\partial\nu})\in C([0,1];C^{2+\alpha}(\partial\Omega)\times
C^{1+\alpha}(\partial\Omega))$, we obtain
$$
(\widehat
u-u_{2,t_0})\vert_{\partial\Omega\setminus\Gamma_0}=\frac{\partial
(\widehat
u-u_{2,t_0})}{\partial\nu}\vert_{\partial\Omega\setminus\Gamma_0}=0.
$$
Therefore $\widehat w=\widehat u-u_{2,t_0}$ satisfies
$$
L_{q_*}(x,D)\widehat w=0 \quad\mbox{in}\,\,\Omega, \quad \widehat
w\vert_{\partial\Omega\setminus\Gamma_0}=\frac{\partial \widehat
w}{\partial\nu}\vert_{\partial\Omega\setminus\Gamma_0}=0.
$$

By the classical uniqueness result for the Cauchy problem for the
second-order elliptic equation (see e.g., Chapter XXVIII, \S 28.3 of
\cite{Ho1}, Corollary 2.9, Chapter XIV of \cite{Taylor}) we have
$\widehat w\equiv 0.$ This contradicts (\ref{contradiction}).

We claim that
\begin{equation}\label{eb1}
u_{1,t}\equiv u_{2,t}, \quad \forall t\in [0,1].
\end{equation}
Our proof is by contradiction. Suppose that for some $t_0\in (0,1]$,
this equality fails. Let $t_*$ be the infimum over such $t_0$ when
$u_{1,t} = u_{2,t}$ holds.  Since $u_{1,0}=u_{2,0}$, such infinum exists.

Setting $u_t=u_{2,t}-u_{1,t}$, we have
\begin{equation}\label{ippolit}
\Delta u_t - q_0(t,x)u_t=-f_1(x,u_{1,t})+f_2(x,u_{1,t})\quad
\mbox{in}\,\,\Omega,\quad u_{t}\vert_{\partial \Omega}=0,
\quad\frac{\partial u_t}{\partial\nu}\vert_{\partial\Omega\setminus
\Gamma_0}=0,
\end{equation}
where $q_0(t,x) = -q(x) + \int_0^1\frac{\partial f_2}{\partial y}(x,
(1-s)u_{2,t}(x)+su_{1,t}(x))ds$.

Let $\phi$ be a pseudoconvex function with respect to the principal
symbol of the Laplace operator. Applying the Carleman estimate (\ref{Lolo1})
with boundary term to equation (\ref{ippolit}), there exists $\tau_0$
such that:
$$
\sqrt{\tau}\Vert e^{\tau\phi} u_t\Vert_{H^{1,\tau}(\Omega)}
\le C\Vert e^{\tau\phi}
(f_1(\cdot,u_{1,t})-f_2(\cdot,u_{1,t}))\Vert_{L^2(\Omega)},\quad
\forall \tau\ge \tau_0.
$$
Fixing a large $\tau > 0$ arbitrarily, we have
$$
\Vert u_t\Vert_{H^1(\Omega)}\le C \Vert
f_1(\cdot,u_{1,t})-f_2(\cdot,u_{1,t})\Vert_{L^2(\Omega)},\quad
\forall t \in [0,1],
$$
and by the elliptic estimate, we obtain
\begin{equation}\label{k1}
\Vert u_t\Vert_{H^2(\Omega)}\le C \Vert
f_1(\cdot,u_{1,t})-f_2(\cdot,u_{1,t})\Vert_{L^2(\Omega)},\quad
\forall t \in [0,1],
\end{equation}
where the constant $C > 0$ depends on fixed $\tau$.

Consider the boundary value problem
\begin{eqnarray*}
&&\Delta v_{k,t} + q(x)v_{k,t} - \frac{\partial f_k}{\partial
y}(x,u_{k,t})v_{k,t}
-\widetilde f_k(x,v_{k,t})\\
&=& \Delta v_{k,t} + q(x)v_{k,t} - f_k(x,v_{k,t}+u_{k,t}) +
f_k(x,u_{k,t}) =0\quad\mbox{in}\,\,\Omega,\quad
v_{k,t}\vert_{\Gamma_0}= 0,
\end{eqnarray*}
where $\widetilde f_k(x,w)=f_k(x,w+u_{k,t})-f_k(x,u_{k,t})
-\frac{\partial f_k}{\partial y}(x,u_{k,t})w.$ Obviously the
functions $\widetilde f_k$ satisfy (\ref{00}), (\ref{01}) and
(\ref{02}). Moreover
\begin{equation}\label{gemoroi}
\Lambda_{q-\frac{\partial f_1}{\partial y}(x,u_{1,t}),\widetilde f_1}
=\Lambda_{q-\frac{\partial f_2}{\partial y}(x,u_{2,t}),\widetilde f_2}.
\end{equation}

Indeed, consider the pair  $(w_1,w_2)$ such that $w_2=\Lambda_{q-
\frac{\partial f_1}{\partial y}(x,u_{1,t}),\widetilde f_1}(w_1)$. Let $w
\in H^1(\Omega)$ be the solution to the boundary value problem
$$
\Delta w + qw - \frac{\partial f_1}{\partial y}(x,u_{1,t})w -\widetilde
f_1(x,w)=0\quad\mbox{in}\,\,\Omega,\quad w\vert_{\Gamma_0}= 0,\quad
w\vert_{\partial\Omega\setminus \Gamma_0}=w_1
$$
such that $\frac{\partial w}{\partial\nu}\vert
_{\partial\Omega\setminus \Gamma_0}=w_2$.

On the other hand, the function $w+u_{1,t}$ solves the boundary
value problem
$$
\Delta(w+u_{1,t}) + q(w+u_{1,t}) - f_1(x, w+u_{1,t})=0\quad\mbox{in}
\,\,\Omega,\quad (w+u_{1,t})\vert_{\Gamma_0}= 0.
$$
Let $\widetilde u$ satisfy
\begin{equation}\label{gem}
\Delta \widetilde{u} + q\widetilde{u} - f_2(x, \widetilde u) =0
\quad\mbox{in}\,\,\Omega,\quad \widetilde u\vert_{\Gamma_0}= 0
\end{equation}
and
\begin{equation}\label{gem1}
\widetilde{u} = w + u_{1,t} \quad \mbox{on $\partial\Omega\setminus
\Gamma_0$}.
\end{equation}
In general, a solution to problem (\ref{gem}) and (\ref{gem1}) is not
unique, but thanks to the assumption $\Lambda_{q,f_1} =
\Lambda_{q,f_2}$, we can assume that
$$
\frac{\partial\widetilde{u}}{\partial\nu} =
\frac{\partial(w+u_{1,t})}{\partial\nu} \quad \mbox{on
$\partial\Omega\setminus \Gamma_0$}.
$$
Setting $\widetilde w = \widetilde u - u_{2,t}$, we obtain
$$
\Delta\widetilde  w + q\widetilde w - \frac{\partial f_2}{\partial
y}(x,u_{2,t})\widetilde w -\widetilde f_2(x,\widetilde
w)=0\quad\mbox{in}\,\,\Omega,\quad \widetilde w\vert_{\Gamma_0}= 0.
$$
Then on $\partial\Omega\setminus\Gamma_0$ we have
$$
\widetilde{w} - w = (\widetilde{u} - u_{2,t}) - (\widetilde u -
u_{1,t}) = u_{1,t} - u_{2,t} = 0
$$
and
$$
\frac{\partial \widetilde w}{\partial\nu} - \frac{\partial
w}{\partial\nu} = \frac{\partial \widetilde u}{\partial\nu} -
\frac{\partial u_{2,t}}{\partial\nu} - \frac{\partial
w}{\partial\nu} = \frac{\partial w}{\partial\nu} + \frac{\partial
u_{1,t}}{\partial\nu} - \frac{\partial u_{2,t}}{\partial\nu} -
\frac{\partial w}{\partial\nu} = 0.
$$
Therefore $\widetilde w = w_1$ and $\frac{\partial \widetilde w}
{\partial\nu} = w_2$ on $\partial\Omega\setminus\Gamma_0$.  Hence
 for the pair $(w_1,w_2)$  we have $  w_2
= \Lambda_{q-\frac{\partial f_2}{\partial
y}(x,u_{2,t}), \widetilde f_2}(w_1)$.
We can similarly prove the reverse inclusion, that is,
if $(w_1,w_2)$ is a pair such that
 $w_2=\Lambda_{q-\frac{\partial f_2}{\partial y}
(x,u_{2,t}),\widetilde f_2}(w_1)$, then
there exists a function $w_* \in H^1(\Omega)$ that solves  the boundary
value problem
$$
\Delta w_* + qw_* - \frac{\partial f_2}{\partial y}(x,u_{2,t})w_* -\widetilde
f_2(x,w_*)=0\quad\mbox{in}\,\,\Omega,\quad w_*\vert_{\Gamma_0}= 0,\quad
w_*\vert_{\partial\Omega\setminus \Gamma_0}=w_1.
$$
such that $\frac{\partial
w_*}{\partial\nu}\vert_{\partial\Omega\setminus \Gamma_0}=w_2$.
Therefore we have proved (\ref{gemoroi})

\vspace{0.4cm}

Therefore we can apply Theorem \ref{liska} to this equation.  Hence
we have the uniqueness for the potential, that is,
\begin{equation} \label{eb}
\frac{\partial f_1}{\partial y}(x,u_{1,t}) = \frac{\partial
f_2}{\partial y}(x, u_{2,t}) \quad\mbox{in}\,\,\Omega, \quad \forall
t\in [0,1].
\end{equation}
Denote $\Xi(t)=\Vert
u_{1,t}-u_{1,t_*}\Vert_{C^0(\overline\Omega)}+\Vert
u_{2,t}-u_{2,t_*}\Vert_{C^0(\overline\Omega)}.$ Since $u_{1,t_*}
=u_{2,t_*}$ in $\Omega$, we have $f_1(x,u_{1,t_*})=\Delta u_{1,t_*}
= \Delta u_{2,t_*} = f_2(x,u_{1,t_*})$ in $\Omega.$ Therefore
$$
f_1(x,
u_{1,t}(x))-f_2(x,u_{1,t}(x))=\int_{{u_{1,t_*}(x)}}^{u_{1,t}(x)}\left
( \frac{\partial f_1}{\partial y}(x,s)-\frac{\partial f_2}{\partial
y}(x,s)\right ) ds.
$$
If $s\in (u_{1,t_*}(x),u_{1,t}(x))$, then, by the continuity of
$u_{1,t}(x)$ with respect to $t$ and the intermediate value theorem,
there exists $t_0(s,x)\in [0,t]$ such that $s=u_{1,t_0(s,x)}(x).$
Hence
$$
f_1(x, u_{1,t}(x))-f_2(x,u_{1,t}(x)) =
\int_{{u_{1,t_*}(x)}}^{u_{1,t}(x)}\left( \frac{\partial
f_1}{\partial y}(x,u_{1,t_0(s,x)}(x)) - \frac{\partial f_2}{\partial
y}(x,u_{1,t_0(s,x)}(x))\right)ds.
$$
Applying (\ref{eb}) and (\ref{-10}), we have
\begin{eqnarray}\label{ooooo}
f_1(x,
u_{1,t}(x))-f_2(x,u_{1,t}(x))=\int_{{u_{1,t_*}(x)}}^{u_{1,t}(x)}
\left(\frac{\partial f_2}{\partial y}(x,u_{2,t_0(s,x)}(x)) -
\frac{\partial f_2}{\partial y}(x,u_{1,t_0(s,x)}(x))\right)ds
                                                 \nonumber\\
\le \left\Vert \frac{\partial^2 f_2}{\partial
y^2}\right\Vert_{C^0(\overline\Omega\times [-K,K])} \sup_{\widetilde
t\in
(0,t)}\vert (u_{1,\widetilde t}-u_{2,\widetilde t})(x)\vert \Xi( t)\nonumber\\
\le \left\Vert \frac{\partial^2 f_2}{\partial
y^2}\right\Vert_{C^0(\overline\Omega\times [-K,K])} \sup_{\widetilde
t\in (t_*,t)}\vert (u_{1,\widetilde t}-u_{2,\widetilde t})(x)\vert
\Xi ( t).
\end{eqnarray}
In order to obtain the last inequality, we used the fact that
$u_{1,\widetilde t}-u_{2,\widetilde t}\equiv 0$ for all $\widetilde
t$ from $[0,t_*].$ Therefore inequality (\ref{ooooo}) implies
\begin{equation}\label{k2}
\sup_{\widetilde t\in (t_*,t)}\Vert f_1(x, u_{1,\widetilde
t})-f_2(x,u_{1,\widetilde t})\Vert_{L^2(\Omega)}\le C\Xi (t)
\sup_{\widetilde t\in (t_*,t)}\Vert u_{1,\widetilde
t}-u_{2,\widetilde t}\Vert_{L^2(\Omega)}.
\end{equation}
From (\ref{k1}) and (\ref{k2}), we obtain
$$
\Vert u_t\Vert_{H^{2}(\Omega)}\le  C\Xi (t) \sup_{\widetilde t\in
(t_*,t)}\Vert u_{1,\widetilde t}-u_{2,\widetilde
t}\Vert_{L^2(\Omega)}, \quad \forall\widetilde{t}\in (t_*,t).
$$
This implies that
\begin{equation} \label{popka}
\sup_{\widetilde t\in (t_*,t)}\Vert u_{\widetilde
t}\Vert_{H^{2}(\Omega)}\le  C\Xi (t) \sup_{\widetilde t\in
(t_*,t)}\Vert u_{\widetilde t} \Vert_{L^2(\Omega)}.
\end{equation}
From (\ref{popka}) and the fact that $\Xi (t)$ goes to zero as
$t\rightarrow t_*$, we obtain that there exists $\widehat t>t_*$ such
that $u_{1,t}=u_{2,t}$ for all $t$ from $(t_*,\widehat t)$. We reach
a contradiction.  Equality (\ref{eb1}) is proved and the statement
of the theorem follows from it and (\ref{eb}).  $\blacksquare$
%
%
%
%
%
%
%
%
\section{Uniqueness by Dirichlet-to-Neumann maps for the
Lam\'e equations and the Navier-Stokes equations} We discussed the
uniqueness for inverse boundary value problems for systems for
elliptic equations with the same principal parts in Section 4.
In addition to such elliptic
systems, there are other important elliptic systems in mathematical
physics. In this section, we survey recent results for for the
Lam\'e equations and the Navier-Stokes equations.
\subsection{Three dimensional Lam\'e equations}
Let $\Omega \subset \Bbb R^3$ be a bounded domain with smooth
boundary $\partial\Omega$.  Let $\nu=(\nu_1,\nu_2,\nu_3)$ be the outward
unit normal vector to $\partial\Omega$.

Assume that
$$
\mu(x)>0, \thinspace (3\lambda+2\mu)(x)>0\quad
\mbox{on}\,\,\overline\Omega
$$
and
set
$$
C_{ijk\ell}=\lambda(x)\delta_{ij}\delta_{k\ell}
+\mu(x)(\delta_{ik}\delta_{j\ell}
+\delta_{i\ell}\delta_{jk}),
$$
for $1\le i,j,k,\ell\le 3$, where $\delta_{ij}=0$ if $i\ne j$ and
$\delta_{ii} = 1$.
We call  functions $\lambda$ and $\mu$ the Lam{\' e} coefficients,
$u(x)=(u_1(x), u_2(x), u_3(x))$ is the displacement.
We set
$$
\mathcal L_{\lambda,\mu}(x,D) u =
\left(\sum_{j,k,\ell=1}^3\frac{\partial}{\partial
x_j}\left(C_{1jk\ell}\frac{\partial u_k}{\partial x_\ell}\right),
\sum_{j,k,\ell=1}^3\frac{\partial}{\partial
x_j}\left(C_{2jk\ell}\frac{\partial u_k}{\partial x_\ell}\right),
\sum_{j,k,\ell=1}^3\frac{\partial}{\partial
x_j}\left(C_{3jk\ell}\frac{\partial u_k}{\partial x_\ell}\right)\right).
$$

Let $\Gamma_0$ be an arbitrarily fixed subboundary. We define the
Dirichlet-to-Neumann map $\Lambda_{\lambda,\mu,\Gamma_0}$ on
$\Gamma_0$ as follows.
$$
\Lambda_{\lambda,\mu,\Gamma_0}f =
\left(\sum_{j,k,\ell=1}^3\nu_jC_{1jk\ell}\frac{\partial u_k}{\partial
x_\ell},
\sum_{j,k,\ell=1}^3\nu_jC_{2jk\ell}\frac{\partial u_k}{\partial x_\ell},
\sum_{j,k,\ell=1}^3\nu_jC_{3jk\ell}\frac{\partial u_k}{\partial x_\ell}
\right)\Biggl\vert_{\partial\Omega\setminus\Gamma_0}
$$
where
$$
\mathcal L_{\lambda,\mu}(x,D)u = 0 \quad\mbox{in $\Omega$}, \quad
u\vert_{\Gamma_0} = 0, \quad
u\vert_{\partial\Omega\setminus\Gamma_0} = f.
$$
We are concerned with the uniqueness in determining $\lambda, \mu$ by
$\Lambda_{\lambda,\mu,\Gamma_0}$.

Then we can prove
\begin{theorem}\label{lame}
Let $\Omega \in \Bbb R^3$ be a bounded domain with smooth boundary
and let us assume that
$$
\mbox{$\mu_1,\mu_2$ are some positive constants}
$$
and that $\lambda_1,\lambda_2\in  C^{\infty}(\overline\Omega)$
satisfy $\lambda_1 = \lambda_2$ on $\Gamma_0$.
Then $\Lambda_{\lambda_1,\mu_1,\Gamma_0}
= \Lambda_{\lambda_2,\mu_2,\Gamma_0}$ implies
that $\lambda_1 = \lambda_2$ and $\mu_1 =\mu_2$ in
$\Omega$.
\end{theorem}
For the proof, one refers to Imanuvilov, Uhlmann and Yamamoto
\cite{IUYlame}.

We can similarly formulate the two dimensional case and
Imanuvilov and Yamamoto \cite{IYlame}
recently proved a result similar to Theorem \ref{lame}:

\begin{theorem}\label{lame1}
Let $\Omega \in \Bbb R^2$ be a bounded domain with smooth boundary
and let us assume that
$$
\mbox{$\mu_1,\mu_2$ are some positive constants}
$$
and that $\lambda_1,\lambda_2\in  C^{4}(\overline\Omega).$
Then $\Lambda_{\lambda_1,\mu_1,\Gamma_0}
= \Lambda_{\lambda_2,\mu_2,\Gamma_0}$ implies
that $\lambda_1 = \lambda_2$ and $\mu_1 =\mu_2$ in
$\Omega$.
\end{theorem}

The result of Theorem \ref{lame1} is stronger than Theorem \ref{lame}
for the three dimensional case: no information on
the trace of the Lam\'e coefficients $\lambda_j$ is required on
$\Gamma_0$ and only the finite-order regularity of the Lam\'e
coefficients is assumed.

This inverse problem has been studied since the 90's.
Ikehata \cite{IK} discussed a linearized version of this inverse
problem for the Dirichlet-to-Neumann map on the whole boundary
(i.e., $\Gamma_0 = \emptyset$), and in two dimensions, Akamatsu, Nakamura and
Steinberg \cite{ANS} proved that the Dirichlet-to-Neumann map on the
whole boundary can recover the Lam\'e
coefficients and its normal derivatives of arbitrary orders on
the boundary provided that the Lam\'e coefficients are
$C^{\infty}$-functions.
As for higher dimensional case, see Nakamura and Uhlmann \cite{NU3}.
In \cite{NU1} Nakamura and Uhlmann proved that the Dirichlet-to-Neumann
map on the whole boundary in two dimensions uniquely determines
the Lam\'e coefficients, assuming that they are sufficiently close to
a pair of positive constants.

In the three dimensional case, Eskin and Ralston \cite{Es1} proved the
following uniqueness by Dirichlet-to-Neumann map
$\Lambda_{\lambda,\mu,\emptyset}$ on the whole
boundary:
\begin{theorem}
Let $\lambda_j, \mu_j, \mu_j^{-1}$, $j=1,2$, be in a bounded set $B$ in
$C^k(\overline{\Omega})$ with sufficiently large $k \in \Bbb N$.
Then there exists $\epsilon(B) > 0$ such that
$\Lambda_{\lambda_1,\mu_1,\emptyset}
= \Lambda_{\lambda_2,\mu_2,\emptyset}$ implies
$\lambda_1 = \lambda_2$ and $\mu_1 = \mu_2$ in $\Omega$ provided that
$\Vert \nabla\mu_j\Vert_{C^{k-1}(\overline{\Omega})} < \epsilon(B)$,
$j=1,2$.
\end{theorem}
See also \cite{Es2}.  The proof relies on
construction of complex geometric optics solutions (e.g., Eskin \cite{Es0}).
The proof of Theorem  \ref{lame} is based on \cite{Es1}.
Similar attempt has been done in Nakamura and Uhlmann \cite{NU2}.
We note that all the above works except for \cite{IYlame} needs
the Dirichlet-to-Neumann map on the whole boundary.

\subsection{Navier-Stokes equations}
Let $\Omega \subset \Bbb R^2$ be a bounded domain with smooth
boundary.
We define
$$
P_\mu(u,p) \equiv \biggl(\sum_{j=1}^2(-2\partial_j
(\mu(x)\epsilon_{1j}(u)) + u_j\partial_j u_1
+ \partial_1 p,
\sum_{j=1}^2(-2\partial_j
(\mu(x)\epsilon_{2j}(u)) + u_j\partial_j u_2
+ \partial_2 p\biggr),
$$
where $\epsilon_{ij}(u)=\frac 12 (\partial_j u_i
+ \partial_iu_j)$, $1 \le i,j, \le 2$, and we assume that
$$
\mu \in C^4(\overline\Omega), \quad \mu > 0
\quad \mbox{on $\overline{\Omega}$}.
$$
We define the Dirichlet-to-Neumann map on the whole
boundary by
$$
\widetilde{\Lambda_{\mu}}f = \frac{\partial u}{\partial \nu} \quad
\mbox{on $\partial\Omega$}
$$
where $u \in H^2(\Omega)$ and $p \in H^1(\Omega)$ satisfy
$P_\mu(u,p) = 0$, div $u= 0$ in $\Omega$ and $u\vert_{\partial
\Omega} = f$.

Then we can prove the uniqueness in determining the viscosity by
the Dirichlet-to-Neumann map.
\begin{theorem}
We assume that $\partial_x^{\alpha}\mu_1 = \partial_x^{\alpha}\mu_2$
on $\partial\Omega$ for each multi-index $\alpha$ with
$\vert \alpha\vert \le 1$.
If $\widetilde{\Lambda_{\mu_1}}=\widetilde{\Lambda_{\mu_2}}$, then
$\mu_1=\mu_2$ in $\Omega$.
\end{theorem}
The proof is given in a forthcoming paper.
In the three dimensional case, the uniqueness is proved in
Heck, Li and Wang \cite{Heck} for the Stokes system and
in Li and Wang \cite{Li} for the Navier-Stokes equations.
\section{Appendix}
Here we prove several technical propositions used in the previous
sections.
Let $G\subset \Bbb R^2$ be a bounded domain  with smooth boundary,
$\phi\in C^\infty(\Omega)$ be some function, and $\lambda\in \Bbb
R^1$ be a parameter. Consider the following integral
$$
I(\lambda)=\int_G ge^{i\lambda\phi(x)}dx.
$$
\\
{\bf Definition.} {\it Let $A$ be a symmetric $n\times n$ square
matrix, $A^{-1}$ exists and $\lambda_1, \dots, \lambda_n$ be the eigenvalues of
this matrix counted with the multiplicities. Then}
$$
sgn\, A=\mbox{[number of positive eigenvalues] $-$ [number of negative
eigenvalues].}
$$

Let a function $\phi$ have a finite number of critical points on
$\overline \Omega.$ We denote these points as $\widetilde x_1,\dots ,\widetilde
x_\ell.$ Assume that
\begin{equation}\label{zemlja}
\mbox{det}\, H_{\phi}(x)\ne 0\quad \forall x\in\{ x_1,\dots ,
x_\ell\}.
\end{equation}
The following is proposition proved in \cite{BH}:
\begin{proposition}\label{jabloko}
Let (\ref{zemlja}) hold true. If $g\in C^\infty_0(G)$, then
\begin{equation}\label{ux}
I(\lambda)=\frac{2\pi}{\lambda}\sum_{j=1}^\ell \frac{g(\widetilde
x_j)e^{i\lambda \phi(\widetilde x_j)+\frac{\pi i}{4}\mbox {sgn}\,
H_\phi(\widetilde x_j)}}{\root\of{\mbox{det}\, H_\phi(\widetilde
x_j)}}+o(\frac 1\lambda)\quad \mbox{as}\quad \lambda \rightarrow
+\infty .
\end{equation}
If a function $\phi$ does not have critical points
on $\partial\Omega$ and $g\in C^\infty(\overline G)$, then
\begin{equation}\label{ux}
I(\lambda)=\frac{2\pi}{\lambda}\sum_{j=1}^\ell \frac{g(\widetilde
x_j)e^{i\lambda \phi(\widetilde x_j)+\frac{\pi i}{4}\mbox {sgn}\,
H_\phi(\widetilde x_j)}}{\root\of{\mbox{det}\, H_\phi(\widetilde
x_j)}}+\frac{1}{i\lambda}\int_{\partial G}\frac{g}{\vert\nabla
\phi\vert^2}\frac{\partial\phi}{\partial\nu}
e^{i\lambda\phi(x)}d\sigma+o(\frac 1\lambda)\quad \mbox{as}\quad
\lambda \rightarrow +\infty.
\end{equation}
\end{proposition}

Using Proposition \ref{jabloko}, we prove the following
asymptotic formula.
\begin{proposition}\label{gandonnal1} Let $\Phi$ satisfy (\ref{1'})
and (\ref{22}). For every $g\in C^\infty_0(\Omega)$, we have
\begin{equation}\label{uxx}
\int_\Omega ge^{\tau(\Phi-\overline
\Phi)}dx=\sum_{j=1}^\ell\frac{\pi g(\widetilde x_j)e^{2{\tau}
i\psi(\widetilde x_j)}} {{\tau} \vert(\mbox{det}\thinspace
H_\psi)(\widetilde x_j)\vert^\frac 12}+o(\frac{1}{\tau}) \quad
\mbox{as}\quad \tau\rightarrow +\infty.
\end{equation}
\end{proposition}

{\bf Proof.} Since the function $\Phi$ is holomorphic, the real part
$\phi$ and the imaginary part $\psi$ of $\Phi$ satisfy the Cauchy-Riemann
equations:
$$
\frac{\partial \phi}{\partial x_1}=\frac{\partial \psi}{\partial
x_2}\quad\mbox{and}\quad \frac{\partial \phi}{\partial
x_2}=-\frac{\partial \psi}{\partial x_1}.
$$

Hence $\frac{\partial^2 \phi}{\partial x_1^2}=-\frac{\partial^2
\phi}{\partial x_2^2} $ and the Hessian matrix has the form
$$
H_\phi=\left (\begin{matrix} \frac{\partial^2 \phi}{\partial x_1^2}
&\frac{\partial^2 \phi}{\partial x_1\partial x_2}\\ \frac{\partial^2
\phi}{\partial x_1\partial x_2} & -\frac{\partial^2 \phi}{\partial
x_1^2}
\end{matrix} \right)
$$
and $\mbox{det}\,H_\phi=-(\frac{\partial^2 \phi}{\partial x_1^2})^2
-(\frac{\partial^2
\phi}{\partial x_1\partial x_2})^2.$  Since all the critical points of
the function $\Phi$ are nondegenerating, we have
$$
\mbox{det}\,H_\phi(\widetilde x)<0
$$ if $\widetilde x$ is a critical point of the function $\phi.$ Then
the eigenvalues of the matrix $H_\phi$ are $\pm\root\of{-\mbox{
det}\, H_\phi}.$ Hence $\mbox{sgn}\, H_\phi=0.$
Hence, applying formula
(\ref{ux}) with $\lambda=2\tau$, we obtain (\ref{uxx}).
$\blacksquare$

\begin{proposition}\label{gandonnal}
Let $\Phi$ satisfy (\ref{1'}) and (\ref{22}).
For every $g\in L^1(\Omega)$, we have
$$
\int_\Omega ge^{\tau(\Phi-\overline \Phi)}dx\rightarrow \,\,0\quad
\mbox{as}\quad \tau\rightarrow +\infty .
$$
\end{proposition}

{\bf Proof.} The space $C^\infty_0(\Omega)$ is dense in
$L^1(\Omega)$, and so for any $\epsilon>0$ there exists  a function
$g_\epsilon\in C^\infty_0(\Omega)$ such that $$ \Vert
g-g_\epsilon\Vert_{L^1(\Omega)}\le \epsilon/2.
$$
On the other hand by Proposition \ref{gandonnal1}, we have
$$
\int_\Omega (g-g_\epsilon)e^{\tau(\Phi-\overline \Phi)}dx\rightarrow
\,\,0  \quad
\mbox{as}\quad \tau\rightarrow +\infty .
$$
Then for any positive $\epsilon$ there exists $\tau_\epsilon$ such that
$$
\left\vert\int_\Omega ge^{\tau(\Phi-\overline \Phi)}dx\right\vert
\le \left\vert\int_\Omega g_\epsilon e^{\tau(\Phi-\overline
\Phi)}dx\right\vert
+ \left\vert\int_\Omega (g-g_\epsilon)e^{\tau(\Phi-\overline
\Phi)}dx\right\vert
\le \frac{\epsilon}{2}+\frac{\epsilon}{2}.
$$
The proof of the proposition is complete. $\blacksquare$

We have
\begin{proposition}\label{granata}
Let $g\in C^1 (\partial\Omega\setminus\Gamma_0^*)$ and a holomorphic
function $\Phi$ satisfy (\ref{1'})-(\ref{kk}). Then
\begin{equation}\label{LM}
\int_{\partial\Omega\setminus\Gamma_0^*}ge^{\tau(\Phi-\overline
\Phi)}dx=0\quad\mbox{as}\quad\tau\rightarrow +\infty.
\end{equation}
\end{proposition}
{\bf Proof.} Without loss of generality, using the partition of unity
if necessary, we can assume that
$\partial\Omega\setminus\Gamma_0^*$ is a segment $[c,d].$
Moreover, since $C_0^\infty(c,d)$ is dense in $L^1(c,d)$, we can
assume that $g\in C_0^\infty(c,d).$  Since the function $\Phi$
belongs to $C^2(\overline \Omega)$, the set $\mathcal J$ is closed on
$\partial\Omega\setminus\Gamma_0^*.$ For any positive $\epsilon$,
consider the set $\mathcal J_\epsilon=\{x\in
\partial\Omega\setminus\Gamma_0^*\vert dist(x,\mathcal J)\le \epsilon\}$.
Observe that
$$
\lim _{\epsilon\rightarrow +0}mes (\mathcal J_\epsilon)=0.$$

Then
$$
\int_{\mathcal J_\epsilon} g e^{\tau(\Phi-\overline
\Phi)}dx\le \Vert g\Vert_{C^0(\partial\Omega)}mes (J_\epsilon).
$$

Let $\epsilon$ be sufficiently small. Consider the set
$(\partial\Omega\setminus\Gamma_0^*)\setminus\mathcal
J_{2\epsilon}.$ This set is the union of non-intersecting open
intervals where the distance between any two intervals is greater than or
equal to $2\epsilon.$
Consider an arbitrary interval $(a,b)\subset
(\partial\Omega\setminus\Gamma_0^*)\setminus\mathcal J_{2\epsilon}$
such that
\begin{equation}\label{bruni}
a\in \mathcal
J_{2\epsilon}\cup\{x_-\}\quad \mbox{and} \quad b\in \mathcal
J_{2\epsilon}\cup\{x_+\}.
\end{equation}
Consider a function
$e_{a,b}\in C_0^\infty(a-\epsilon/4, b+\epsilon/4)$ such that
$$
0\le e_{a,b}(x)\le 1\,\,\quad\forall x\in (a-\epsilon/4,
b+\epsilon/4), \quad e_{a,b}\vert_{(a,b)}=1,\quad \vert
e_{a,b}'\vert \le K(\epsilon),
$$
where $K$ is independent of $a,b.$
Then we construct a function  $g_\epsilon$  in the following way:
for any interval $(a,b)$ which satisfies (\ref{bruni}), we set
$g_\epsilon=e_{a,b}g.$ The function $g_\epsilon$ has the following
properties

\begin{equation}\label{soika}
\vert g_\epsilon(x)\vert\le \vert g(x)\vert \quad \forall x\in
\partial\Omega\setminus\Gamma_0^*, \quad g=g_\epsilon
\quad\mbox{on}\quad
(\partial\Omega\setminus\Gamma_0^*)\setminus\mathcal J_{2\epsilon}
\end{equation}
and
\begin{equation}\label{soroka}
g_\epsilon\in
C^1((\partial\Omega\setminus\Gamma_0^*)\setminus\mathcal
J_{2\epsilon}),\quad \mbox{supp}\,g_\epsilon\subset
(\partial\Omega\setminus\Gamma_0^*)\setminus\mathcal
J_{\frac{7\epsilon}{4}}.
\end{equation}

By (\ref{soika}) and (\ref{soroka}), we have

\begin{eqnarray}\label{yoyo}
\int_{\partial\Omega\setminus\Gamma_0^*}ge^{\tau(\Phi-\overline
\Phi)}dx=\int_{(\partial\Omega\setminus\Gamma_0^*)\setminus \mathcal
J_{2\epsilon}}ge^{\tau(\Phi-\overline \Phi)}dx+\int_{ \mathcal
J_{2\epsilon}}ge^{\tau(\Phi-\overline \Phi)}dx\nonumber\\
= \int_{(\partial\Omega\setminus\Gamma_0^*)\setminus \mathcal
J_{2\epsilon}}g_\epsilon e^{\tau(\Phi-\overline \Phi)}dx+\int_{
\mathcal J_{2\epsilon}}ge^{\tau(\Phi-\overline \Phi)}dx \nonumber\\
= \int_{(\partial\Omega\setminus\Gamma_0^*)\setminus \mathcal J_{\frac
74\epsilon}}g_\epsilon e^{\tau(\Phi-\overline
\Phi)}dx+\int_{\mathcal J_{2\epsilon}\setminus \mathcal J_{\frac
74\epsilon}}g_\epsilon e^{\tau(\Phi-\overline \Phi)}dx+\int_{
\mathcal J_{2\epsilon}}ge^{\tau(\Phi-\overline \Phi)}dx .
\end{eqnarray}
By (\ref{soika}) we have
\begin{eqnarray}\label{LLL}
\left\vert\int_{\mathcal J_{2\epsilon}\setminus \mathcal J_{\frac
74\epsilon}}g_\epsilon e^{\tau(\Phi-\overline \Phi)}dx
+ \int_{\mathcal J_{2\epsilon}}ge^{\tau(\Phi-\overline \Phi)}dx \right\vert
\le \left\vert\int_{\mathcal J_{2\epsilon}\setminus \mathcal J_{\frac
74\epsilon}}g_\epsilon e^{\tau(\Phi-\overline
\Phi)}dx\vert+\vert\int_{ \mathcal
J_{2\epsilon}}ge^{\tau(\Phi-\overline \Phi)}dx \right\vert\nonumber\\
\le \int_{\mathcal J_{2\epsilon}\setminus \mathcal J_{\frac
74\epsilon}}\vert g \vert dx+\int_{ \mathcal J_{2\epsilon}}\vert
g\vert dx \le 2\int_{\mathcal J_{2\epsilon}}\vert g \vert dx\le
2\Vert g\Vert_{C^0(\partial\Omega)}mes (J_{2\epsilon}).
\end{eqnarray}
Observe that by (\ref{soroka}) we see
\begin{equation}\label{soika1}
\partial_{\vec\tau}\left(\frac{g_\epsilon\partial_{\vec\tau} \varphi}
{2\tau\vert
\partial_{\vec\tau} \varphi\vert^2}\right )\in
L^1(\partial\Omega\setminus\Gamma_0^*).
\end{equation}
Now we estimate the first term on the right-hand side of
(\ref{yoyo}):
\begin{eqnarray}
\int_{(\partial\Omega\setminus\Gamma_0^*)\setminus \mathcal J_{\frac
74\epsilon}}g_\epsilon e^{\tau(\Phi-\overline
\Phi)}dx=\int_{\partial\Omega\setminus\Gamma_0^*}g_\epsilon
e^{\tau(\Phi-\overline
\Phi)}dx=\int_{\partial\Omega\setminus\Gamma_0^*}g_\epsilon
\frac{\partial_{\vec\tau}
\varphi\partial_{\vec\tau}e^{\tau(\Phi-\overline \Phi)}}{2\tau\vert
\partial_{\vec\tau} \varphi\vert^2}dx             \nonumber\\
= \int_{\partial\Omega\setminus\Gamma_0^*}\partial_{\vec\tau}
\left(\frac{g_\epsilon\partial_{\vec\tau}
\varphi} {2\tau\vert
\partial_{\vec\tau} \varphi\vert^2}\right )e^{\tau(\Phi-\overline
\Phi)}dx .
\end{eqnarray}

By (\ref{soika1}) and Proposition \ref{gandonnal}, we have
\begin{eqnarray}\label{LL}
\int_{(\partial\Omega\setminus\Gamma_0^*)\setminus \mathcal J_{\frac
74\epsilon}}g_\epsilon e^{\tau(\Phi-\overline \Phi)}dx\rightarrow
0\quad\mbox{as}\quad\tau\rightarrow +\infty .
\end{eqnarray}
From (\ref{LL}) and (\ref{LLL}) we obtain (\ref{LM}). $\blacksquare$

\begin{proposition}\label{nikita}
There exists a holomorphic function $w_{0}\in
C^{6+\alpha}(\overline \Omega)$ such that
\begin{equation}\label{xoxo1u}
\lim_{x\rightarrow y}\frac{ \vert w_0(x)\vert}{\vert x-y\vert^{98}}
=0, \quad\forall y\in \mathcal H\cap\Gamma_0^*,\quad w_0(\tilde x)\ne 0.
\end{equation}
\end{proposition}

{\bf Proof.} Let us fix a point $\widetilde x$ from $\mathcal H .$
In order to prove this proposition, it suffices to construct some
holomorphic function $a(z)\in C^7(\overline\Omega)$ which is not
identically equal to any constant, satisfies $Im\, a\vert_{\Gamma_0}=0$
and vanishes at each point of the set $ \mathcal
H\cap\Gamma_0^*$.  Then we set $w_0=a^{100}$ and this is the desired
function.

Let $b(z)$ be a holomorphic function in $\Omega$ such that
$\mbox{Re}\, b\vert_{\Gamma_0}=0$, $\Vert b -N\Vert_{C^0(\overline
{\Gamma_0})}<<1$ and $\vert b(\tilde x)-1\vert\le 14.$  Such a
function exists for any positive $N$ by Proposition 5.1 of
\cite{IUY}. If $b(\widetilde x_{\ell+1})\ne 0$, then we consider the
new function $b_1=b-b^3/b^2(\widetilde x_{\ell+1}).$ Obviously
$$\mbox{Re}\, b_1\vert_{\Gamma_0}=0, \quad b_1(\widetilde
x_{\ell+1})=0$$ and $b_1$ is not identically equal to some constant.
If $b(\widetilde x_{\ell+1})= 0$, then we set $b_1=b.$

If $b_1(\widetilde x_{\ell+2})\ne 0$, then we consider the new function
$b_2=b_1-b_1^3/b_1^2(\widetilde x_{\ell+2}).$ Obviously
$$\mbox{Re}\,
b_2\vert_{\Gamma_0}=0, \quad  b_2(\widetilde x_{\ell+1})=b_2(\widetilde
x_{\ell+2})=0
$$
and $b_2$ is not identically equal to any constant. If
$b_1(\widetilde x_{\ell+2})= 0$, then we set $b_2=b_1.$ Repeating
this procedure $\ell'-2$ times and as the result, we obtain the
holomorphic function $a$ with the prescribed properties, provided
that $N$ is sufficiently large.
 $\blacksquare$

\begin{proposition}\label{govno}
Let $\Omega\subset \Bbb R^2$ be a bounded domain with the smooth boundary,  $\mathcal V\in C^0(\partial\Omega)$ satisfy
$$
\int_{\partial\Omega} \mathcal VP(\nu_1-i\nu_2)d\sigma=0, \quad
\forall P(\overline z)\in H^\frac 12(\Omega).
$$
Then there exist an  antiholomorphic function $\Theta\in H^\frac
12(\Omega)$  such that
$
\Theta\vert_{\partial\Omega}  =\mathcal V.
$

\end{proposition}
{\bf Proof.} Consider the extremal problem:
\begin{equation}\label{AX}
J(\widetilde \Psi)=\Vert \mathcal V
-{\widetilde\Psi}\Vert^2_{L^2(\partial\Omega)}\rightarrow \inf,
\end{equation}
\begin{equation}\label{voron}
 \quad \frac{\partial\widetilde \Psi}{\partial
z}=0\quad\mbox{in}\,\Omega.
\end{equation}
 Denote the unique solution to this extremal problem (\ref{AX}),
(\ref{voron}) by $\widehat{ \widetilde \Psi}$. Applying Lagrange's
principle, we obtain
\begin{equation}\label{ipoa}
\mbox{Re} (\mathcal V -\widehat {\widetilde \Psi}, \widetilde
\delta)_{L^2(\partial\Omega)}=0\quad
\end{equation}
for any $\widetilde\delta$  from $ H^\frac 12(\Omega)$ such that
$$
\frac{\partial\widetilde \delta}{\partial z}= 0\quad\mbox{in}
\,\Omega
$$
and there exists a function $\widetilde P\in H^\frac
12(\Omega)$ such that
\begin{equation}\label{020}
 \frac{\partial\widetilde P}{\partial z}=0\quad\mbox{in}\,\,\Omega,
\end{equation}
\begin{equation}\label{021}
 (\nu_1-i\nu_2) \widetilde P =\overline{\mathcal V   -\widehat
{\widetilde\Psi}}\quad\mbox{on}\,\partial\Omega .
\end{equation}
From (\ref{ipoa}), taking  $\widetilde \delta =\widehat{\widetilde
\Psi}$, we have
\begin{equation}\label{Pona}
\mbox{Re} (\mathcal V -\widehat {\widetilde \Psi},\widehat
{\widetilde \Psi}) _{L^2(\partial\Omega)} =0.
\end{equation}
By (\ref{020}), (\ref{021}) and the assumption of the proposition,
we obtain
\begin{eqnarray*}
\mbox{Re} (\mathcal V -\widehat {\widetilde \Psi},\mathcal V )
_{L^2(\partial\Omega)}
=\mbox{Re} ((\nu_1+i\nu_2)\overline{\widetilde P}, \mathcal V )
_{L^2(\partial\Omega)}
= \mbox{Re} (\overline{\widetilde P}, (\nu_1-i\nu_2)\mathcal V)
_{L^2(\partial\Omega)}=0.
\end{eqnarray*}
By (\ref{lob}) and (\ref{Pona}) we see that
$$
J(\widehat{\widetilde\Psi})=0.
$$ The proof of the proposition is complete.
 $\blacksquare$

\end{document}